\documentclass{aa}  

\usepackage{graphicx}
\usepackage{txfonts}
\usepackage{subfigure}
\usepackage{longtable}
\usepackage{color}
\usepackage{icomma}
\usepackage{bm}
\usepackage{marginnote}

\usepackage{amstext}
\usepackage[normalem]{ulem}

\begin{document}

   \title{CSI 2264: Accretion process in classical T Tauri stars in the young cluster NGC 2264\thanks{Tables 2 and 3 are only available in electronic form at the CDS via anonymous ftp to cdsarc.u-strasbg.fr (130.79.128.5) or via http://cdsweb.u-strasbg.fr/cgi-bin/qcat?J/A+A/}}


   \author{A. P. Sousa
          \inst{1}
          \and
          S. H. P. Alencar\inst{1}
          \and 
           J. Bouvier\inst{2,3}
          \and
           J. Stauffer\inst{4}
          \and
           L. Venuti\inst{2,3}
          \and 
           L. Hillenbrand\inst{5}
          \and
          A.M. Cody\inst{6}
          \and
          P. S. Teixeira\inst{7}
          \and
          M. M. Guimarães\inst{8}
          \and
          P. T. McGinnis\inst{1}
          \and
          L. Rebull\inst{4}
          \and
          E. Flaccomio\inst{9}
          \and 
          G. Fürész\inst{10}
          \and
          G. Micela\inst{9}
          \and
          J. F. Gameiro\inst{11}          
          }

   \institute{Departmento de F\'isica-Icex-UFMG 
               Ant\^onio Carlos, 6627, 31270-90. Belo Horizonte, MG, Brazil\\
              \email{alana@fisica.ufmg.br}
    \and          
    Univ. Grenoble Alpes, IPAG, F-38000 Grenoble, France     
    \and
    CNRS, IPAG, F-38000 Grenoble, France 
    \and
    Spitzer Science Center, California Institute of Technology, 1200 East California Boulevard, Pasadena, CA 91125, USA
    \and
    Astronomy Department, California Institute of Technology, Pasadena, CA 91125, USA 
    \and
    NASA Ames Research Center, Kepler Science Office, Mountain View, CA 94035, USA
    \and
    Universität Wien, Institut für Astrophysik, Türkenschanzstrasse 17, 1180 Vienna, Austria
    \and
    Departmento de F\'isica, Universidade Federal de Sergipe, Aracaju, SE, Brazil
    \and
    INAF–Osservatorio Astronomico di Palermo, Piazza del Parlamento 1, I-90134 Palermo, Italy
    \and
    MIT Kavli Institute for Astrophysics and Space Research, 77 Mass Ave 37-582f, Cambridge, MA 02139, USA
    \and
    Instituto de Astrofísica e Ciências Espaciais and Faculdade de Ciências Universidade do Porto, CAUP, Rua da Estrelas, PT4150-762 Porto, Portugal    
    }
    
   \date{Received May 26, 2015; accepted September 04, 2015}

 
  \abstract
   {NGC 2264 is a young stellar cluster ($\sim 3$ Myr) with 
    hundreds of low-mass accreting stars that allow a detailed 
    analysis of the accretion process taking place in the pre-main sequence.}
   {Our goal is to relate the  photometric and spectroscopic variability of classical 
   T Tauri stars to the physical processes acting  
   in the stellar and circumstellar environment, 
   within a few stellar radii from the star.} 
   {NGC 2264 was the target of a multiwavelength observational campaign  
   with CoRoT, MOST, {\it Spitzer,} and Chandra satellites and  
   photometric and spectroscopic observations from the ground.
   We classified the CoRoT light curves of accreting systems according to their morphology and compared
   our classification to several accretion diagnostics and disk parameters.}
   {The morphology of the CoRoT light curve reflects the evolution of the accretion process and of 
   the inner disk region. Accretion burst stars 
   present high mass-accretion rates and optically thick inner disks. AA Tau-like systems, 
   whose light curves are dominated by circumstellar dust obscuration, 
   show intermediate mass-accretion rates and are located in the transition of thick
   to anemic disks. Classical T Tauri stars with spot-like light curves 
   correspond mostly to systems with a low mass-accretion rate and low mid-IR excess. About $30\%$ of the classical T Tauri 
   stars observed in the 2008 and 2011 CoRoT runs changed their light-curve morphology.
   Transitions from AA Tau-like and spot-like to aperiodic light curves and vice versa were common.  
   The analysis of the $\mathrm{H}\alpha$ emission line variability of 58 accreting stars   
   showed that $8$ presented a periodicity that in a few cases was coincident with the  
   photometric period. The blue and red wings of the $\mathrm{H}\alpha$ line profiles often do not 
   correlate with each other, indicating that they are strongly influenced 
   by different physical processes.  
   Classical T Tauri stars have a dynamic stellar and circumstellar environment 
   that can be explained by magnetospheric accretion and outflow models, including variations from 
   stable to unstable accretion regimes on timescales of a few years.}   
   {}


   \maketitle
\section{Introduction}

Classical T Tauri stars (CTTSs) are young, low-mass stars ($\mathrm{M}_* \leq 2\,\mathrm{M}\odot$),
with spectral types from F to M. They are 
surrounded by a circumstellar disk from which they are still accreting material.
They present strong and broad emission lines in their spectra
and show emission excess with respect to the stellar photosphere that 
goes from the radio to the ultraviolet \citep{2007prpl.conf..479B}.  CTTSs have strong magnetic fields 
\citep{1999ApJ...516..900J,2014MNRAS.437.3202J} that disrupt the accretion disk at a 
few stellar radii from the star and channel the accreting material,
forming accretion columns. The accreting gas hits the stellar surface and creates 
hot spots. CTTSs also present cold spots 
at the stellar surface; these are caused by magnetic activity \citep{1995A&A...299...89B}.
Part of the gas in the inner disk region is ejected as a wind from 
the star-disk system along open magnetic field lines that may form 
collimated jets \citep[e.g.,][]{2006A&A...453..785F}. In a few million years, before reaching 
the main sequence, CTTSs lose their disks and become weak-lined T Tauri 
stars (WTTSs), which no longer show detectable signs of accretion \citep[e.g.,][]{2009IAUS..258..111M}.

A characteristic of CTTSs is the photometric and spectroscopic variability at 
various wavelengths. The photometric variability occurs from X-ray to 
infrared on a timescale from a few minutes to several 
years and is usually irregular \citep[e.g.,][]{1989A&ARv...1..291A}. Some stars, however, 
show periodic behavior, which may 
be due to the presence of stable cold and hot spots on the stellar surface 
or to circumstellar dust extinction, as observed in the classical 
T Tauri star AA Tau \citep{2007A&A...463.1017B,2010A&A...519A..88A}. 
 The analysis of the photometric variability of CTTSs allows the determination
of cold or hot spot characteristics and the typical timescale of the physical 
processes that cause each type of variability (dynamo, accretion, star-disk interaction).
We can also estimate the line-of-sight dust distribution in the inner disk
in favorable star-disk inclinations, when the inner disk occults the star as
the system rotates.
Spectroscopic variations are also present, and emission lines can vary in shape 
and intensity on a timescale of hours to days 
\citep[e.g.,][]{1995AJ....109.2800J,1998AJ....116..455M,2014MNRAS.440.3444C}. 
The study of the spectroscopic variability of CTTSs can be related to the predictions
of magnetospheric accretion models and magnetohydrodynamical simulations,
which may include the accretion and the wind components, as well
as the star-disk interaction.

CTTS also show infrared excess emission that indicates the presence of a circumstellar disk 
\citep[e.g.,][]{2012A&A...540A..83T}. This emission can be used to estimate 
the amount of dust in the system and relate disk and accretion evolution. 
The more evolved the star-disk system, the 
smaller the amount of circumstellar material available, and consequently, the lower 
the emission excess in the infrared. 

Magnetospheric accretion is the standard model to describe the accretion 
process in CTTS \citep{1994ApJ...429..781S,1994ApJ426669H}. Magnetohydrodynamic simulations 
predict that magnetospheric accretion can occur in stable and unstable regimes 
\citep{2008MNRAS.385.1931K, 2008MNRAS.386..673K, 2009MNRAS.398..701K, 2013MNRAS.431.2673K}. 
In the stable regime, accretion occurs through two main accretion 
funnels (one in each hemisphere), and periodic spectroscopic and photometric variations 
are expected \citep{2013MNRAS.431.2673K}, since there is a global organization 
of the accretion geometry.
In the unstable regime, several accretion streams are formed
that appear at random 
locations, creating multiple hot spots on the surface of the star. This accretion 
regime can be maintained by the Rayleigh-Taylor instability, which acts at the 
magnetosphere-disk interface \citep{2013MNRAS.431.2673K} and causes 
stochastic photometric and spectroscopic variability. If the 
Rayleigh-Taylor instability is weak, both the unstable and stable 
accretion regimes may coexist. Accretion will then occur mainly 
through two accretion funnels, but random accretion funnels can also appear
in the magnetosphere \citep{2008MNRAS.386..673K}.  These predictions can be checked
with photometric and spectroscopic observations of CTTSs that span different
timescales, from days to years, as the ones discussed in this work.

We analyze young stars belonging to NGC 2264, a young stellar cluster ($\lesssim3\mathrm{Myr}$)
located at a distance of $\sim760\,\mathrm{pc}$ 
from the Sun \citep{1997AJ....114.2644S}. This cluster shows evidence of ongoing
star formation, such as the presence of molecular outflows and Herbig-Haro objects. 
In the pioneering work of \cite{1954ApJ...119..483H}, 
84 pre-main sequence stars were found in NGC 2264 with $\mathrm{H}\alpha$ emission 
and were classified as T Tauri stars. Since then, and because of its proximity and low extinction in our 
line of sight, NGC 2264 has been the subject of many observational campaigns, from the radio to 
X-rays \citep[e.g.,][]{2002AJ....123.1528R,2004A&A...417..557L,2005AJ....129..829D,2008hsf1.book..966D,2012A&A...540A..83T}. 
About 1000 pre-main sequence stars have already been confirmed as members of NGC 2264 \citep{2009AJ....138.1116S}.

NGC 2264 was observed with the Convection Rotation and planetary Transits (CoRoT) 
satellite during 23 consecutive days in 2008 
(from March 7 to 30), and photometric data with high cadence and high signal-to-noise ratio were obtained for $\sim$ 300 known cluster members. 
The analysis of these data yielded studies in different
research areas such as astroseismology \citep{2013A&A...550A.121Z}, stellar rotation \citep{2013MNRAS.430.1433A}, 
binary systems \citep{2014A&A...562A..50G}, and accretion \citep{2010A&A...519A..88A}. In the latter, it was shown 
that the light curve exhibited by the classical T Tauri star AA Tau, which is due to obscuration of the photosphere 
by circumstellar material in the inner disk region, is common among
the cluster members. This opens the possibility of studying the inner
disk evolution with photometric variability at different wavelengths. It 
was also shown that the CoRoT light curve morphology is related to the inner 
disk evolution and consequently to the accretion process.

A second observational campaign of NGC 2264 was organized in 2011 \citep{Cody2013}, including
optical, infrared (IR), and X-ray simultaneous observations of the accreting and non-accreting 
members of the cluster. We present this in the next section. Using this new data set, we 
discuss the observed photometric and spectroscopic variability and its relation to magnetospheric 
accretion model predictions. 
We classify the CoRoT light curves of accreting systems according 
to their morphology and compare this classification with accretion diagnostics 
such as $\mathrm{H}\alpha$ and $\mathrm{HeI}\,6678\mathring{\mathrm{A}}$ emission from 
FLAMES at the Very Large Telescope (VLT), UV excess from MegaCam
at the Canada-France-Hawaii Telescope (CFHT), 
and disk parameters such as IR excess. A set of non-accreting members of NGC 2264 is 
used as a control sample to define the role of accretion in the observed 
correlations. Results from previous observations of NGC 2264 with the CoRoT satellite in 2008 are compared with
the 2011 campaign to analyze the dynamical nature of accretion and the inner disk evolution. 
We analyze the variability in light-curve morphology between the two CoRoT campaigns 
and relate it to the proposed stable and unstable accretion scenarios in the literature. 
\cite{2014AJ....147...82C} presented a detailed description of the photometric data sets and  
morphological classification of the various types of CTTS light curves observed with CoRoT and
Spitzer in the 2011 campaign.

The paper is organized as follows. In Sect. \ref{sec:data}, we present the 2011 observational data 
of NGC 2264 and the reduction procedures. The CTTS sample selection criteria are presented in 
Sect. \ref{sec:sample}. In Sect. \ref{sec:corot} we discuss the morphological classification 
of the CoRoT light curves, following the classification scheme proposed by \citet{2010A&A...519A..88A}. 
In Sect. \ref{sec:CFHT} we compare the CoRoT light-curve classification to the CFHT photometric analysis 
undertaken by \cite{2014A&A...570A..82V}. In Sect. \ref{sec:AccHa} we analyze the accretion rate obtained 
directly from the $\mathrm{H}\alpha$ line flux.
In Sect. \ref{sec:Hatype} we morphologically classify the $\mathrm{H}\alpha$ line, as proposed by 
\cite{1996A&AS..120..229R}. In Sects. \ref{sec:per_halpha} and \ref{sec:heI} we obtain the periodicities 
of the $\mathrm{H}\alpha$ and HeI $6678\,\mathring{\mathrm{A}}$ lines.
We discuss the presence of unstable and stable accretion regimes in Sect. \ref{sec:Inst}, and we analyze the 
correlation between the different $\mathrm{H}\alpha$ line structures (emissions and absorptions) in Sect. \ref{sec:Matcorr}. 
In Sect. \ref{sec:IntObj} we discuss some interesting objects to be analyzed individually in a future work. 
Section \ref{sec:conclusion} presents our conclusions.                                                                                                                                                                             

\section{Data and reduction}\label{sec:data}

NGC 2264 was observed with the CoRoT satellite during 40 days, from December 2, 
2011 to January 10, 2012, providing exquisite photometry for $\sim$ 500 probable cluster 
members. Simultaneously with the 2011 CoRoT run, the 
cluster was observed with the {\it Spitzer} satellite at 3.6 and 4.5 $\mu m$ for 
30 days and for 3.5 days with the Chandra satellite. This campaign was called 
\textit{Coordinated Synoptic Investigation of NGC 2264} (CSI 2264)\footnote{All of the CoRoT 
and {\it Spitzer} light curves can be viewed and downloaded from 
\textit{http://irsa.ipac.caltech.edu/data/SPITZER/CSI2264/}} , and details about 
the data acquisition and reduction can be found in \citet{2014AJ....147...82C}.   
We also obtained $\it{u-}$ and $\it{r}$ -band observations of the cluster with MegaCam/CFHT
from February 14 to 28, 2012. They were combined with observations of a 
first MegaCam campaign in 2010 to yield photometry in the $\it{ugri}$ bands for
a large sample of stars in the NGC 2264 region. A detailed description of the CFHT
observations is given in \citet{2014A&A...570A..82V}. 

As a complement to the CSI 2264 campaign, we obtained FLAMES spectra at VLT/ESO 
of $58$ CTTSs and $34$ WTTSs. These stars were selected based 
on the CoRoT light-curve classification of 2008 \citep{2010A&A...519A..88A}, to include most of the AA Tau and spot-like systems.
FLAMES is a high- and medium-resolution multi-object spectrograph, covering 
a field of view of $25$' in diameter. We obtained 20 to 22 spectra of each target in two different fields,
distributed in 20 days from December 2011 to February 2012, part of
them simultaneously with the CoRoT observations.
We used the HR15N setup for the FLAMES/GIRAFFE spectrograph, centered at $6650\,\mathring{\mathrm{A}}$ and covering the
$6470\,\mathring{\mathrm{A}}<\lambda<6790\,\mathring{\mathrm{A}}$ spectral region
with a resolution of $R=17,000$. 
This spectral region includes $\mathrm{H}\alpha$, and the $\mathrm{HeI}\,6678\,\mathring{\mathrm{A}}$ 
spectral lines that are analyzed in detail in the next sections.
The observed spectra were reduced using the GASGANO reduction package from ESO. We performed the standard 
procedure of bias subtraction, flat-field correction, wavelength calibration, and barycentric 
velocity correction for all spectra and later normalized their
continuum.
In Table \ref{tab:obs}, we summarize the observations of the CSI 2264
campaign that are discussed in the following sections.

 \begin{table*}
 \centering
 \caption{Observations from the Coordinated Synoptic Investigation of NGC 2264}
 \label{tab:obs}
 \begin{tabular}{llll}
  \hline
  \hline
  Telescope & Instrument & Dates & Bands \\
  \hline
  CoRoT             & E2 CCD         & 2011-Dec-1 to 2012-Jan-3      & $3000-10000\,\mathring{\mathrm{A}}$  \\
  VLT               & FLAMES/GIRAFFE & 2011-Dec-4 to 2012-Feb-29     & $6470-6790\,\mathring{\mathrm{A}}$   \\
  CFHT              & MegaCam        & 2012-Feb-14 to 2012-Feb-28    & \textit{u} and \textit{r}            \\  \hline
 \end{tabular} 
\end{table*}

NGC 2264 is located in front of an emission nebula that is seen spectroscopically
as a series of emission lines, such as [NII], [SII], and $\mathrm{H}\alpha$, superposed 
on the stellar spectra. The $\mathrm{H}\alpha$ nebular contribution is very narrow and in general much
smaller than the accretion contribution to the $\mathrm{H}\alpha$ emission of most CTTSs, presenting
a mean equivalent width of the $\mathrm{H}\alpha$ nebular contribution of $10.1\pm0.5\,\mathring{\mathrm{A}}$, averaged over all the sky fibers.
However, in a few spectra, the nebular emission is completely blended with the emission from the star, 
making it impossible to identify whether a component is of nebular or accretion origin. This sometimes makes 
removing the nebular emission from the total spectrum challenging.
In general, this is a difficult task when using 
fiber spectrographs, as is the case of FLAMES. The nebular contribution is quite variable
across the sky, and the sky fiber closest to the stellar fiber does not always present 
the same nebular contribution as the stellar fiber. Another difficulty is that some targets 
are very weak and present a spectrum comparable in intensity to the sky contribution.

\begin{figure}
 \begin{center}
\includegraphics[width=4.4cm]{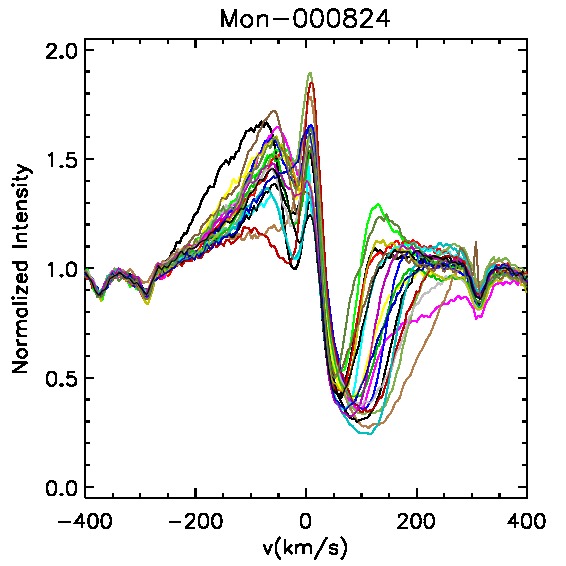}
\includegraphics[width=4.4cm]{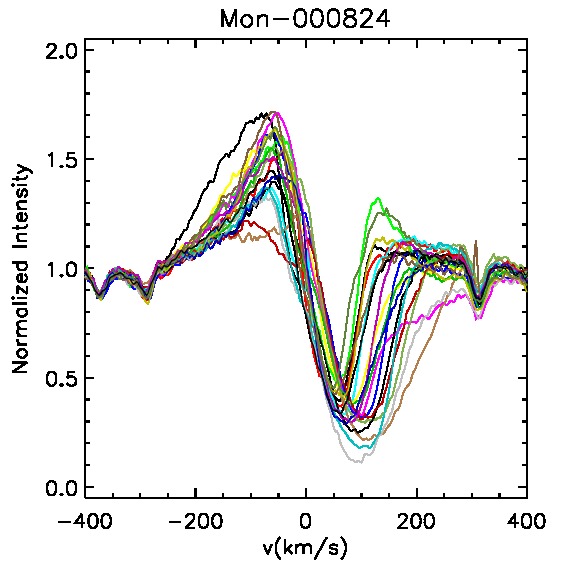}
 \end{center}
\caption{\label{fig:specsky} 
$\mathrm{H}\alpha$ emission lines of a CTTS before (left) and after (right) sky subtraction
that show the removal of the nebular emission contribution.
The nebular component is the very sharp, narrow emission seen near the line center in the left panel. 
Different colors correspond to different observing nights.} 
\end{figure}

To remove the nebular contribution from the stellar spectra, we subtracted the sky
spectrum of the nearest sky fiber from the stellar spectrum of each star (see Fig. \ref{fig:specsky}). 
The subtraction of the nebular emission is sometimes faulty, and a residual 
nebular line often remains. 
Even when the sky subtraction leaves residuals in the stellar spectrum, the contamination
from the nebula is normally distinguishable from the stellar emission because the sky emission 
is narrower and weaker than the emission of most CTTSs.

\section{CTTS and WTTS samples}\label{sec:sample}

We selected the CTTSs in NGC 2264 from the known cluster members according to 
photometric and spectroscopic criteria. 
We measured the $\mathrm{H}\alpha$ equivalent width ($\mathrm{EW}_{\mathrm{H}\alpha}$) 
and $\mathrm{H}\alpha$ width at $10\%$ of maximum intensity ($\mathrm{W}10\%_{\mathrm{H}\alpha}$) 
in the FLAMES spectra after removing the nebular contribution.
We used published values of these parameters for the stars we did not observe
spectroscopically, as shown in Table \ref{tab:ctts}.
To test the influence of the nebular emission on the measured values of $\mathrm{EW}_{\mathrm{H}\alpha}$, 
we calculated the ${\mathrm{H}\alpha}$ equivalent width before and after subtracting the nebular part, from which we
obtained a mean difference of about $10\%$. This difference does not affect our 
CTTS classification.
 
According to \cite{2003ApJ...582.1109W}, CTTS have an $\mathrm{EW}_{\mathrm{H}\alpha}$ 
higher than a threshold that depends on the stellar spectral type. A young star is 
considered a CTTS if it presents an $\mathrm{EW}_{\mathrm{H}\alpha}$ higher than $3 \, \mathring{\mathrm{A}}$ 
for K0-K5, $10 \, \mathring{\mathrm{A}}$ for K7-M2.5, $20 \, 
\mathring{\mathrm{A}}$ for M3-M5.5, and $40 \, \mathring{\mathrm{A}}$ for M6-M7.5. 
Stars that show $\mathrm{W}10\%_{\mathrm{H}\alpha}$ higher than 270 km/s are also considered CTTSs, 
following \cite{2003ApJ...582.1109W}. \cite{2009A&A...504..461F} proposed more stringent
limits to separate WTTSs from CTTSs that extend to late-M spectral type. 
Since we do not have stars in this spectral type region, we decided to use the 
criteria proposed by \cite{2003ApJ...582.1109W} to select CTTS. We selected 
$220$ CTTSs in NGC 2264 that satisfy at least one of the above spectroscopic criteria. 

\begin{figure}
 \begin{center}
\includegraphics[scale=0.28]{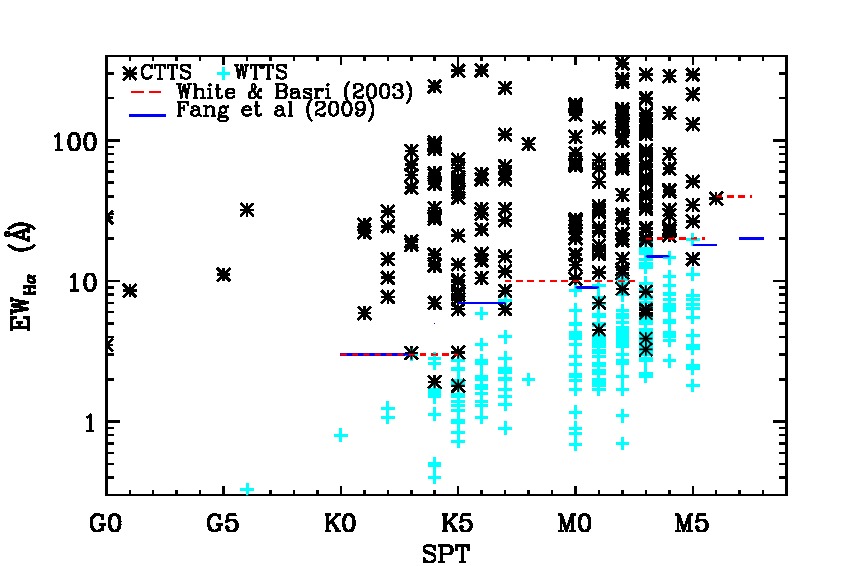}
\includegraphics[scale=0.28]{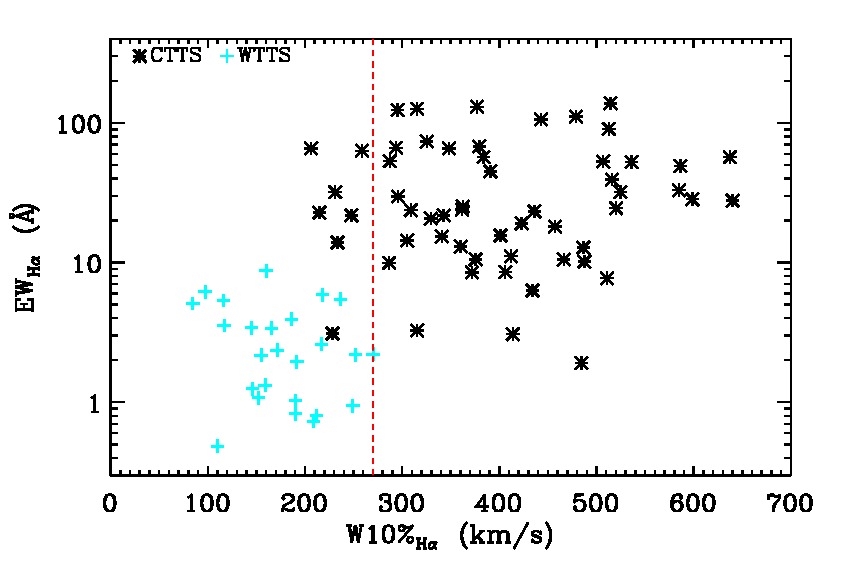}
 \end{center}
\caption{\label{fig:spt} Spectroscopic criteria used to select CTTSs and WTTSs in NGC 2264.
CTTSs are shown as asterisks and WTTSs as plus signs.
Top: $\mathrm{H}\alpha$ equivalent width vs. spectral type for our sample of stars. 
Dashed horizontal lines represent the \cite{2003ApJ...582.1109W} classification
threshold and solid horizontal lines the \cite{2009A&A...504..461F} proposed criteria.
Bottom: $\mathrm{H}\alpha$ equivalent width vs. $\mathrm{H}\alpha$ width at 
$10\%$ of maximum intensity. The dashed vertical line is drawn at $270\,\mathrm{km/s}$,
the separation between CTTSs and WTTSs, according to \cite{2003ApJ...582.1109W}. 
A few stars do not satisfy one spectroscopic CTTS criterion, but they 
were classified as CTTSs either based on the other spectroscopic criterion or on the UV excess.}
\end{figure}

Using UV excess obtained with MegaCam/CFHT, \citet{2014A&A...570A..82V} found $66$ new 
CTTS candidates in NGC 2264. These stars do not have FLAMES spectra and
were not previously classified as CTTS in the literature or are unknown NGC 2264 members, but since 
they present considerable UV excess with respect to the WTTS locus, they were added to our CTTS list. 
This means that based on photometric and spectroscopic accretion criteria, $286$ stars are 
thought to be CTTSs in NGC 2264. 

Mon-000056 and Mon-000358 were initially not included in our CTTS list, but they 
show CoRoT light curves in 2008 and/or 2011 that
resemble the AA Tau light curve (see Sect. \ref{sec:corot}). We would therefore
expect them to be CTTSs, but they were not in the fields of our FLAMES 
observations and show little UV excess.
These stars were nevertheless observed with FLAMES/GIRAFFE as part of the 
Gaia/ESO public spectroscopic survey \citep{2013Msngr.154...47R,gilmore12}. We 
downloaded the spectra from the ESO science archive and reduced them
following the same steps as listed in Sect. \ref{sec:data}.
The Gaia/ESO spectra show that Mon-000056 presents $\mathrm{EW}_{\mathrm{H}\alpha}=2.9\,\mathrm{\mathring{A}}$ 
and $\mathrm{W}10\%_{\mathrm{H}\alpha}=276\,\mathrm{km}/\mathrm{s}$, while
Mon-000358 shows $\mathrm{EW}_{\mathrm{H}\alpha}=32.8\,\mathrm{\mathring{A}}$
and $\mathrm{W}10\%_{\mathrm{H}\alpha}=395\,\mathrm{km}/\mathrm{s}$, which
allows us to classify them as CTTSs. Our final CTTS list therefore contains $288$ stars; it is shown in Table \ref{tab:ctts}. Of these, 
$58$ stars were observed by FLAMES and $157$ were observed by CoRoT in the 2008 and/or 2011 campaign, 
including $12$ CTTSs observed by CoRoT only in 2008, $63$ only in 2011, and $84$ CTTSs that were observed in both campaigns.
\begin{longtab}[!htb]
\tiny
\addtolength{\tabcolsep}{-1pt}  
\begin{longtable}{llllllllllllll}
\caption{\label{tab:ctts} Parameters of the CTTSs observed by CoRoT and/or FLAMES/VLT.}\\
 \hline\hline
Mon ID\tablefootmark{a} & 2Mass ID\tablefootmark{b} & SpT\tablefootmark{c}& $\mathrm{EW}_{\mathrm{H}\alpha}$\tablefootmark{d} & $\mathrm{Er}_\mathrm{EW}$\tablefootmark{d} & $\mathrm{W}10\%_{\mathrm{H}\alpha}$\tablefootmark{d} & $\mathrm{EW}_{\mathrm{H}\alpha}$\tablefootmark{e}  & FUR06\tablefootmark{f} & $\alpha_{\mathrm{IRAC}}$ \tablefootmark{g} & $\mathrm{LC}$\tablefootmark{h} & $\mathrm{P}$\tablefootmark{h} & $\mathrm{LC}$\tablefootmark{h}  &  $\mathrm{P}$\tablefootmark{h} & Q\tablefootmark{i} \\
 & &  & $(\mathring{\mathrm{A}})$ & $(\mathring{\mathrm{A}})$ & $(\mathrm{km}\mathrm{s}^{-1})$  &$(\mathring{\mathrm{A}})$ & & $(2008)$ & $(\mathrm{days})$ & $(2011)$  & $(\mathrm{days})$ & \\
\hline
\endfirsthead
\caption{Continued.} \\
\hline
Mon ID\tablefootmark{a} & 2Mass ID\tablefootmark{b} & SpT\tablefootmark{c}& $\mathrm{EW}_{\mathrm{H}\alpha}$\tablefootmark{d} & $\mathrm{Er}_\mathrm{EW}$\tablefootmark{d} & $\mathrm{W}10\%_{\mathrm{H}\alpha}$\tablefootmark{d} & $\mathrm{EW}_{\mathrm{H}\alpha}$\tablefootmark{e} & FUR06\tablefootmark{f} & $\alpha_{\mathrm{IRAC}}$ \tablefootmark{g} & $\mathrm{LC}$\tablefootmark{h} & $\mathrm{P}$\tablefootmark{h} & $\mathrm{LC}$\tablefootmark{h}  &  $\mathrm{P}$\tablefootmark{h} & Q\tablefootmark{i} \\
 & &  & $(\mathring{\mathrm{A}})$ & $(\mathring{\mathrm{A}})$ & $(\mathrm{km}\mathrm{s}^{-1})$  &$(\mathring{\mathrm{A}})$ & & $(2008)$ & $(\mathrm{days})$ & $(2011)$  & $(\mathrm{days})$ & \\
\hline
\endhead
\hline
\endfoot
\hline
\endlastfoot
 CSIMon-000007  &  06415304+0958028  &  K7     &        &       &    &        &  c  &  -1.47  &  3 (A) &         &  3 (A) &         & 0.75 \\
 CSIMon-000011  &  06411725+0954323  &  K7     &        &       &    &  58.3  &     &         &  3 (A) &         &  3 (A) &         &      \\
 CSIMon-000017  &  06413199+1000244  &  K5     &        &       &    &  13.1  &     &  -2.69  &     1  &   4.78  &     1  &   4.73  &-0.03 \\
 CSIMon-000021  &  06405944+0959454  &  K5     &        &       &    &   7.4  &  c  &         &     2  &   3.23  &     2  &   3.16  & 0.15 \\
 CSIMon-000056  &  06415315+0950474  &  K5     &        &       &    &   1.8  &  w  &  -1.36  &     2  &   5.71  &     2  &   5.86  & 0.14 \\
 CSIMon-000058  &  06420870+0941212  &  K4.5   &        &       &    &  94.0  &  c  &         &     -  &         &     1  &   2.14  & 0.22 \\
 CSIMon-000063  &  06411193+0959412  &  M2.5   &        &       &    &  19.4  &     &  -1.37  &     -  &         &     3  &         &      \\
 CSIMon-000090  &  06410896+0933460  &  M3     &        &       &    &  51.0  &  c  &  -0.88  &     3  &         &     3  &         &      \\
 CSIMon-000103  &  06405954+0935109  &  K6     &  15.6  &   0.9 & 401&   6.4  &  c  &  -1.14  &     1  &   1.67  &     1  &   3.35  & 0.21 \\
 CSIMon-000117  &  06405413+0948434  &  M2.5   &        &       &    & 353.0  &     &  -2.34  &     -  &         &  3 (A) &         & 0.72 \\
 \hline                                                                                            
\end{longtable}
\tablefoot{Only a portion of this table is shown here. A full version is available in the online journal. This table is ordered according to the Mon ID.}\\
\tablefoottext{a}{``CSIMon'' is an internal identification of the CSI 2264 campaign. Elsewhere in the text "CSI" was omitted for brevity.}
\tablefoottext{b}{2MASS identification.}
\tablefoottext{c}{Spectral type obtained by \cite{2014A&A...570A..82V,2005AJ....129..829D,2002AJ....123.1528R,1956ApJS....2..365W}.}
\tablefoottext{d}{$\mathrm{H}\alpha$ parameters obtained in this work using FLAMES spectra. We used the convention that positive $\mathrm{H}\alpha$ equivalent width indicates $\mathrm{H}\alpha$ in emission, and negative values indicate $\mathrm{H}\alpha$ in absorption. The uncertainties were obtained assuming a Poisson distribution.}
\tablefoottext{e}{$\mathrm{H}\alpha$ equivalent width obtained by \cite{2005AJ....129..829D}.}
\tablefoottext{f}{Classification as CTTS (c) and WTTS (w) by \cite{2006ApJ...648.1090F}.}
\tablefoottext{g}{$\alpha_{IRAC}$ is the slope of the spectral energy distribution between $3.6\,\mu\mathrm{m}$ and $8\,\mu\mathrm{m}$ obtained by \cite{2012A&A...540A..83T}.}
\tablefoottext{h}{CoRoT light curve morphology and photometric period obtained in this work: ``1''= spot-like, ``2''= AA Tau-like, ``3''= non-periodic light curves. Accretion bursts and aperiodic extinction light curves are identified by (A) and (E), respectively. ``UNC'' = unclassified light curve and ``B''= Binary.
\tablefoottext{i}{Parameter defined by \cite{2014AJ....147...82C} to distinguish periodic from aperiodic light curves. According to \cite{2014AJ....147...82C}, when Q>0.6 the star is not periodic photometrically.}
}
\end{longtab}
\begin{longtab}[!htb]
\tiny
\begin{center}
\begin{longtable}{lllllllllllll}
\caption{\label{tab:wtts}  Parameters of the WTTSs observed by CoRoT and/or FLAMES/VLT.}\\
\hline\hline
Mon ID\tablefootmark{a} & 2Mass ID\tablefootmark{b} & SpT\tablefootmark{c}& $\mathrm{EW}_{\mathrm{H}\alpha}$\tablefootmark{d} & $\mathrm{Er}_\mathrm{EW}$\tablefootmark{d} & $\mathrm{W}10\%_{\mathrm{H}\alpha}$ \tablefootmark{d} & $\mathrm{EW}_{\mathrm{H}\alpha}$\tablefootmark{e} & FUR06\tablefootmark{f} & $\alpha_{\mathrm{IRAC}}$\tablefootmark{g} \\ 
&   &  & $(\mathring{\mathrm{A}})$ & $(\mathring{\mathrm{A}})$ & $(\mathrm{km}\mathrm{s}^{-1})$  &$(\mathring{\mathrm{A}})$ & \\ 
\hline
\endfirsthead
\caption{Continued.} \\
\hline
Mon ID\tablefootmark{a} & 2Mass ID\tablefootmark{b} & SpT\tablefootmark{c}& $\mathrm{EW}_{\mathrm{H}\alpha}$\tablefootmark{d} &  $\mathrm{Er}_\mathrm{EW}$\tablefootmark{d} &$\mathrm{W}10\%_{\mathrm{H}\alpha}$ \tablefootmark{d} & $\mathrm{EW}_{\mathrm{H}\alpha}$\tablefootmark{e} & FUR06\tablefootmark{f} & $\alpha_{\mathrm{IRAC}}$\tablefootmark{g} \\ 
&   &  & $(\mathring{\mathrm{A}})$  & $(\mathring{\mathrm{A}})$ & $(\mathrm{km}\mathrm{s}^{-1})$ & $(\mathring{\mathrm{A}})$& \\ 
\hline
\endhead
\hline
\endfoot
\hline
\endlastfoot
 CSIMon-000008  &  06414859+0954115  &  K5     &        &      &      &         &     &         \\
 CSIMon-000009  &  06420914+0948048  &  F5     &        &      &      &         &     &         \\
 CSIMon-000014  &  06420664+0941317  &  K7:M0  &        &      &      &         &     &         \\
 CSIMon-000015  &  06420911+0959027  &  K7:M0  &        &      &      &         &     &         \\
 CSIMon-000018  &  06411322+0955086  &  K3:K4  &        &      &      &    3.0  &     &  -2.73  \\
 CSIMon-000020  &  06420924+0944034  &  K7     &        &      &      &         &  w  &         \\
 CSIMon-000022  &  06410457+1000426  &  M4     &        &      &      &         &     &         \\
 CSIMon-000024  &  06415684+0947451  &  M1     &        &      &      &         &     &         \\
 CSIMon-000026  &  06410518+1000189  &  M4     &        &      &      &    6.5  &     &         \\
 CSIMon-000029  &  06410328+0957549  &  K7     &        &      &      &    1.5  &     &  -2.85  \\

  \hline
\end{longtable}
\end{center}
\tablefoot{Only a portion of this table is shown here. A full version is available in the online journal. This table is ordered according to the Mon ID.}
\tablefoottext{a}{``CSIMon'' is an internal identification of the CSI 2264 campaign.}
\tablefoottext{b}{2MASS identification.}
\tablefoottext{c}{Spectral type obtained by \cite{2014A&A...570A..82V,2005AJ....129..829D,2002AJ....123.1528R,1956ApJS....2..365W}.}
\tablefoottext{d}{$\mathrm{H}\alpha$ parameters obtained in this work using FLAMES spectra. We used the convention that positive $\mathrm{H}\alpha$ equivalent width indicates $\mathrm{H}\alpha$ in emission, and negative values indicate $\mathrm{H}\alpha$ in absorption. The uncertainties were calculated assuming a Poisson distribution.}
\tablefoottext{e}{$\mathrm{H}\alpha$ equivalent width obtained by \cite{2005AJ....129..829D}.}
\tablefoottext{f}{Classification as CTTS (c) and WTTS (w) by \cite{2006ApJ...648.1090F}.}
\tablefoottext{g}{$\alpha_{IRAC}$ is the slope of the spectral energy distribution between $3.6\,\mu\mathrm{m}$ and $8\,\mu\mathrm{m}$ obtained by \cite{2012A&A...540A..83T}.}
\end{longtab}

We also selected $431$ WTTSs among the confirmed members of NGC 2264 that we used as a comparison sample to the accreting CTTS. We added
$257$ WTTSs to the list based on the $\mathrm{H}\alpha$ equivalent 
width or $\mathrm{W}10\%_{\mathrm{H}\alpha}$ criteria of \cite{2003ApJ...582.1109W},
previously discussed, and $174$ WTTSs were selected based on the criteria discussed 
in \cite{2014A&A...570A..82V}. 
Information about the WTTS sample can be found in Table \ref{tab:wtts}. Of the 
$431$ WTTSs, $34$ were observed by FLAMES and $308$ stars were observed by CoRoT 
in the 2008 and/or 2011 campaign. We did not analyze the CoRoT data for WTTSs; that
is beyond the scope of this paper.

In Fig. \ref{fig:spt}, we plot the spectroscopic criteria used to distinguish CTTSs 
from WTTSs, as proposed by \cite{2003ApJ...582.1109W} and \cite{2009A&A...504..461F}, together
with the $\mathrm{H}\alpha$ data for the $58$ CTTS and $34$ WTTS
observed with FLAMES. The spectroscopic data of the stars that were not observed with FLAMES 
come from \cite{2005AJ....129..829D}.
A few stars do not satisfy the $\mathrm{EW}_{\mathrm{H}\alpha}$ CTTS criteria, but they were classified as CTTSs 
based on  $\mathrm{W}10\%_{\mathrm{H}\alpha}$ (Fig. \ref{fig:spt}, bottom panel) and/or UV excess. 

\section{Morphology of the CoRoT light curves} \label{sec:corot}

\begin{figure}
 \begin{center}
\subfigure[]{\label{fig:2a}\includegraphics[width=4.4cm]{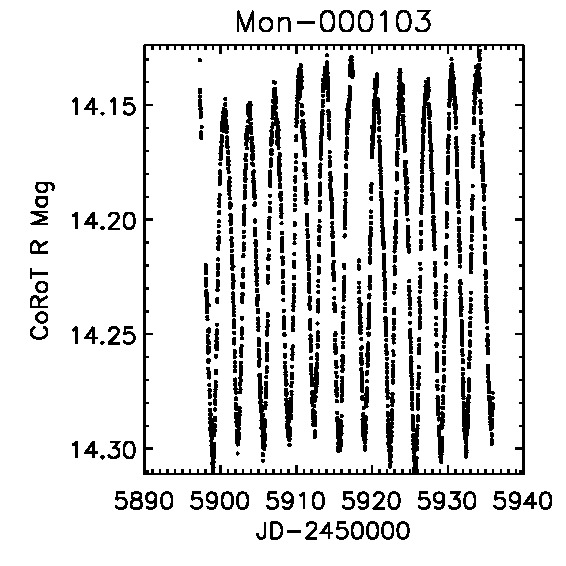}}
\subfigure[]{\label{fig:2b}\includegraphics[width=4.4cm]{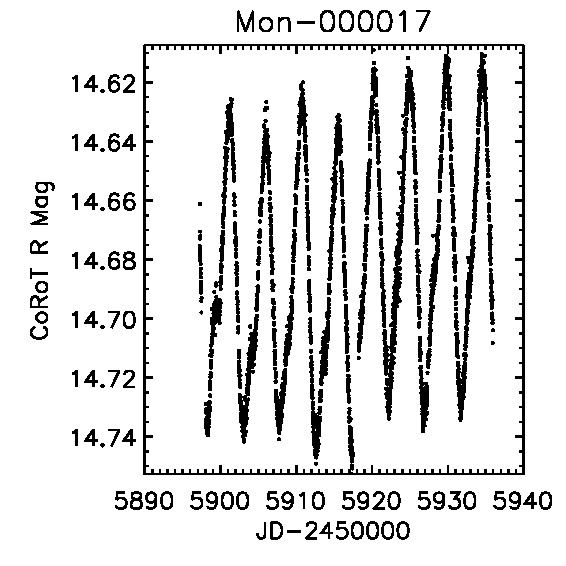}}\\
\subfigure[]{\label{fig:2c}\includegraphics[width=4.4cm]{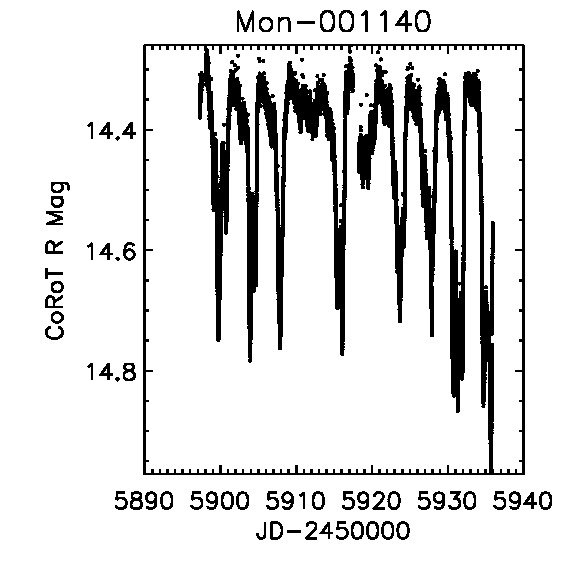}}
\subfigure[]{\label{fig:2d}\includegraphics[width=4.4cm]{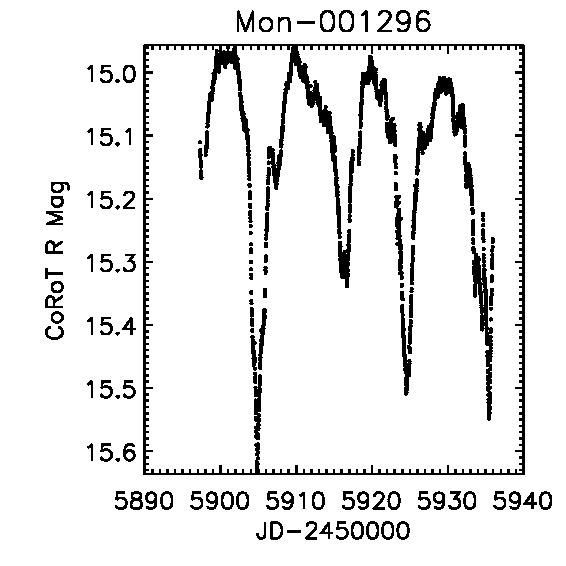}}\\
\subfigure[]{\label{fig:2e}\includegraphics[width=4.4cm]{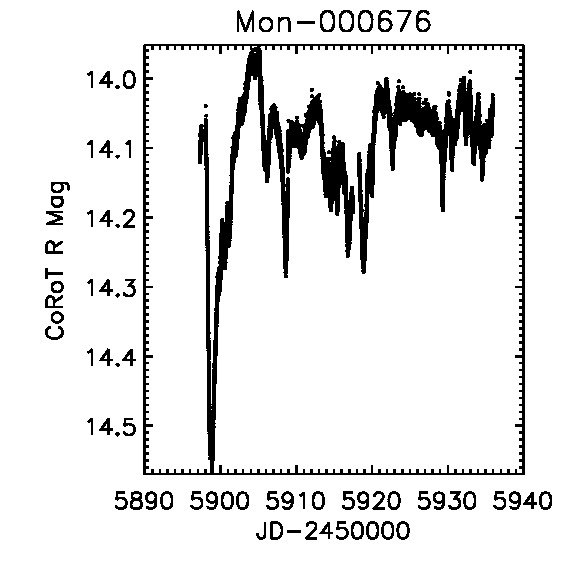}}
\subfigure[]{\label{fig:2f}\includegraphics[width=4.4cm]{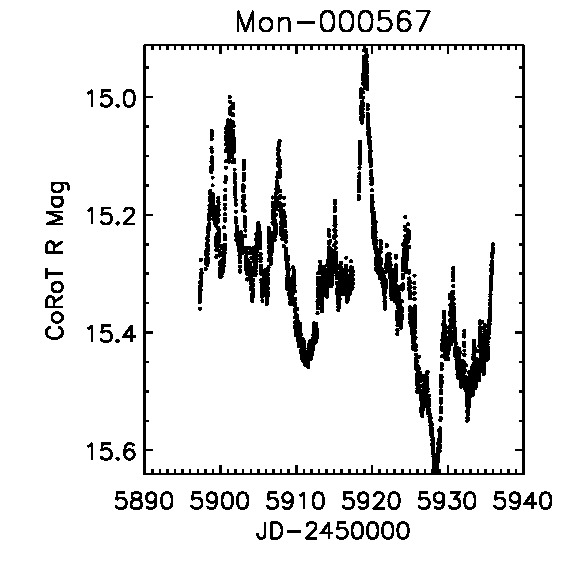}}
 \end{center}
\caption{\label{fig:LC} CoRoT light curves of CTTSs from 2011 showing different photometric 
variabilities. Panels (a) and (b) correspond to spot-like light curves, (c) and (d) to 
AA Tau-like light curves, (e) is a non-periodic light 
curve caused by obscuration by circumstellar material, and (f) is a non-periodic 
light curve dominated by accretion bursts. The light-curve magnitude was calibrated 
using the $R$ filter with a zero point of $26.74$ mag, as described in \citet{2014AJ....147...82C}.} 
\end{figure}

The 2008 CTTS light curves were previously classified morphologically as spot-like, AA Tau-like, or non-periodic by
\cite{2010A&A...519A..88A}. The spot-like light curves display sinusoidal 
variations that are due to large, cold spots at the stellar surface, which are very stable 
on the timescale of the observations (about three weeks). The AA Tau-like light curves show a well -defined maximum interrupted by periodic minima that vary in width and depth from 
cycle to cycle, like AA Tau itself. They correspond to systems whose light-curve variability is 
caused by a periodic occultation of the stellar photosphere
by circumstellar dust \cite[see][for more information about 
AA Tau-like stars in NGC 2264 and \cite{Fonsecapreparation} for a specific AA Tau-like 
system, V354 Mon - Mon-000660]{2015A&A...577A..11M,2010A&A...519A..88A}. Non-periodic 
light curves may be due, for example, to accretion bursts \citep{2014AJ....147...83S}
or random circumstellar dust obscuration. Some non-periodic light curves are difficult to assign to a major physical phenomenon, however, and are probably
the result of different superposed variability processes, such
as variable accretion, 
dust obscuration, and spot variations. We show below by comparing
our light-curve classification with accretion diagnostics and inner disk 
parameters that the
morphological 
classification of the CoRoT light curves is related to the evolution
of the 
accretion process and of the inner disk region.

\begin{figure}
 \begin{center}
\subfigure[]{\includegraphics[width=4.3cm]{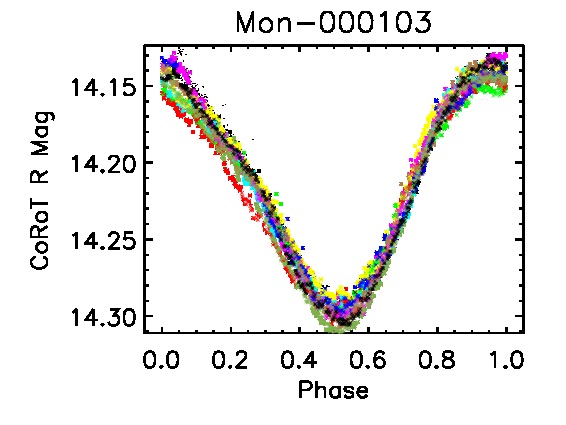}}
\subfigure[]{\includegraphics[width=4.3cm]{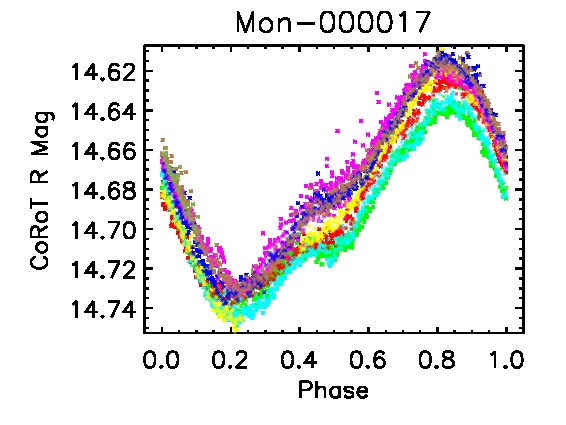}}\\
\subfigure[]{\includegraphics[width=4.3cm]{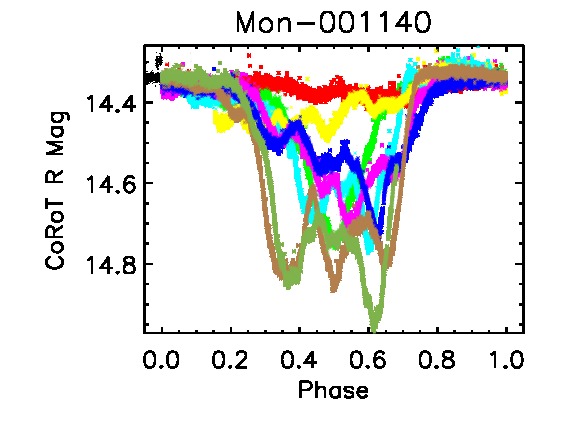}}
\subfigure[]{\includegraphics[width=4.3cm]{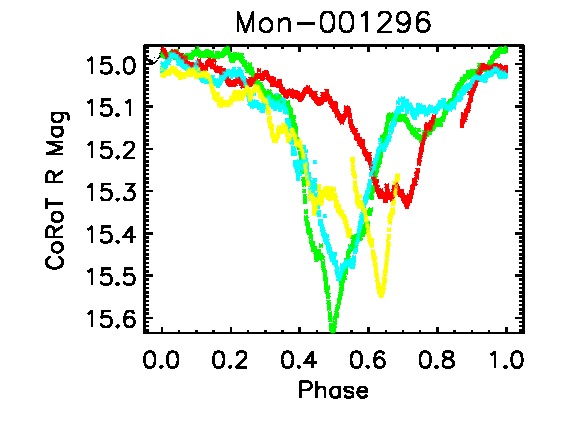}}
 \end{center}
\caption{\label{fig:fase} Periodic CoRoT light curves from Fig. \ref{fig:LC} 
folded in phase. The spot-like light curves present stable variations (top) that 
contrast with the high variability of the AA Tau-like light curves (bottom). 
Different colors represent different rotational cycles.} 
\end{figure}

We morphologically classified the $145$ CTTSs light curves observed in 2011 by CoRoT, and
in Fig. \ref{fig:LC} we show examples of these light curves. In Fig. \ref{fig:fase}, we present the 
periodic light curves of Fig. \ref{fig:LC} folded in phase with the period obtained 
with the Scargle periodogram, as modified by \cite{1986ApJ...302..757H}. The 
stability of the spot-like light curve modulation is evident, which contrasts with the high variability of the typical 
AA Tau-like light curves. A detailed statistical and morphological analysis 
of the light curves of stars that present infrared excess observed in our 
campaign with CoRoT and {\it Spitzer} are discussed in \cite{2014AJ....147...82C}, where
the light curves are separated into twelve different classes that represent diverse 
physical processes and geometric effects. We list in the last column of Table \ref{tab:ctts}
the $Q$ parameter defined by \cite{2014AJ....147...82C} that can be used as a 
metric to distinguish periodic and quasi-periodic ($Q < 0.6$) from aperiodic ($Q > 0.6$) light curves.

Table \ref{tab:Estatis} shows some morphological statistics of the CTTS light curves
observed with CoRoT in $2008$ and $2011$. The number of periodic CTTSs did not 
vary considerably from one run to the other. However, the number of 
CTTSs observed with CoRoT increased in 2011, and more non-periodic 
stars were observed in 2011 than in 2008. This was expected, since most 
of the known periodic variable systems were already included in the 2008 run. 
The number of AA Tau-like systems identified in 
2008 differs from that of \citet{2010A&A...519A..88A}, since we were able here to use the simultaneous FLAMES spectroscopy to refine our classification.
In addition, \cite{2015AJ....149..130S} identified a new subclass of stars with variable extinction light curves 
that show periodic, shallow, and short-duration flux dips that are approximately Gaussian in shape. There is some
overlap between this new class and the classical AA Tau systems, since a few 
AA Tau-like stars also present narrow dips superposed on the broad deep minima, 
as explained in their paper. 
Light curves labeled ``unclassified'' in Table \ref{tab:Estatis} correspond to 
the light curves that we were unable to clearly associate with one major physical phenomenon.

\begin{table}[ht]
 \centering
 \caption{Morphological classification of CTTS light curves observed with CoRoT in $2008$ and $2011$.}
 \label{tab:Estatis}
 \small          
 \begin{tabular}{lll}
  \hline
  \hline
  Light curve type                & Nº of stars-$2008$  & Nº of stars-$2011$ \\
  \hline
  \multicolumn{3}{c}{Periodic or quasi-periodic} \\ \hline  \cline{1-3}  
  Spot                                             & $24$    & $19$ \\
  AA Tau                                           & $15$    & $18$ \\
  Eclipsing binary                                 & $2$     & $2$  \\ 
  Unclassified                                     & $0$     & $4$  \\ \hline
  \multicolumn{3}{c}{Non-periodic} \\ \hline  \cline{1-3} 
  Random extinction                                & $13$    & $13$ \\ 
  Accretion bursts                                 & $12$    & $20$ \\
  Unclassified                                     & $30$    & $69$ \\ \hline
 Total                                             & $96$    & $145$\\ \hline
 \end{tabular} 
\end{table}

\subsection{Morphological changes from 2008 to 2011}\label{sec:MorChange}

\begin{figure}
 \begin{center}
\subfigure[]{\includegraphics[width=4.4cm]{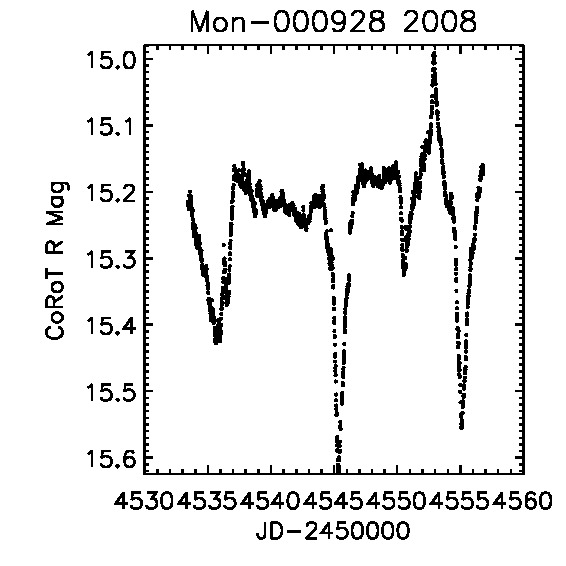}}
\subfigure[]{\includegraphics[width=4.4cm]{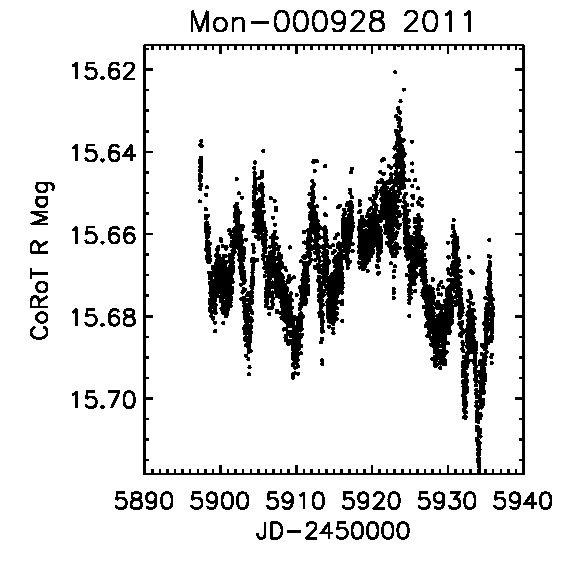}}\\
\subfigure[]{\includegraphics[width=4.4cm]{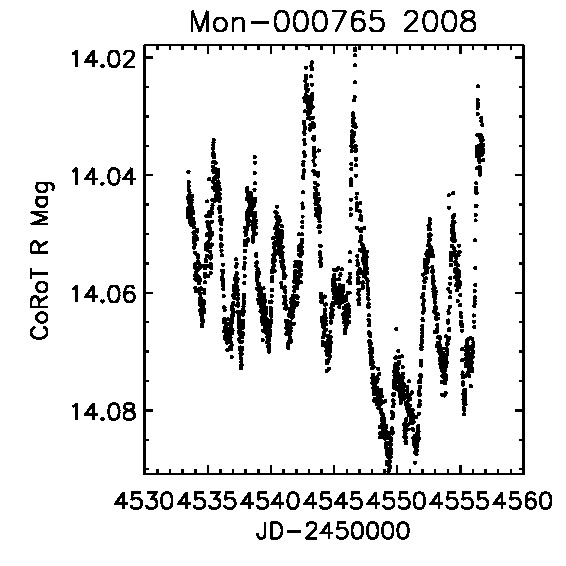}}
\subfigure[]{\includegraphics[width=4.4cm]{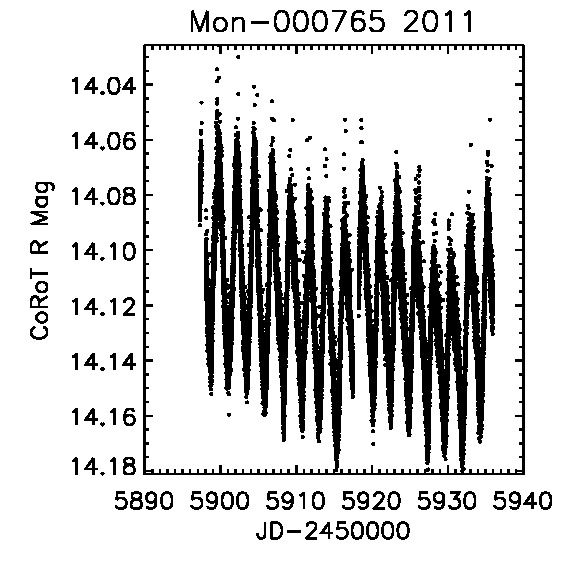}}
\end{center}
\caption{\label{fig:changLC}Examples of CTTSs whose
morphological classification of the light curve varied from 2008 to 2011. Mon-000928 displays an AA-Tau-like light curve in 
2008 (a) with a period of $\mathrm{P}=4.96\,\mathrm{days}$, while in 2011 the 
light curve was aperiodic (b). Mon-000765 was aperiodic in 2008 (c), while in 2011 the light curve 
was spot-like with a period of $\mathrm{P}=2.41\,\mathrm{days}$ (d).} 
\end{figure}

Several stars exhibited the same light-curve morphology in 2008 and 2011, 
but $\sim\,30\%$ of the stars presented variations in the morphology of their light curves. We show in Table \ref{tab:changLC} 
the variation of morphological classification of the 84 CTTS light curves that were observed in both campaigns. 
The transition from AA Tau to non-periodic classification and vice versa is common, but we
did not observe a transition from AA Tau to spot-like or from spot-like to AA Tau. A few spot-like 
systems became irregular, and the opposite also occurred.
In Fig. \ref{fig:changLC} (a) and (b), we show the 2008 and 2011 light curves of the CTTS
Mon-000928, whose light-curve classification changed between the two runs. In 2008, Mon-000928 was classified 
as an AA Tau-like system with a maximum brightness periodically interrupted by minima, while in 2011
it was clearly irregular with no detected periodicity. These variations show that the inner disk structure that is responsible for 
the stellar occultation can move from a well-organized geometry with a stable inner disk warp to a 
random dust distribution in just a few years. These results may also hint at the typical timescale of stability of the 
accretion regime, where the accretion geometry evolves from a main accretion
funnel in each hemisphere, the base of which may correspond to the stable inner disk warp to an
unstable accretion scenario, where random accretion funnels are formed 
\citep{2013MNRAS.431.2673K,2015A&A...577A..11M}.

In young low-mass stars, cold spots often occupy a substantial fraction of 
the stellar surface and may last for months to years
\citep[e.g.,][]{1989AJ.....98.2268H,1999ApJS..121..547V}. 
If the spots indeed did not evolve significantly over a period
of few years, then the light curves would also be substantially the same
in our two epochs. However, in our sample, $9$ of $21$ 
spot-like light curves, observed in both epochs, changed their morphology from 2008 to 2011
and became aperiodic, as seen in Table \ref{tab:changLC}. 
At the same time, $2$ aperiodic light curves in 2008 became spot-like in 2011. 
In a study of the Orion nebula cluster (ONC), \cite{2001AJ....121.1676R} analyzed Cousins $I_C$ 
light curves from young stars and showed that in a one-year interval, about half of the observed
periodic spot-like light curves became aperiodic. They also compared their measured periods
with the results of similar analyses of the ONC by other authors \citep{1999AJ....117.2941S,2000AJ....119..261H} and 
showed that only about $50\%$ of the periodic systems are typically
recovered between the different works.
The results above show that cold spots in young stars can evolve 
substantially and that the observable characteristic features, such as light-curve 
periodicity, can disappear on a timescale of a few years. 
 Cool spots are expected to be present in all young low-mass stars, but only some of the CTTSs
present periodic spot-like light curves, since overlying accretion hot spots and circumstellar dust obscuration
can effectively mask cool spot signatures. When CTTSs are observed to switch between variability
categories, it is generally attributable to changes in accretion behavior on timescales of a few years or less.
In Fig. \ref{fig:changLC} (c) and (d), 
we present the 2008 and 2011 light curves of the CTTS Mon-000765, whose light-curve classification
changed from aperiodic to spot-like.

Figure \ref{fig:changLC} shows that flux variations can also be observed from 2008 to 2011 in the spot-like stars
that presented a variable light-curve periodicity. During the periodic epoch, most spot-like
systems were fainter than in the aperiodic cycles, as seen in Mon-000765 (Fig. \ref{fig:changLC},
lower panels). The brighter phase during the irregular cycles could be due to accretion bursts,
leading to the appearance of randomly distributed bright spots at the stellar surface, which would
at the same time explain the irregular photometric variability. The AA Tau-like stars also show 
flux variability between the periodic and aperiodic
phases, but in the opposite sense as compared to the spot-like systems: they are generally brighter 
when periodic, as seen in Mon-000928 (Fig. \ref{fig:changLC}, top panels). These systems are
observed at high inclinations with respect to our line of sight \citep[$i>60$\degr, see][]{2015A&A...577A..11M}, and the aperiodic cycles may 
correspond to the unstable accretion regime 
described by \citet{2008MNRAS.385.1931K} and \citet{2013MNRAS.431.2673K}, where multiple accretion streams are randomly formed
and occult the star as the system rotates. In this scenario, the star can be partially occulted
at all rotational phases, never exhibiting its maximum flux to the observer, and being
therefore fainter than in the stable and ordered AA Tau phase. This is consistent with the fact that,
in the aperiodic cycles, we generally do not observe a well-defined maximum in the light curves,
while in the periodic AA Tau phase a clear maximum is seen, which should correspond to the 
unocculted photospheric brightness.

\begin{table}[ht]
 \centering
 \caption{Variations of the morphological classification 
 of the 84 CTTS light curves observed with CoRoT in 2008 and 2011.}
 \label{tab:changLC}
 \small
 \begin{tabular}{lc}
  \hline
  \hline
   Light curve 2008 $\rightarrow$ Light curve 2011     & Nº of stars \\
  \hline
  Spot $\rightarrow$ AA Tau                 & $0$  \\
  Spot $\rightarrow$ non-periodic           & $9$ \\
  AA Tau  $\rightarrow$  spot               & $0$  \\
  AA Tau $\rightarrow$ non-periodic         & $5$  \\ 
  Non-periodic $\rightarrow$ AA Tau         & $8$ \\
  Non-periodic $\rightarrow$ spot           & $2$  \\
  Non-periodic $\rightarrow$ periodic unclassified & $1$ \\
  AA Tau $\rightarrow$ AA Tau               & $8$  \\
  Spot   $\rightarrow$ spot                 & $12$ \\
  Non-periodic $\rightarrow$ non-periodic   & $37$ \\
  Binary $\rightarrow$ binary               & $2$  \\
   \hline
 \end{tabular} 
\end{table}

\subsection{Period distribution}\label{sec:Period}

Table \ref{tab:Estatis} shows that $\sim 43\%$ of the $96$ CTTSs observed in 2008
and $\sim 30\%$ of the $145$ CTTSs observed in 2011 were classified as periodic.
We only classified as periodic the light curves
that presented a high-power, detached, and dominant peak in the periodogram analysis.
We established a minimum power level of the periodogram for a reliable detection based on the periodogram 
analysis of the sinusoidal spot-like light curves, which are clearly periodic with a well-determined period 
within our dataset. We also inspected by eye each of the periodic light curves folded in phase at the 
measured period and verified that they presented phase coherence.

\begin{figure}[htb!] 
 \centering
 \includegraphics[scale=0.30]{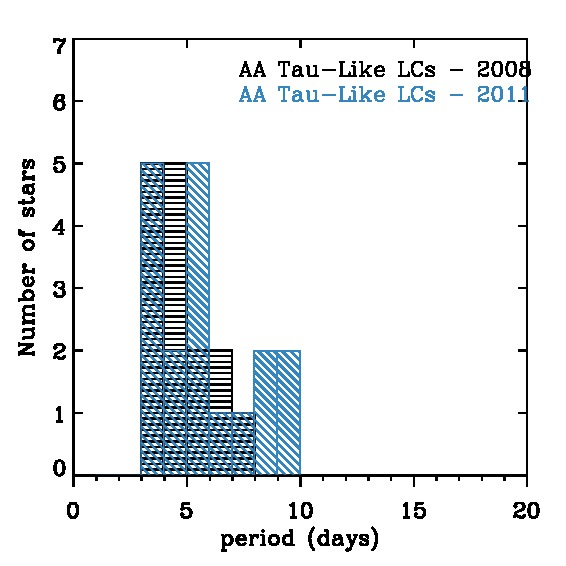}
 \includegraphics[scale=0.30]{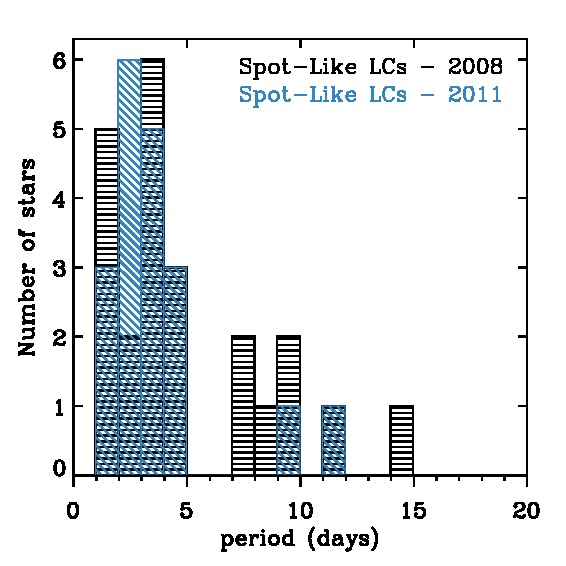}
 \caption{\label{fig:perAASpot}Period distribution of the CTTSs observed 
 in 2008 (black) and 2011 (blue). Top: AA Tau-like light curves.  
 Bottom: Spot-like light curves. AA Tau-like stars clearly present
 periods within the range of spot-like systems.}
\end{figure}

 Many light curves that show some periodic signal, but with low power in 
the periodogram, were conservatively classified as non-periodic in the present work. 
Venuti et al. (in preparation) discuss the peridicity of all the stars observed by
CoRoT in more detail and the number of periodic stars may 
be somewhat different from ours.
We present in Fig. \ref{fig:perAASpot} the period distribution of AA Tau-like 
and spot-like systems observed in 2008 and 2011. Despite  
the observed morphological variations in the light curves from 2008 to 2011 (Table 
\ref{tab:changLC}), the overall period distribution did not vary significantly from 2008 
to 2011. As seen in Fig. \ref{fig:perAASpot}, there are more AA Tau-like stars with 
long photometric periods in the new campaign. This is probably due to the 
the difference in datasets between the two campaigns and to the longer observation time span 
in 2011 (40 days) than in 2008 (23 days), which makes it easier to confidently determine
longer periods in the irregular, though periodic, AA Tau-like light curves. 
Despite the longer observational time span, we did not detect longer spot-like 
periods in 2011. In 2008 and 2011 the spot-like light curves were very regular
and their periods were easily measured with our data. Therefore, the lack of periods
longer than 14 days in 2008 and 10 days in 2011 is probably real
among the observed CTTSs. The longer periods measured were different in 
both campaigns, and this is due to different samples, since the stars observed in 
each run were not exactly the same, and it is due to accretion variability, since some spot-like 
systems became aperiodic and the opposite also occurred between the two observational runs.

The period of a spot-like light curve is the rotational period of the
star, since spots are located on the stellar surface. In both epochs, the AA Tau-like stars 
presented periods within the range of periods measured in the spot-like light curves. Therefore, 
the material that obscures the AA Tau-like stars is probably not located very far from the corotation radius of 
the star-disk system, as pointed out by \citet{2010A&A...519A..88A}. 
A more detailed analysis of the period distribution of CTTSs and WTTSs in NGC 2264 
and the implication for angular momentum evolution will be presented in another paper (Venuti et al., in
preparation).

\section{UV excess and photometric mass-accretion rates}\label{sec:CFHT}

In this section, we compare the CoRoT light-curve morphology with 
accretion diagnostics and disk indicators 
to determine whether the optical light-curve
morphology and accretion and inner dusty disk evolution are related.

\begin{figure}
 \begin{center}
\includegraphics[width=9.5cm]{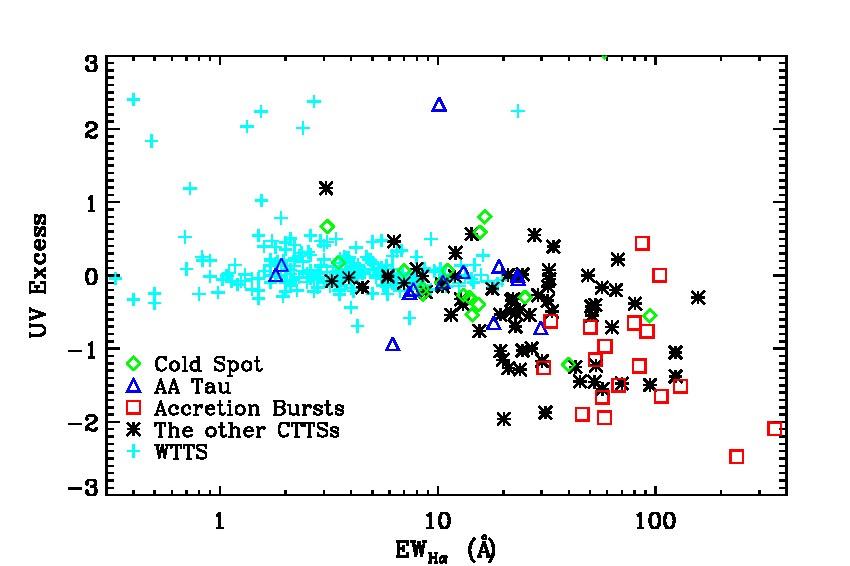}
\includegraphics[width=9.5cm]{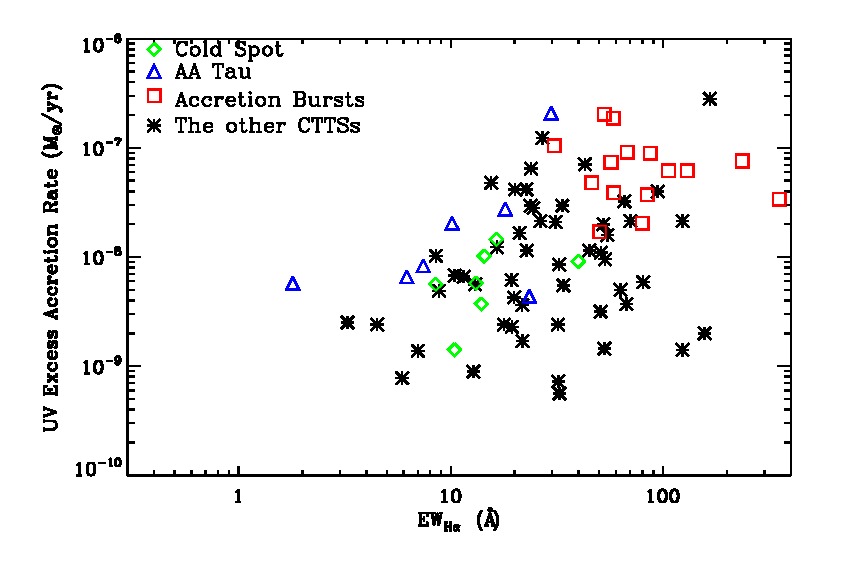}
 \end{center}
\caption{\label{fig:UV} Comparison between the CoRoT light-curve classification 
and UV excess (top) or mass-accretion rates (bottom) calculated by \cite{2014A&A...570A..82V}. 
In the top plot, more negative values indicate higher UV excess.} 
\end{figure}

\cite{2014A&A...570A..82V} obtained the UV excess emission of 
CTTSs and WTTSs in NGC 2264 using \textit{u-} and \textit{r} -band
measurements from CFHT data. They determined the locus of WTTSs 
in the \textit{r} vs. \textit{u-r} color-magnitude diagram that
was fitted with a polynomial function, which was used as a 
reference value to non-accreting systems. The UV excess of a CTTS was 
defined as $\mathrm{E}(u)=(u-r)_\mathrm{obs}-(u-r)_\mathrm{ref}$, 
where $(u-r)_\mathrm{obs}$ is the observed CTTS color, and 
$(u-r)_\mathrm{ref}$ is the reference color at magnitude $r_\mathrm{obs}$. 
The UV excess measurements have an average rms error of 0.16 mag to 
account for the distribution of WTTS around the reference 
sequence \citep{2014A&A...570A..82V}.
With the UV excess, \cite{2014A&A...570A..82V} calculated mass-accretion 
rates ($\mathrm{\dot{\mathrm{M}}_\mathrm{UV}}$) for the CTTSs in NGC 2264. 
We show in Fig. \ref{fig:UV} the UV excess (top) and mass-accretion rate (bottom) 
vs. $\mathrm{H}\alpha$ equivalent width for the stars in our sample. Stars presenting 
accretion burst light curves show some of the highest UV excesses (more negative 
on the scale of the plot) and consequently the highest mass-accretion rates among 
the observed systems, as noted by \citet{2014AJ....147...83S}. 
This is supported by a two-sided Kolmogorov-Smirnov test that shows 
that the mass-accretion rate distribution of accretion burst stars,
when compared to spot-like and AA Tau-like systems, has only a 
$5.8\times10^{-6}$ probability to be equivalent.
AA Tau-like and spot-like stars are an intermediate population between the accretion bursts 
systems and WTTSs, which present low UV excess. 
The mean UV excess values of each light curve group corroborate with 
this analysis: $-1.15$ for the accretion bursts, 
$-0.04$ for the AA Tau systems, $0.05$ for the spot-like systems, and $0.14$ for the WTTSs.
These results suggest that the CoRoT light curve morphology is related to the evolution of
the accretion process.

Many of the stars observed by CoRoT have {\it Spitzer} observations from the literature 
\citep{2012A&A...540A..83T}, which allowed us to compute the $\alpha_{IRAC}$ index, 
as proposed by \cite{Lada2006}. This index is the slope of the spectral energy 
distribution between $3.6\,\mu\mathrm{m}$ and $8\,\mu\mathrm{m}$, and it allows a classification of 
inner disk evolution.  
If $\alpha_{IRAC}\, < \,-2.56$, the disk is classified as a naked photosphere (no dust 
in the inner disk). For systems with $-2.56\, < \alpha_{IRAC}\, <\,-1.80$, 
the inner disk is optically thin and called an anemic disk. For stars with 
$-1.80 \,< \alpha_{IRAC}\, < \,-0.5$, the inner disk is classified as optically thick. 
Stars with $-0.5\, < \, \alpha_{IRAC}\, <\, 0.5$ are classified as flat spectrum sources, 
and $\alpha_{IRAC}\, >\, 0.5$ characterizes class I sources. 

\begin{figure}[htb!] 
 \centering
 \includegraphics[scale=0.31]{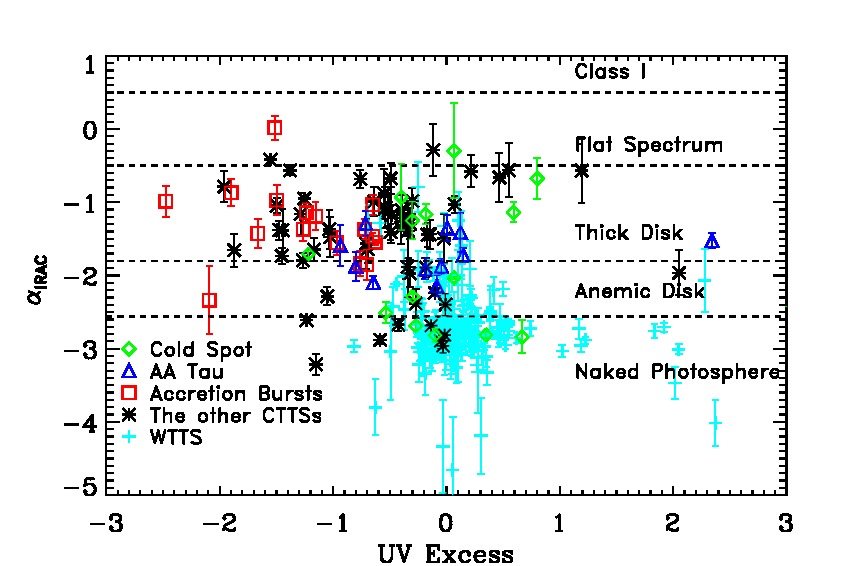}
 \caption{\label{fig:alphaTaxur} Comparison between the CoRoT light-curve classification with the 
UV excess from \cite{2014A&A...570A..82V} and the slope of the spectral energy distribution from 
$3.6\,\mu\mathrm{m}$ to $8\,\mu\mathrm{m}$ ($\alpha_{IRAC}$). These results show that the CoRoT 
light-curve morphology is related to the accretion process and the evolution of dust in the inner disk.} 
\end{figure}

In Fig. \ref{fig:alphaTaxur}, we show the $\alpha_{IRAC}$ index vs. the UV excess 
calculated by \cite{2014A&A...570A..82V}. 
Most of the CTTSs present dust in the inner disk. Spot-like systems are present
in all the inner disk classes, but are mostly found among the anemic disks and naked 
photosphere populations. All AA Tau-like systems 
have dust in the inner disk, which is important, since we assumed their light-curve 
variability to be dominated by circumstellar dust extinction from that region. 
Most of the accretion burst systems present thick disks, which is consistent with an earlier
disk evolution and accretion phase than the AA Tau-like stars. The mean $\alpha_{IRAC}$ values 
of each light-curve group corroborate this analysis: $-1.42$ for the accretion bursts, 
$-1.71$ for the AA Tau systems, $-1.84$ for the spot-like systems, and $-2.76$, for the WTTSs.
These results show that the CoRoT light-curve morphology is related to the 
evolution of the inner disk. It also shows that accretion and disk evolution
are related to each other, since the stars with highest UV excesses (more negative values)
are the ones with the dustiest inner disks, while stars with little 
or no sign of accretion have almost no dust in their inner disks.

Figure \ref{fig:alphaTaxur} also shows some interesting objects that deserve further 
investigation, which beyond the scope of the present paper, however, such as the WTTSs among 
the anemic and thick disk systems. These are non-accreting stars (based on spectra) that still 
show substantial dust in their inner disks. In spite of this, for some as yet unknown reason, accretion has been 
shut off or is at a very low level in these systems, at least at the time of spectroscopic observation. 
There is also a significant number of anemic disk candidates among the CTTSs, with various light-curve 
morphologies that would be interesting targets to investigate the inner disk dispersal in accreting systems. 

\section{Equivalent width and accretion rates from the $\mathrm{H}\alpha$ line} \label{sec:AccHa}

\begin{figure}[htb!] 
 \centering
 \includegraphics[scale=0.40]{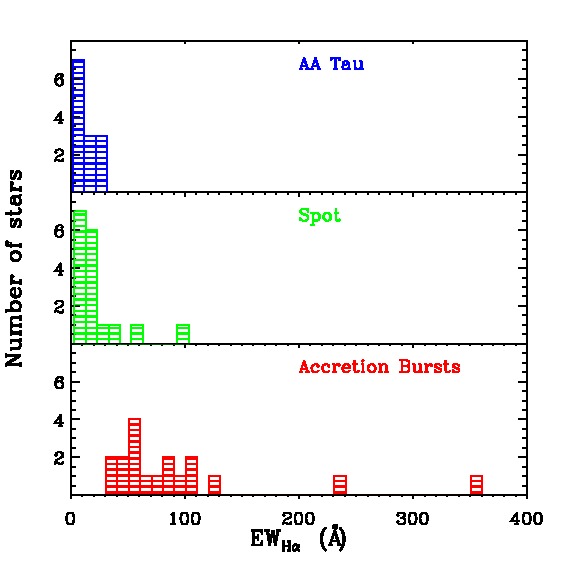}
 \caption{\label{fig:histEW} Distribution of $\mathrm{H}\alpha$ equivalent width of CTTSs that
present AA Tau-like (top), spot-like (middle), and accretion burst (bottom) light curves.} 
\end{figure}

The $\mathrm{H}\alpha$ line of a CTTS is generally quite strong and wide and
is often characterized by its equivalent width. Compared to the morphology 
of the CoRoT light curves, the accretion burst systems show the
highest $\mathrm{EW}_{\mathrm{H}\alpha}$ in our sample, with an average 
value of $96\,\pm\,19\,\mathring{\mathrm{A}}$, clearly distinguishable from
the AA Tau and spot-like systems. AA Tau-like stars 
present an average $\mathrm{EW}_{\mathrm{H}\alpha}$ of 
$13.2\,\pm\,2.4\,\mathring{\mathrm{A}}$, while CTTSs with spot-like light curves present a mean $\mathrm{EW}_{\mathrm{H}\alpha}$ value of 
$21.1\,\pm\,5.6\,\mathring{\mathrm{A}}$. The uncertainties correspond to the 
standard deviations of the measured equivalent width values.
The distribution of $\mathrm{EW}_{\mathrm{H}\alpha}$ among the three CTTS groups 
is shown in Fig. \ref{fig:histEW}, where we included $\mathrm{EW}_{\mathrm{H}\alpha}$ 
measurements from \cite{2005AJ....129..829D} for the stars that we did not observe. 
The same tendency is observed with the $\mathrm{H}\alpha$ width at $10\%$ 
of maximum line intensity.

Mass-accretion rates ($\dot{\mathrm{M}}_{\mathrm{H}\alpha}$) are frequently calculated 
from the $\mathrm{H}\alpha$ line flux, using calibrations such as the ones proposed by 
\citet{2009A&A...504..461F} and \citet{1998ApJ...492..323G}. 
We compared the mass-accretion rates obtained from the UV excess by \cite{2014A&A...570A..82V} with the values obtained with the $\mathrm{H}\alpha$ flux, as described below.  

We calculated the $\mathrm{H}\alpha$ flux ($\mathrm{F}_{\mathrm{H}\alpha}$) from the $\mathrm{EW}_{\mathrm{H}\alpha}$,

\begin{equation}
\mathrm{F}_{\mathrm{H}\alpha}=\mathrm{EW}_{\mathrm{H}\alpha}\mathrm{F}_\mathrm{c}(\mathrm{H}\alpha)
,\end{equation}

where $\mathrm{F}_\mathrm{c}(\mathrm{H}\alpha)$ is the continuum flux in the $\mathrm{H}\alpha$ region.
Since the FLAMES spectra were not 
flux calibrated, we estimated $\mathrm{F}_\mathrm{c}(\mathrm{H}\alpha)$ from 
the CoRoT flux using CoRoT observations obtained simultaneously with the 
FLAMES spectra. The CoRoT flux was converted into $R$ magnitudes using the $R$-band photometry
zero-point of 26.74, as determined by \cite{2014AJ....147...82C}. 
For the stars observed with FLAMES that were not observed with CoRoT in 2011, we used 
the continuum flux from the CFHT \textit{r} band \citep{2014A&A...570A..82V}
taken simultaneously with the FLAMES observations. 
The continuum flux was then corrected for extinction with $\mathrm{A_V}$ 
obtained from the CFHT data \citep{2014A&A...570A..82V}.

The $\mathrm{H}\alpha$ luminosity was then computed as 
$\mathrm{L}_{\mathrm{H}\alpha}=4\pi\mathrm{d}^2\mathrm{F}_{\mathrm{H}\alpha}$, 
where we used $\mathrm{d}=760\,\mathrm{pc}$ \citep{1997AJ....114.2644S} as the distance to the cluster. 
We used the fit proposed by \cite{2009A&A...504..461F} to calculate the accretion luminosity 
($\mathrm{L}_\mathrm{acc}$) from $\mathrm{L}_{\mathrm{H}\alpha}$:

\begin{equation}
\log{\left(\dfrac{\mathrm{L}_\mathrm{acc}}{\mathrm{L}_\odot}\right)}=\left(2.27\pm0.23\right)+\left(1.25\pm0.07\right)\log{\left(\dfrac{\mathrm{L}_{\mathrm{H}\alpha}}{\mathrm{L}_\odot}\right)} 
.\end{equation}

With the accretion luminosity, we obtained the mass-accretion rate,

\begin{equation}
 \dot{\mathrm{M}}_\mathrm{\mathrm{H}\alpha}=\dfrac{\mathrm{L}_\mathrm{acc}\mathrm{R}_\ast}{\mathrm{G}\mathrm{M}_\ast\left(1-\dfrac{\mathrm{R}_\ast}{\mathrm{R}_\mathrm{in}} \right)}
,\end{equation}
where $\mathrm{R}_\ast$ and $\mathrm{M}_\ast$ are the radius and mass of the star, respectively, taken 
from \cite{2014A&A...570A..82V}, and G denotes the gravitational constant. The inner radius was set to 
$\mathrm{R}_\mathrm{in}=5\,\mathrm{R}_\ast$, which is a standard value of the disk truncation radius of CTTSs 
\citep{1998ApJ...492..323G}. The individual values of mass-accretion rates calculated from the
$\mathrm{H}\alpha$ line flux are presented in Table \ref{tab:CttsFlames}.

\begin{figure}[htb!] 
 \centering
 \includegraphics[scale=0.31]{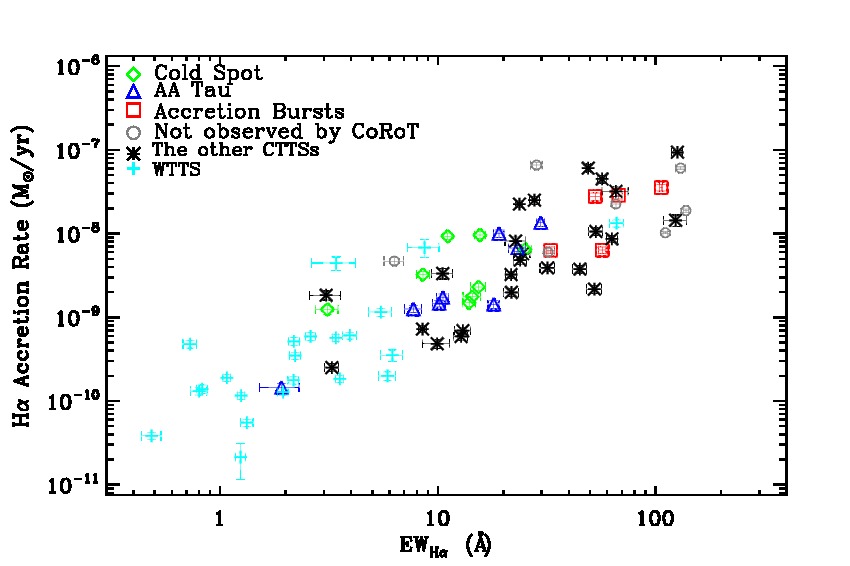}
 \caption{\label{fig:AccRateHa} Comparison between the CoRoT light-curve classification and
mass-accretion rates obtained from $\mathrm{H}\alpha$ flux as a function of $\mathrm{H}\alpha$ 
equivalent width (FLAMES/VLT data). The error bars represent the night-to-night variability of the line flux.} 
\end{figure}

Figure \ref{fig:AccRateHa} shows the mass-accretion rates obtained from the mean $\mathrm{H}\alpha$ 
equivalent width of the four to six FLAMES and CoRoT simultaneous observations, and
also those obtained from simultaneous CFHT observations. We plot the 
mean $\mathrm{EW}_{\mathrm{H}\alpha}$, and the error bar represents the night-to-night variability 
of $\mathrm{EW}_{\mathrm{H}\alpha}$ and $\dot{\mathrm{M}}_\mathrm{\mathrm{H}\alpha}$. 
As observed with the $\mathrm{\dot{\mathrm{M}}_\mathrm{UV}}$ relation (Fig. \ref{fig:UV}), 
stars that present accretion burst light curves have higher $\dot{\mathrm{M}}_{\mathrm{H}\alpha}$.
Since WTTSs are not actually accreting, their $\mathrm{H}\alpha$ equivalent widths
are probably of chromospheric origin, a contribution that is also present in the CTTS
spectra. Therefore, the mass-accretion rate locus of WTTSs represents a lower limit to 
our ability to measure mass-accretion rates of CTTSs from the $\mathrm{H}\alpha$ equivalent
width in our sample, as discussed by \cite{2011ApJ...743..105I} and \cite{2013AA...551A.107M}.  

\begin{figure}[htb!] 
 \centering
 \includegraphics[scale=0.31]{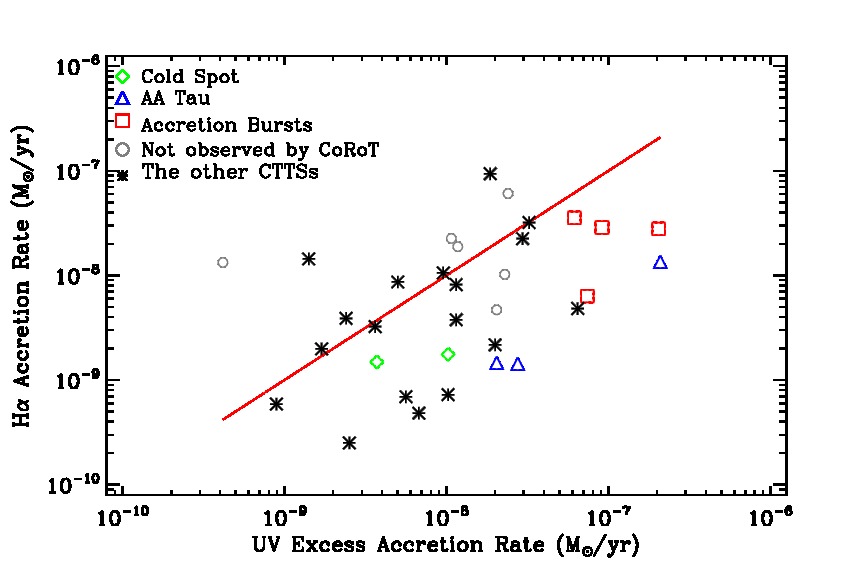}
 \caption{\label{fig:AccRateHaCFHT} Comparison between mass-accretion rates 
obtained from $\mathrm{H}\alpha$ flux with the UV excess mass-accretion rates. 
The red line shows a slope equal to $1$.} 
\end{figure}

Figure \ref{fig:AccRateHaCFHT} shows the relation of the mass-accretion rates calculated with 
the UV excess and $\mathrm{H}\alpha$ equivalent width. The two methods show similar tendencies, but 
the individual value can sometimes be different.
This difference may arise because the 
$\mathrm{H}\alpha$ line does not necessarily originate in the accretion columns alone, but 
can also have contributions from the disk wind and accretion shock, as absorptions superposed 
on the accretion profile. When we calculated the $\mathrm{H}\alpha$ equivalent width, the absorptions
decrease the measured $\mathrm{EW}_{\mathrm{H}\alpha}$ value, leading to lower values of mass-accretion rates. 
At the same time, the errors in $\mathrm{A_V}$ affect the UV excess accretion rates more than the $\mathrm{H}\alpha$ 
ones and can lead to more uncertain values of mass-accretion rates calculated from UV excess.
Although UV excess is clearly a more direct measurement of the
mass-accretion rate, $\mathrm{H}\alpha$ line fluxes
are much easier to obtain and allow for a good estimate of mass-accretion rates for all accreting objects, 
even those that do not present a substantial UV excess, due, for example, to circumstellar dust obscuration
of the hot spot. 

\section{$\mathrm{H}\alpha$ line variability} \label{sec:Hatype}

\begin{figure}[htb!] 
 \centering
 \includegraphics[scale=0.31]{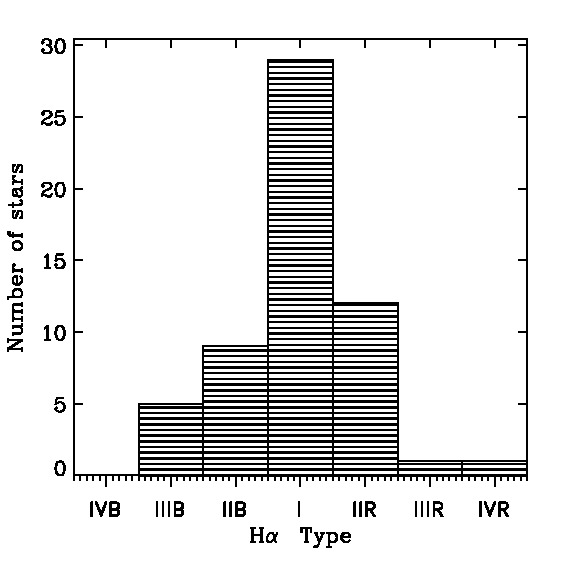}
 \includegraphics[scale=0.31]{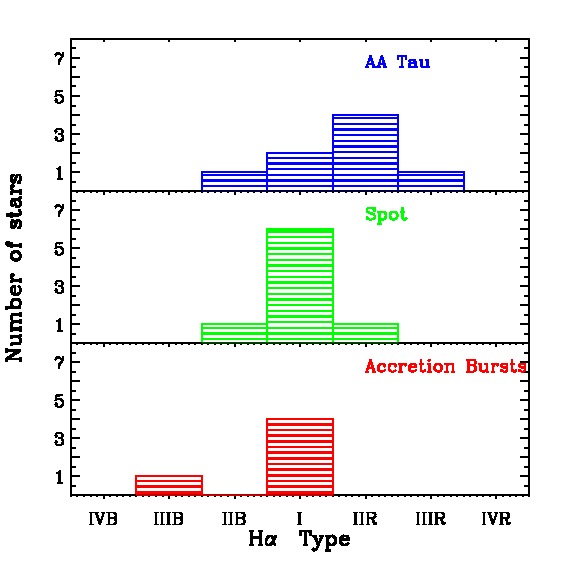}
 \caption{\label{fig:Hatype} Distribution of $\mathrm{H}\alpha$ profile types
according to the classification of \citet{1996A&AS..120..229R}. 
Top: All the CTTSs observed with FLAMES. Bottom: CTTSs observed both
with FLAMES and CoRoT. Light curves classified as spot-like, 
AA Tau-like, and accretion burst are shown as a function
of profile type.} 
\end{figure}

The CTTSs in NGC 2264 present a variety of $\mathrm{H}\alpha$ line profiles.
In this section, we verify how common each $\mathrm{H}\alpha$ profile type is and 
compare its occurrence with the CoRoT light-curve classification. 
We followed the $\mathrm{H}\alpha$ profile classification proposed by
\cite{1996A&AS..120..229R}, which is divided into seven profile types. 
A type I profile is symmetric and shows no absorption feature. A type II profile presents two 
peaks, and the less intense one exceeds half the intensity of the main peak. 
A type III profile also shows two peaks, and the less intense one is 
smaller than half the intensity of the main peak. A type IV profile shows 
a pronounced absorption feature on the blue or red side of the line where 
no emission is seen, like a P Cygni or inverse P Cygni profile. The letters B 
or R are attached to types II, III and IV, depending on the location (in the 
blue (B) or red (R) side of the line) of the secondary peak or absorption 
feature with respect to the main peak.
We have 20 to 22 nights of observation of each target, and the 
$\mathrm{H}\alpha$ line is quite variable both in intensity and shape 
from night to night. Therefore, we morphologically classified only 
the mean $\mathrm{H}\alpha$ profile of each star and present
this classification in Table \ref{tab:CttsFlames} and 
Fig. \ref{fig:Hatype}. The top panel of Fig. \ref{fig:Hatype} shows 
the profile distribution of all CTTSs observed with FLAMES. Type I is the most common 
$\mathrm{H}\alpha$ profile in our sample. This profile corresponds to a main
emission feature without any prominent absorption and is attributed to
magnetospheric accretion. The emission profiles of type I
stars sometimes display some superposed absorption in one or a few
observing nights, but the analyzed mean profile effectively does not present 
absorptions.

Figure \ref{fig:Hatype} (bottom) shows the distribution of the 
$\mathrm{H}\alpha$ profile types according to the CoRoT light-curve
classification. Spot-like stars typically present 
type I $\mathrm{H}\alpha$ profiles, generally showing no evidence of blueshifted
absorption due to wind, or redshifted absorption due to hot spot photons
absorbed by the accretion funnel. 
Moreover, AA Tau-like stars tend to
present type IIR and IIIR profiles, which are characterized by 
redshifted absorptions, compared to the most common type I profile 
seen among the CTTSs in our sample.
 
\cite{2006MNRAS.370..580K} used disk-wind and magnetosphere radiative 
transfer models for CTTS to examine the $\mathrm{H}\alpha$ classification 
proposed by \cite{1996A&AS..120..229R} and with a range of parameters 
reproduced all the $\mathrm{H}\alpha$ profile types. In general, they 
found that the type I profile is the most common among CTTSs,
since it is produced from a range of system inclinations and is dominated 
by magnetospheric emission.
\cite{2006MNRAS.370..580K} found that type IIB and IIR
profiles were equally common, as also found by \cite{1996A&AS..120..229R}
and by us, since they result from the same physical conditions,
with the exception of the system inclination.
Type IIB needs medium inclinations to be produced, while
type IIR results from high inclinations. This is also
consistent with our results, since most AA Tau-like stars, which are
high-inclination systems \citep{2015A&A...577A..11M}, tend to present type IIR profiles 
(see Fig. \ref{fig:Hatype}, bottom). 

\cite{1996A&AS..120..229R} instead found that the $\mathrm{H}\alpha$ profile
distribution in their sample was dominated by type IIIB profiles ($33\,\%$ of 
their sample), while in our sample
only $\sim\,9\,\%$ of the observed CTTSs present type IIIB profiles. 
In the models of \cite{2006MNRAS.370..580K}, only high mass-accretion rate stars presented type IIIB profiles. The difference in profile distribution found by us 
and \cite{1996A&AS..120..229R} might therefore simply be due to a sample selection effect. 
The stars classified as type IIIB by \cite{1996A&AS..120..229R} 
have spectral types between G0 and K7 and a very high $\mathrm{H}\alpha$ equivalent width
(average value of $64\,\mathrm{\mathring{A}}$), which may indeed indicate that
they are high mass-accretion rate systems.
The type IIIB CTTSs in our sample present an average $\mathrm{EW}$ of $47\,\mathrm{\mathring{A}}$ 
and a mean mass-accretion rate of $\sim2.5\times10^{-8}\,\mathrm{M}_\odot/\mathrm{yr}$,
which is not particularly high, but does correspond to one of the highest 
mean mass-accretion rates of the various profile types.

\section{Periodograms of $\mathrm{H}\alpha$ emission line}\label{sec:per_halpha}

As the star-disk system rotates, the observed $\mathrm{H}\alpha$ line of CTTSs 
varies in intensity and shape. This is expected to occur, even in a stable accretion regime, if
the stellar magnetic and rotation axis are not aligned. The misalignment between the axes breaks the 
axisymmetry of the accretion environment and creates individual accretion funnels in each hemisphere.
The projection of the $\mathrm{H}\alpha$ emission region in our line of 
sight will then be different at every rotational phase, which will cause line variability
\citep[see for example][]{2012A&A...541A.116A}.
If most of the $\mathrm{H}\alpha$ line is formed in the accretion columns, we would expect 
to measure a periodic line variability when the star is in a stable accretion regime. 
However, the $\mathrm{H}\alpha$ line may also form in different circumstellar regions, such as 
the chromosphere, disk wind, and jet, and each region may have a different periodicity. This would make a single
period difficult to  detect. 
With $20$ to $22$ nights of observation of each target with the FLAMES spectrograph, we can check 
whether the $\mathrm{H}\alpha$ line variations were periodic, and compare these periods 
with the periodicities measured photometrically. This will allow 
us to infer if the $\mathrm{H}\alpha$ line-forming region is directly related to 
the physical process causing the photometric variability. 

We analyzed the line periodicity by calculating Scargle periodograms, 
as modified by \cite{1986ApJ...302..757H}, along the line profile.
The Scargle periodograms fit the data with sinusoidal functions through
a least-squares method. The code requires a range of periods
to be tested,  and the best fit corresponds
to the highest power in the periodogram. 
To obtain the periodicity in the $\mathrm{H}\alpha$ line, 
we divided each normalized $\mathrm{H}\alpha$ line profile into 
small velocity intervals ($0.5\,\mathrm{km}/\mathrm{s}$). For each 
velocity interval, Scargle periodograms were independently calculated. 
Each periodogram was grouped side by side in velocity, to form a diagram 
of period vs. velocity. The power of the periodogram is represented by colors, 
varying from zero (white) to the maximum power intensity (black). With this 
procedure, we can check how each period spreads
across the $\mathrm{H}\alpha$ line. Periodic signals only in a few velocity bins 
may not be significant, but periodic signals that are spread across several velocity 
channels indicate periodicity in an extended region of line formation. 
We can also verify if there are different periodicities in different 
line regions (like in emission and absorption).

\begin{figure*}
\begin{center}
 \subfigure[]{\label{fig:AA2}\includegraphics[width=4.5cm]{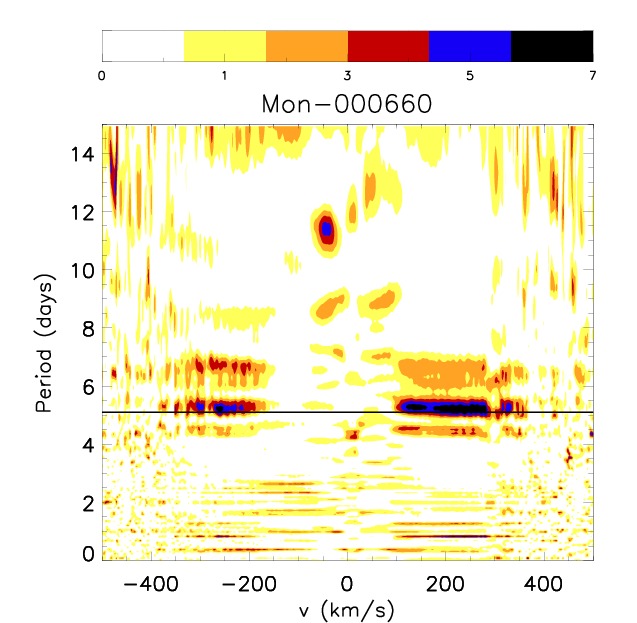}}
 \subfigure[]{\label{fig:S-AA2}\includegraphics[width=4.5cm]{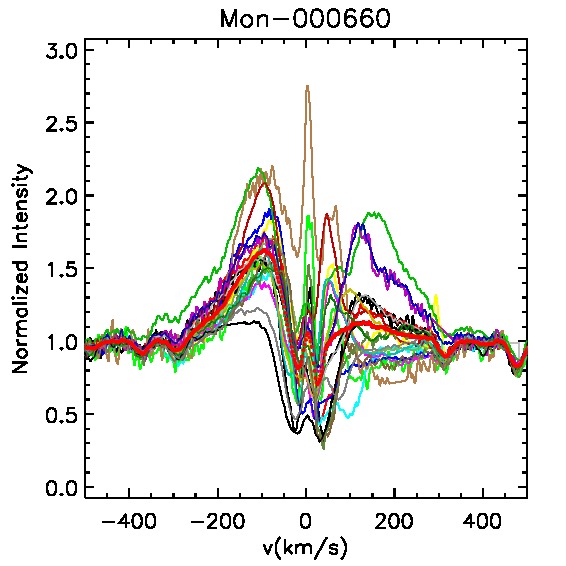}}
 \subfigure[]{\label{fig:C-AA2}\includegraphics[width=4.5cm]{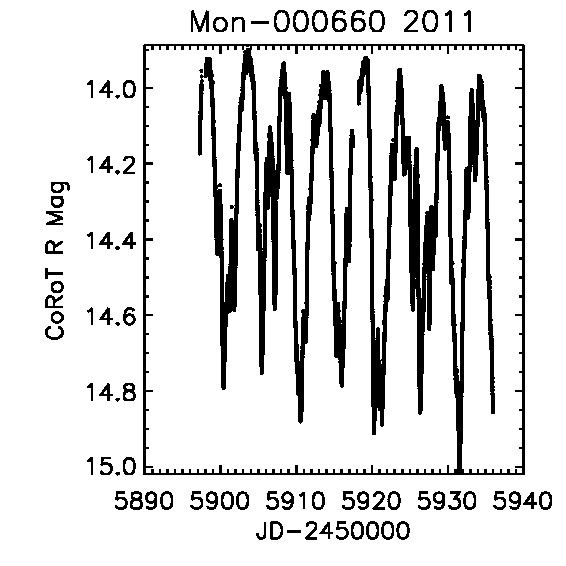}}\\
 \subfigure[]{\label{fig:AA1}\includegraphics[width=4.5cm]{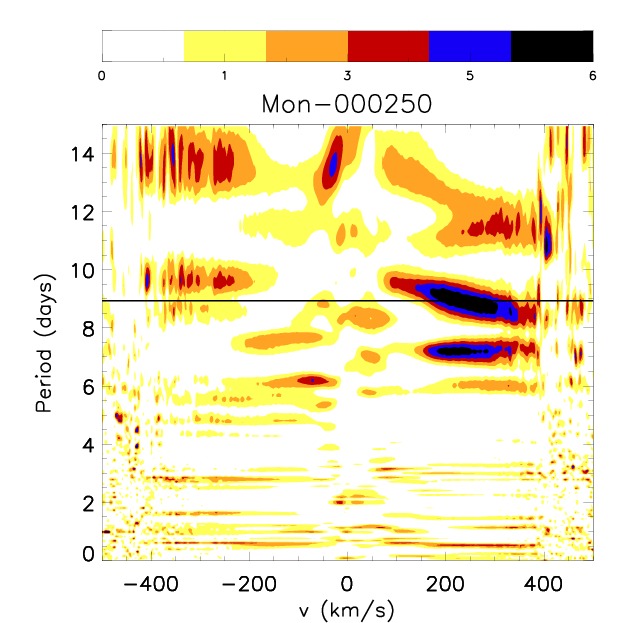}}
 \subfigure[]{\label{fig:S-AA1}\includegraphics[width=4.5cm]{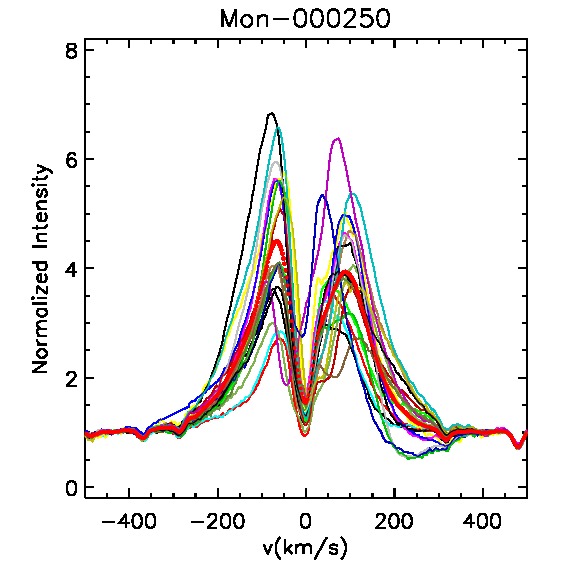}}
 \subfigure[]{\label{fig:C-AA1}\includegraphics[width=4.5cm]{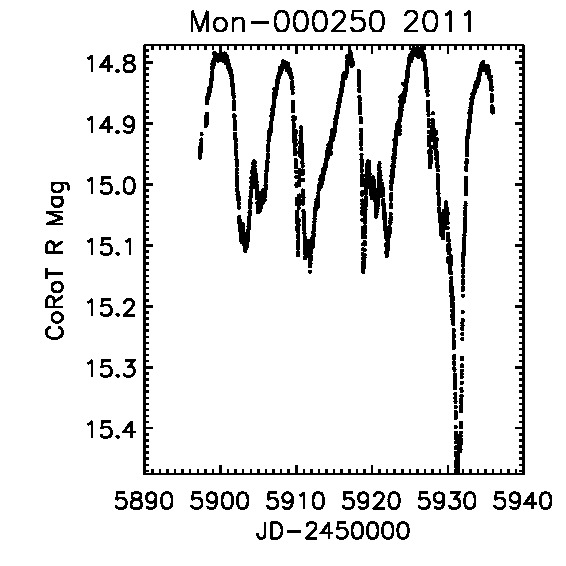}}\\
 \subfigure[]{\label{fig:AA3}\includegraphics[width=4.5cm]{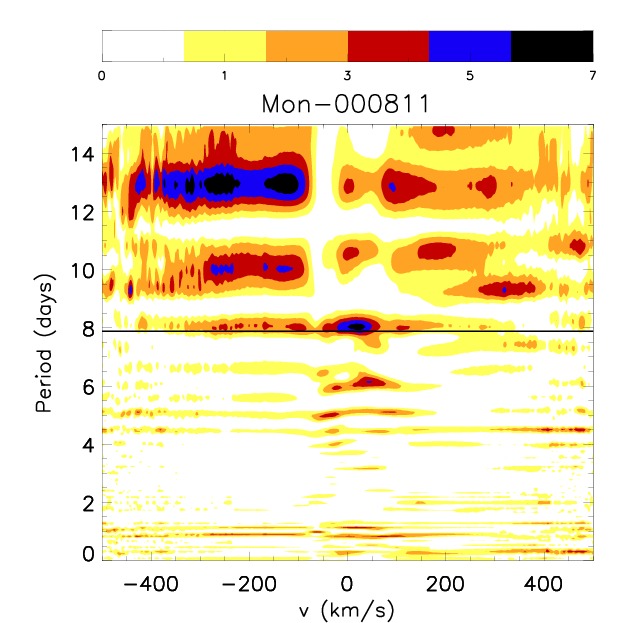}}
 \subfigure[]{\label{fig:S-AA3}\includegraphics[width=4.5cm]{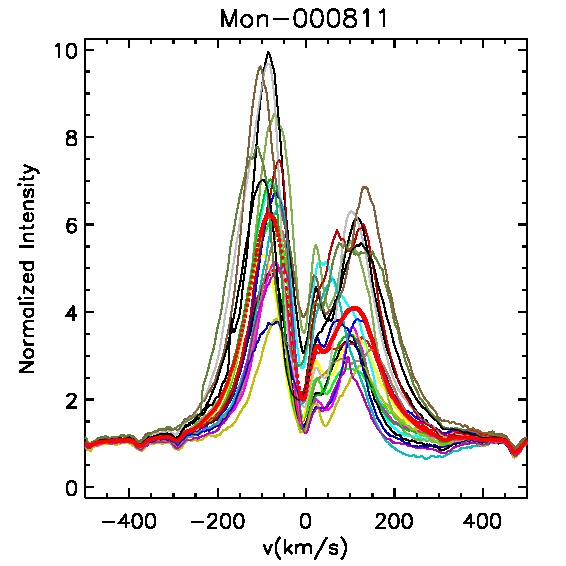}}
 \subfigure[]{\label{fig:C-AA3}\includegraphics[width=4.5cm]{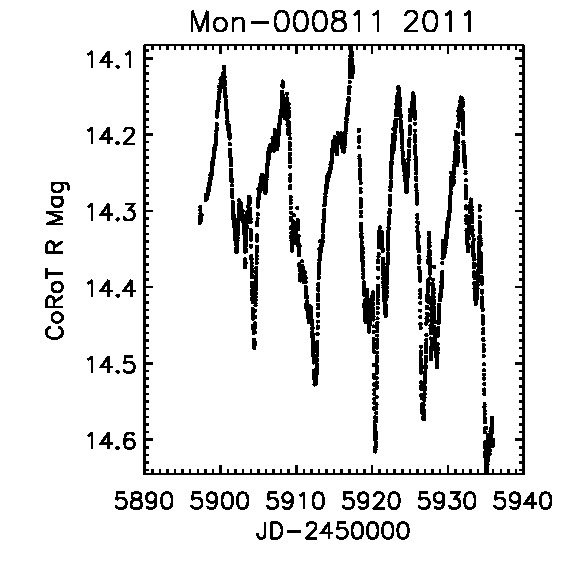}}\\
 \end{center}
 \caption{\label{fig:periodAA} AA Tau-like stars in 2011 that present a periodicity in the $\mathrm{H}\alpha$ line. Left column: Bidimensional periodograms of the $\mathrm{H}\alpha$ line. The color code represents the power of periodogram, varying from zero (white) to the maximum power intensity (black). The black horizontal lines correspond to the photometric period of the 2011 CoRoT light curves. Middle column: $\mathrm{H}\alpha$ line profiles. Different colors correspond to different observation nights, and the thick red line is the average line profile. Right column: CoRoT light curves from 2011.}
\end{figure*}

Only $\text{eight}$ CTTSs observed with the FLAMES spectrograph
presented a periodic signal in the $\mathrm{H}\alpha$ line (see Table \ref{tab:CttsFlames}),
and their periodograms are shown in Figs. \ref{fig:periodAA}, \ref{fig:periodNPAcc}, and \ref{fig:periodNPC}.
These figures are discussed individually in the following subsections, where we show that most of the periodicity
in the $\mathrm{H}\alpha$ line occurs in the red wing,
which comes from gas moving away from the observer
and probably is related to the accretion process.
These periodic stars presented organized and stable accretion regions rotationally
modulated by the stellar magnetic field. 

Our spectroscopic target selection in 2011 was largely based on the CoRoT light-curve
classification from 2008, and we included as many AA Tau-like and spot-like systems as
possible in our two pointings to study periodic systems. We did not anticipate, however, that a large number of
systems would change their light-curve variability
from periodic to aperiodic between the two runs. 
Of the $58$ stars observed with FLAMES, only $16$ presented periodic CoRoT light curves in 2011, 
$8$ of which were classified as AA Tau-like and the other $8$ as spot-like systems. 
Probably because of our unfortunate selection, the $\mathrm{H}\alpha$ line of most of the $58$ 
observed CTTSs showed no periodicity in our data. 

In the next subsections, we analyze
the $\mathrm{H}\alpha$ periodicity compared to the CoRoT light-curve morphology discussed in
Sect. \ref{sec:corot}.
Because of the large number of stars observed spectroscopically with FLAMES that presented a
non-periodic CoRoT light curve in 2011, our main conclusion is that the stars that
presented non-periodic CoRoT light curves are generally not periodic in $\mathrm{H}\alpha$ either. We show in Sect. \ref {sec:Inst} that
the lack of periodicity in $\mathrm{H}\alpha$ and in the photometry may be caused by an unstable
accretion regime, as proposed by \cite{2013MNRAS.431.2673K}.

\begin{table*}[htb!]
\tiny
\caption{\label{tab:CttsFlames} Periods of the classical T Tauri stars observed with the FLAMES spectrograph
obtained with photometric and spectroscopic diagnostics.}
\begin{center}
\begin{tabular}{llllllllll}
 \hline\hline
Mon ID\tablefootmark{a} & $\mathrm{LC}\_2008$\tablefootmark{b} & $\mathrm{P}_{CoRoT}\_2008$\tablefootmark{b} & $\mathrm{LC}\_2011$\tablefootmark{b}  &  $\mathrm{P}_{CoRoT}\_2011$\tablefootmark{b} & $\mathrm{P}_{\mathrm{H}\alpha}$\tablefootmark{c} & $\mathrm{P}_{\mathrm{HeI}}$\tablefootmark{d} & Regime\tablefootmark{e} & $\mathrm{H}\alpha$ Type\tablefootmark{f} & $\dot{\mathrm{M}}_{\mathrm{H}\alpha}$\tablefootmark{g} \\
 & & $(\mathrm{days})$ &   & $(\mathrm{days})$  & $(\mathrm{days})$ & $(\mathrm{days})$ & & & $(\mathrm{M}_\odot\,/\mathrm{yr})$\\
\hline
CSIMon-001099  &     1  &   3.31  &     1  &   3.38   & NP       & -      & S  & IIB   &  3.23e-09 \\  
CSIMon-000103  &     1  &   1.67  &     1  &   3.35   & NP       & -      &    & I     &  9.61e-09 \\  
CSIMon-000177  &     1  &   3.01  &     1  &   3.01   & NP       & -      &    & I     &  9.29e-09 \\  
CSIMon-000335  &     1  &   4.51  &     1  &   4.58   & NP       & -      &    & I     &  2.31e-09 \\  
CSIMon-000965  &     1  &   9.38  &     1  &   9.68   & NP       & 9.4    & S  & IIR   &  1.76e-09 \\  
CSIMon-000810  &     1  &   2.93  &     1  &   2.93   & NP       & -      &    & I     &  1.24e-09 \\  
CSIMon-000964  &     1  &   3.34  &     1  &   3.32   & NP       & -      &    & I     &  1.49e-09 \\  
CSIMon-000326  &     1  &   7.05  &     3  &          & NP       & NP     &    & I     &  1.99e-09 \\  
CSIMon-000220  &     1  &   0.76  &     3  &          & NP       & -      &    & I     &  2.50e-10 \\  
CSIMon-000804  &     1  &   3.23  &     3  &          & NP       & -      &    & I     &  4.83e-10 \\  
CSIMon-001033  &     1  &  14.15  &     -  &          & 6.30     & -      & S  & IIR   &  4.69e-09 \\  
CSIMon-000250  &     2  &   4.16  &     2  &   8.93   & 7.5/8.9  & 8.6    & S  & IIR   &  1.44e-09 \\  
CSIMon-000660  &     2  &   5.25  &     2  &   5.10   & 5.2      & -      & S  &IIIR   &  1.45e-10 \\  
CSIMon-000498  &     2  &   4.23  &     2  &   4.28   & NP       & -      &    & I     &  1.00e-08 \\  
CSIMon-001199  &     2  &   3.75  &     2  &   3.61   & NP       & -      &    & IIB   &  1.46e-09 \\  
CSIMon-000297  &     2  &   3.16  & 3 (E)  &          & NP       & -      &    & IIR   &  5.75e-09 \\  
CSIMon-000928  &     2  &   4.96  &     3  &          & NP       & 10.3   &    & I     &  6.91e-10 \\  
CSIMon-000654  &     2  &   4.66  & 3 (E)  &          & NP       & NP     &    & I     &  4.82e-09 \\  
CSIMon-000441  &     2  &   4.06  & 3 (E)  &          & NP       & Noise  &    & IIB   &  3.24e-09 \\  
CSIMon-000824  &     2  &   7.05  &     -  &          & NP       & -      & U  & IVR   &  3.28e-09 \\  
CSIMon-001144  & 3 (E)  &         & 3 (E)  &          & NP       & Noise  &    & II    &  2.17e-09 \\  
CSIMon-001275  &     3  &         &     3  &          & NP       & -      &    & IIIB  &  2.51e-08 \\  
CSIMon-000119  &     3  &         &     3  &          & NP       & NP     &    & IIIB  &  3.34e-09 \\  
CSIMon-000558  & 3 (A)  &         &     3  & 11.70?   & 10.5     & 10.7   &    & IIB   &  6.06e-08 \\  
CSIMon-000168  & 3 (E)  &         & 3 (E)  &          & NP       & Noise  &    & II    &  3.77e-09 \\  
CSIMon-000926  &     3  &         &     3  &          & NP       & NP     &    & I     &  8.66e-09 \\  
CSIMon-000681  & 3 (E)  &         & 3 (E)  &          & NP       & NP     &    & IIB   &  4.49e-08 \\  
CSIMon-000280  &     3  &         &     3  &          & NP       & -      &    & IIR   &  5.90e-10 \\  
CSIMon-000290  &     3  &         &     3  &          & NP       &8.3/11.4&    & I     &  2.25e-08 \\  
CSIMon-000328  &     3  &         &     3  &          & NP       & NP     &    & I     &  3.90e-09 \\  
CSIMon-000448  &     3  &         &     3  &          & NP       & Noise  &    & I     &  8.16e-09 \\  
CSIMon-000314  & 3 (E)  &         & 3 (E)  &          & NP       & Noise  &    & I     &  1.06e-08 \\  
CSIMon-000951  &     3  &         &     3  &          & NP       & Noise  &    & I     &  3.21e-08 \\  
CSIMon-000937  &     3  &         &     3  &          & NP       & -      &    & I     &  7.24e-10 \\  
CSIMon-000667  & 3 (E)  &         &  3 (E) &          & 5.9      & -      & S  & IIB   &  1.84e-09 \\  
CSIMon-000945  & 3 (A)  &         &  3 (A) &          & NP       & 7.4    &    & I     &  6.35e-09 \\  
CSIMon-001022  & 3 (A)  &         &  3 (A) &          & NP       &8.9/12.5&    & IIIB  &  6.31e-09 \\  
CSIMon-000996  & 3 (A)  &         &  3 (A) &          & 8.3      & 8.3    &    & I     &  2.79e-08 \\  
CSIMon-000510  & 3 (A)  &         &  3 (A) &          & NP       & NP     &    & I     &  2.87e-08 \\  
CSIMon-000341  & 3 (A)  &         &  3 (A) &          & NP       & NP     &    & I     &  3.55e-08 \\  
CSIMon-000370  &     3  &         &     3  &  11.82?  & NP       & NP     &    & I     &  2.55e-08 \\ 
CSIMon-000765  &     3  &         &     1  &   2.41   & NP       & -      &    & I     &  6.48e-09 \\  
CSIMon-000379  &  3 (E) &         &     2  &   3.68   & NP       & -      & S  & IIR   &  1.26e-09 \\  
CSIMon-001054  &  3 (E) &         &     2  &4.08/8.17*& NP       & 8.1    &    & I     &  1.35e-08 \\  
CSIMon-000296  &  3 (E) &         &     2  &   3.91   & NP       & -      & S  & IIR   &  1.73e-09 \\ 
CSIMon-000811  &  3 (E) &         &     2  &   7.88   & 12.5     & 10.5   & S  & IIR   &  6.75e-09 \\  
CSIMon-000846  &     3  &         &     -  &          & NP       & NP     &    & I     &  -        \\  
CSIMon-000795  &     3  &         &     -  &          & NP       & NP     &    & IIB   &  6.59e-08 \\  
CSIMon-000457  &     3  &         &     -  &          & NP       & -      &    & IIIB  &  6.01e-09 \\  
CSIMon-000260  &  3 (E) &         &     -  &          & NP       & 9.2    &    & IIB   &  2.26e-08 \\  
CSIMon-000893  &     3  &         &     -  &          & NP       & NP     &    & I     &  1.02e-08 \\  
CSIMon-000131  &     -  &         &     3  &          & NP       & 9.1    &    & IIR   &  -        \\  
CSIMon-000423  &     -  &         &     3  &          & NP       & Noise  &    & I     &  1.44e-08 \\ 
CSIMon-000632  &     -  &         &     -  &          & NP       & Noise  &    & IIB   &  1.33e-08 \\ 
CSIMon-001287  &     -  &         &     -  &          & NP       & Noise  &    & IIR   &  9.38e-08 \\ 
CSIMon-000603  &     -  &         &     -  &          & NP       & Noise  &    & IIIB  &  6.07e-08 \\ 
CSIMon-000239  &     -  &         &     -  &          & 5.6      & -      &    & ?     &  -        \\ 
CSIMon-001128  &     -  &         &     -  &          & NP       & Noise  &    & I     &  1.88e-08 \\ 
\hline
\end{tabular}
\end{center}
\tablefoot{This table is ordered by the classification of the CoRoT light-curve morphology from the 2008 campaign.}
\tablefoottext{a}{CSIMon is an internal identification of the CSI 2264 campaign.}
\tablefoottext{b}{CoRoT light-curve morphology and photometric period obtained in this work: 1= spot-like, 2= AA Tau-like, 3= non-periodic light curves. Accretion bursts stars and aperiodic extinction light curves are identified by (A) and (E), respectively.}
\tablefoottext{c}{Period obtained with the $\mathrm{H}\alpha$ line from FLAMES spectra. NP denotes a non-periodic line.}
\tablefoottext{d}{Period obtained with HeI line $6678\mathring{\mathrm{A}}$ from FLAMES spectra.  Noise means that the HeI line is present on the spectrum but it is dominated by noise.}
\tablefoottext{e}{Accretion regime: S-stable and U-unstable, as discussed in Sect. \ref{sec:Inst}.}
\tablefoottext{f}{Morphology of $\mathrm{H}\alpha$ line by \cite{1996A&AS..120..229R} discussed in Sect. \ref{sec:Hatype}.}
\tablefoottext{g}{Mass-accretion rates calculated with $\mathrm{H}\alpha$ equivalent width.}
\tablefoottext{*}{The photometric period was obtained using Scargle periodogram and auto-correlation function, as explained in Sect. \ref{sec:AAHe}.}
\end{table*}

\subsection{$\mathrm{H}\alpha$ periodicity in the AA Tau-like stars}

AA Tau-like stars are expected to be in a stable accretion regime, where a major
funnel flow is formed in each hemisphere, and the base is associated with an inner disk warp that periodically occults the star as the system rotates \citep{2007A&A...463.1017B,2008MNRAS.385.1931K}. 
The photometric period of AA Tau-like stars corresponds to the Keplerian period of the occulting
circumstellar material, and it can be different from the stellar 
rotation period, unless the structure responsible for the occultation of the star is at the
disk corotation radius. As shown in Sect. \ref{sec:Period}, at least in our sample,
the photometric periods of AA Tau-like systems is not expected
to differ substantially from the stellar 
rotation period, indicating that the inner disk warp is indeed located close to the disk corotation radius. 
Eight AA Tau-like stars were observed 
with the FLAMES spectrograph, but only $\text{three}$ are periodic in 
$\mathrm{H}\alpha$ (see Fig. \ref{fig:periodAA}). 
This is somewhat unexpected. If the $\text{eight}$ stars were in a stable
accretion regime, it would naturally produce periodic variability in $\mathrm{H}\alpha$
if the line were mostly formed in the accretion funnel. However, if the accretion flow
is significantly time variable, it could mask the periodicity imposed by the stellar rotation. 
We will explore evidence for the time variability of accretion during
funnel-flow accretion in a forthcoming paper (Stauffer et al., in preparation).

The star Mon-000660 (V354 Mon), Fig. \ref{fig:AA2}bc, is an AA Tau-like analog that presents the same period in 
$\mathrm{H}\alpha$ and the photometry. Assuming that $\mathrm{H}\alpha$ is formed 
mainly in the accretion funnel and that the measured period corresponds to the stellar rotational 
period, it confirms that the occulting disk structure, which causes the photometric variability, is 
located at the corotation radius, as in AA Tau itself \citep{2007A&A...463.1017B}. 
Another AA Tau-like star, Mon-000250, Fig. \ref{fig:AA1}ef, is also periodic in $\mathrm{H}\alpha$ 
and presents the same period in the photometry. In the $\mathrm{H}\alpha$ periodogram, this star 
shows two periodic signals in the red wing, the stronger power peak coincident with 
the photometric period ($\sim 8.9\,\mathrm{days}$) and the weaker peak 
around $\sim 7.5\,\mathrm{days}$. This second period is also detected in the CoRoT light curve, but 
with a lower power than the main periodic signal, and is due to the periodicity of a small structure
present inside the deepest dips of the light curve (see Fig. \ref{fig:C-AA1}).

On the other hand, star Mon-000811, Fig. \ref{fig:AA3}hi, which is also AA Tau-like, only shows a faint hint of periodicity
in $\mathrm{H}\alpha$ at the photometric period ($\sim 7.9\,\mathrm{days}$),
and presents a strong periodic detection at about $13$ days in the $\mathrm{H}\alpha$
blue wing. This star also presents two periodic signals in the HeI $6678\,\mathring{\mathrm{A}}$
line, one at the photometric period and another at about $11$ days, as shown in Sect. \ref{sec:heI}.
The HeI line is expected to be formed near the hot spot, close to the stellar photosphere, so one
of its periodicities is probably close to the stellar rotation period. If the approximately
$ \text{eight}$-day period of HeI corresponds to the stellar rotational period,
the structure that occults the star could be located close to the corotation radius
of the star-disk system. This is the only star that presents a periodicity 
only in the blue wing of the $\mathrm{H}\alpha$ line. Mon-000811 can be compared to the CTTS
SU Aur, which exhibited a blueshifted absorption in $\mathrm{H}\alpha$ that is periodic 
\citep{1994ASPC...64..190J}, and thought to be due to a wind \citep{1993ApJS...89..321G}. 
Its variability is explained as rotational modulation of 
the wind by the magnetic field of SU Aur. Mon-000811 shows no 
absorption in the blue wing of $\mathrm{H}\alpha$ (as seen in Fig. \ref{fig:S-AA3}), and we cannot 
easily associate the observed periodicity in the blue wing with a wind. However, the blue-wing period is longer 
than the HeI period (which we associate with the rotation period), so the material responsible for 
the $\mathrm{H}\alpha$ blue-wing periodicity is probably located outside the corotation radius 
and could indeed come from a disk wind. 

As discussed above, three AA Tau-like stars present the same periodicity in the photometric and spectroscopic 
analysis. Since the spectroscopic periods most likely correspond to the stellar rotational periods, we
can assume that, at least in these three cases, the inner disk warp is located at the
corotation radius. Following \cite{2008A&A...478..155B}, we computed the star-disk corotation radius 
and calculated the magnetic field strength needed to enforce disk truncation at the corotation radius. 
The results are presented in Table \ref{tab:AApar}. At a few stellar radii from the star, the star-disk 
interaction is expected to occur mostly through the dipole component of the stellar magnetic field, and
the calculated magnetic fields in Table \ref{tab:AApar} are within the range of values of the few dipole 
components measured in CTTSs ($\mathrm{B}_{\mathrm{dip}}=0.02\,\mathrm{kG}$ to $1.9\,\mathrm{kG}$) \citep[see Table 2 from][]{2012ApJ...755...97G}. 

\begin{table*}[htb!]
\tiny
\caption{\label{tab:AApar}Parameters used to calculate corotation radius ($\mathrm{R}_C$)
and magnetic field strength ($\mathrm{B}_*$) of the three AA Tau-like stars that present observational evidence
that the inner disk warp is located at the corotation radius.}
\begin{center}
\begin{tabular}{llllllll}
 \hline\hline
Mon ID\tablefootmark{a} & $\mathrm{M}_*$\tablefootmark{b}& $\mathrm{R}_*$\tablefootmark{c} & $\dot{\mathrm{M}}_{\mathrm{H}\alpha}$\tablefootmark{d} & $\mathrm{P}_{CoRoT}\_2011$\tablefootmark{e} & $\mathrm{P}_{\mathrm{H}\alpha}$\tablefootmark{f} & $\mathrm{R}_C$\tablefootmark{g} & $\mathrm{B}_*$\tablefootmark{h} \\
 & $(\mathrm{M}_\odot)$ & $(\mathrm{R}_\odot)$ & $(\mathrm{M}_\odot\,/\mathrm{yr})$ & $(\mathrm{days})$ & $(\mathrm{days})$ & $(\mathrm{R}_*)$ & (kG)  \\
\hline
CSIMon-000250  & 1.63   & 1.35 &  1.44e-09 &   8.93   & 7.5/8.9 & 12.3 & 0.81 \\  
CSIMon-000660  & 1.86   & 1.4  &  1.45e-10 &   5.10   & 5.2     & 7.5  & 0.07 \\  
CSIMon-000811  & 0.91   & 1.97 & 6.75e-09  &   7.88   &7.9/12.5*& 8.2  & 0.44 \\  
\hline
\end{tabular}
\end{center}
\tablefoot{}
\tablefoottext{a}{CSIMon is an internal identification of the CSI 2264 campaign.}
\tablefoottext{b}{Stellar mass obtained by \cite{2014A&A...570A..82V}.}
\tablefoottext{c}{Stellar radius obtained by \cite{2014A&A...570A..82V}.}
\tablefoottext{d}{Mass-accretion rates calculated with $\mathrm{H}\alpha$ equivalent width.}
\tablefoottext{e}{Photometric period obtained from the 2011 CoRoT light curve.}
\tablefoottext{f}{Period obtained with the $\mathrm{H}\alpha$ line from FLAMES spectra.}
\tablefoottext{g}{Calculated corotation radius.}
\tablefoottext{h}{Calculated stellar magnetic field.} 
\tablefoottext{*}{This star shows only a faint hint of periodicity in $\mathrm{H}\alpha$ at the 
photometric period ($\sim 7.9\,\mathrm{days}$) that we used as the stellar rotational period.}
\end{table*}

\subsection{$\mathrm{H}\alpha$ periodicity in stars classified as non-periodic with the CoRoT photometry} \label{sec:haAcc}

\begin{figure*}
\begin{center}
\subfigure[]{\label{fig:Acc1}\includegraphics[width=4.5cm]{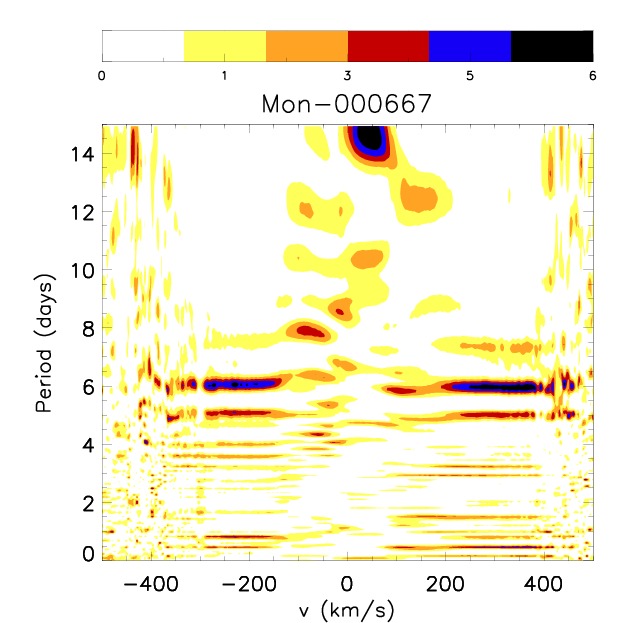}}
 \subfigure[]{\label{fig:S-Acc1}\includegraphics[width=4.5cm]{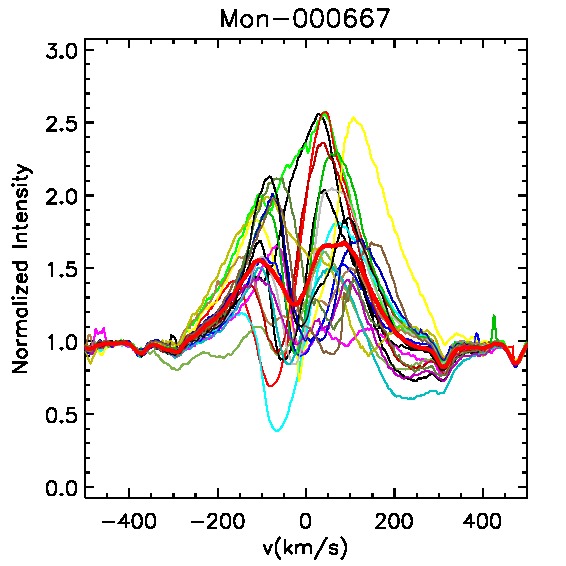}}
 \subfigure[]{\label{fig:C-Acc1}\includegraphics[width=4.5cm]{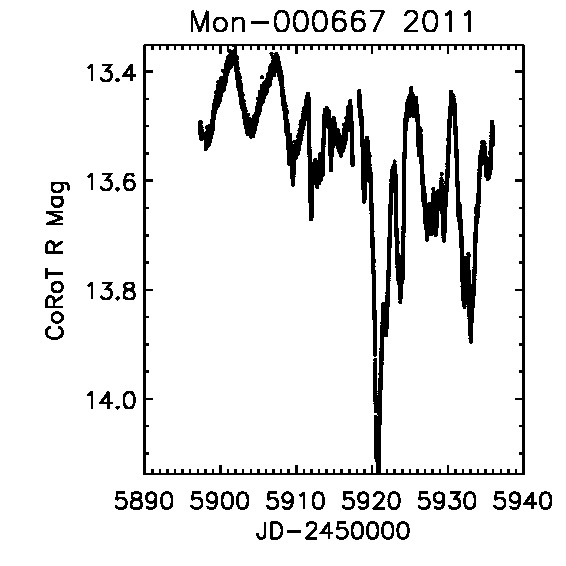}}\\
 \subfigure[]{\label{fig:Acc2}\includegraphics[width=4.5cm]{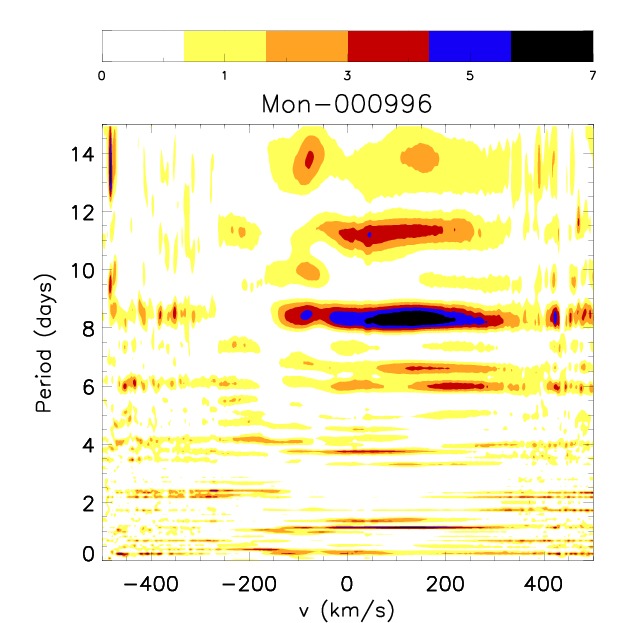}}
 \subfigure[]{\label{fig:S-Acc2}\includegraphics[width=4.5cm]{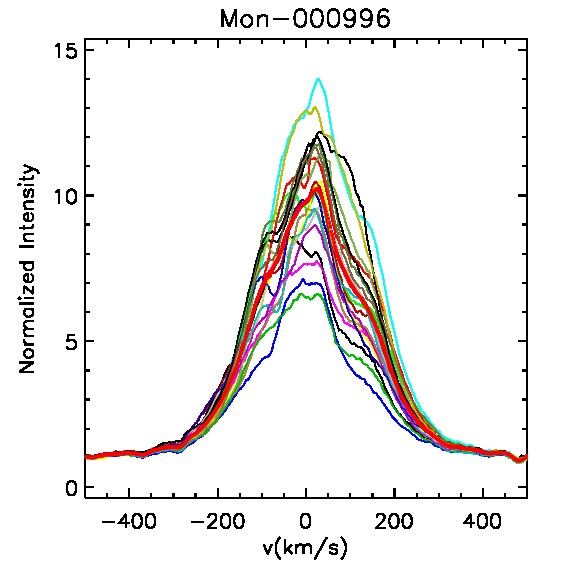}}
 \subfigure[]{\label{fig:C-Acc2}\includegraphics[width=4.5cm]{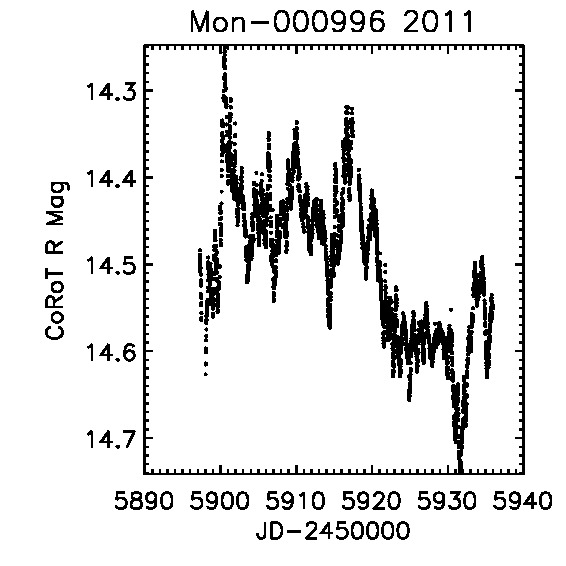}}\\
 \subfigure[]{\label{fig:NP1}\includegraphics[width=4.5cm]{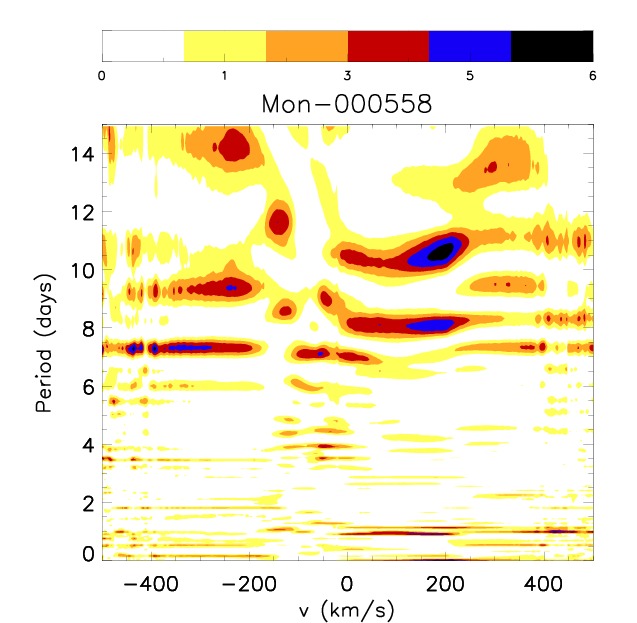}}
 \subfigure[]{\label{fig:S-NP1}\includegraphics[width=4.5cm]{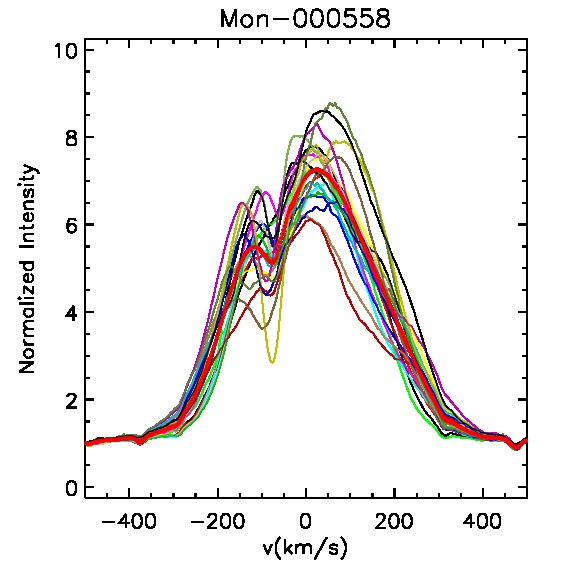}}
 \subfigure[]{\label{fig:C-NP1}\includegraphics[width=4.5cm]{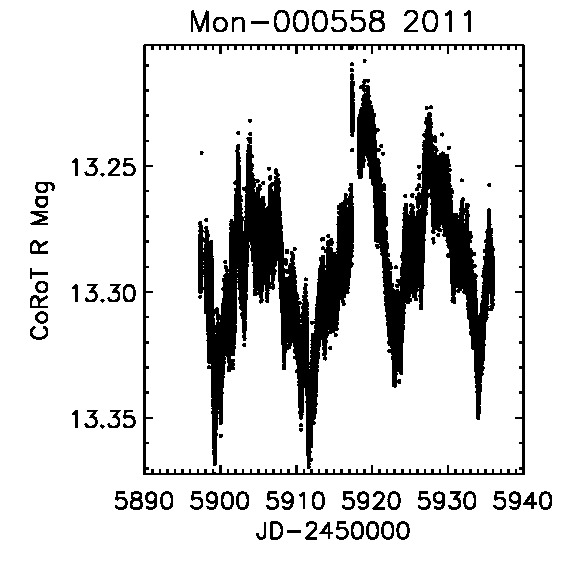}}\\
\end{center}
 \caption{\label{fig:periodNPAcc}Periodic stars in the $\mathrm{H}\alpha$ line that are not periodic in the CoRoT photometry. Left column: Bidimensional periodograms of the $\mathrm{H}\alpha$ line. The color code represents the power of periodogram, varying from zero (white) to the maximum power intensity (black). Middle column: $\mathrm{H}\alpha$ line profiles. Different colors correspond to different nights of observation, and the thick red line is the average line profile. Right column: CoRoT light curves from the 2011 campaign.}
\end{figure*}

The stars Mon-000667 (random occultation by circumstellar material), Mon-000996 (accretion burst), and Mon-000558 (irregular 
variability, but not associated with a main physical process) showed no well-defined periodicity in their CoRoT light curves 
in 2008 and in 2011, as shown in Table \ref{tab:CttsFlames}. These stars are periodic in $\mathrm{H}\alpha$, however, 
as seen in Figs. \ref{fig:Acc1}, \ref{fig:Acc2}, and \ref{fig:NP1}.

Mon-000667 has a period of $5.92\,\mathrm{days}$ 
measured by \cite{2004A&A...417..557L}, which is very similar to our detected period 
($\sim 5.9 \, \mathrm{days}$) and
could indeed correspond to the stellar rotational period. 
In this case, the accreting gas is apparently organized,
but the inner dust environment does not show a clear correlation with the gas structure.

We could not find any information about the rotational period of Mon-000996 in the literature, 
but it presents an $\mathrm{H}\alpha$ period ($\sim 8.3 \,\mathrm{days}$) that is
equal to the period found in the 
HeI $6678\,\mathring{\mathrm{A}}$ line (see also Fig. \ref{fig:Acc2He}; we discuss this in Sect. \ref{sec:heI}). This is quite
unexpected, since its light curve is dominated by random hot spot variability, which would
suggest unstable accretion and not a globally organized accretion environment. 
This star also presents a similar periodicity ($\sim8.1\,\mathrm{days}$) in the {\it Spitzer} (IRAC) 
light curve at $3.6\,\mu\mathrm{m}$ and $4.5\,\mu\mathrm{m}$ \citep{2014AJ....147...82C}. 
Recently, \cite{2015arXiv150101948B} divided the unstable accretion regime into two subclasses: 
ordered unstable (one or two unstable tongues) and chaotic unstable (several unstable tongues). 
A star in the ordered unstable accretion regime may retain 
some periodicity in its accretion bursts, as may be the case for Mon-000996, which presents 
only a low-power period detection in its light-curve periodogram at about $8.7$ days.

Mon-000558 was classified by \cite{2014AJ....147...82C}, using the same CoRoT data, 
as quasi-periodic, since its light curve includes low-amplitude stochastic variability superposed 
on some slightly periodic signal. 
The possible photometric period is $\sim 11.7\,\mathrm{days}$. In the literature, it also presents a period of 
$0.88\,\mathrm{days}$ \citep{2006A&A...455..903F}, which is very different from the $\mathrm{H}\alpha$ 
and the photometric period from our survey, however. We show in the next subsection that the $\mathrm{H}\alpha$ 
period of $\sim 10.5\,\mathrm{days}$ is also seen in the HeI line (see Table \ref{tab:CttsFlames}), which may indicate 
that it corresponds to the stellar rotation period. This star had an accretion burst CoRoT light curve in 
2008, while in 2011 it has an irregular, possibly hot-spot light curve. This is another system that is a
candidate to be in the ordered unstable regime described by \cite{2015arXiv150101948B}.

\subsection{$\mathrm{H}\alpha$ periodicity in stars not observed by CoRoT in 2011}

\begin{figure*}
\begin{center}
 \subfigure[]{\label{fig:NO}\includegraphics[width=4.5cm]{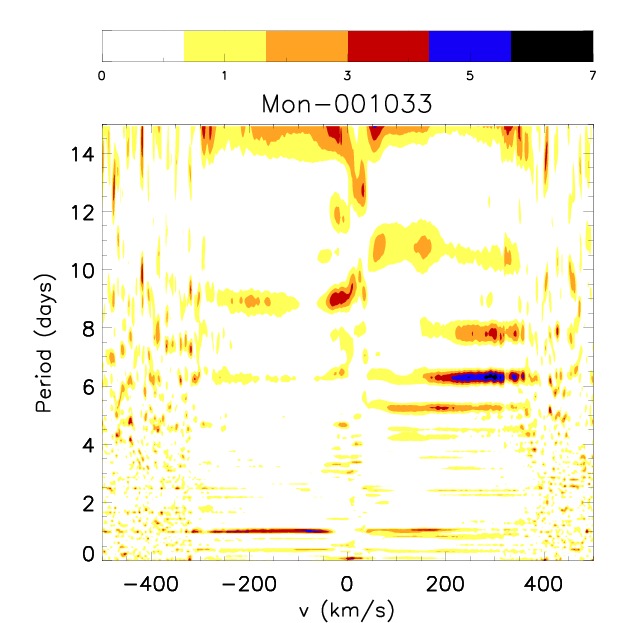}}
 \subfigure[]{\label{fig:S-NO}\includegraphics[width=4.5cm]{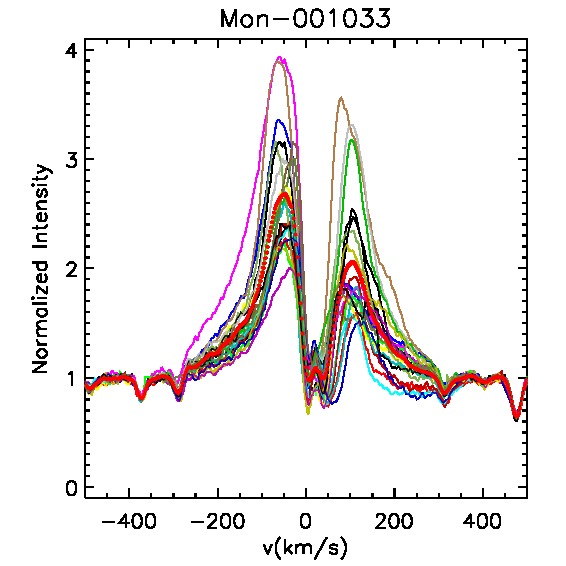}}
 \subfigure[]{\label{fig:C-NO}\includegraphics[width=4.5cm]{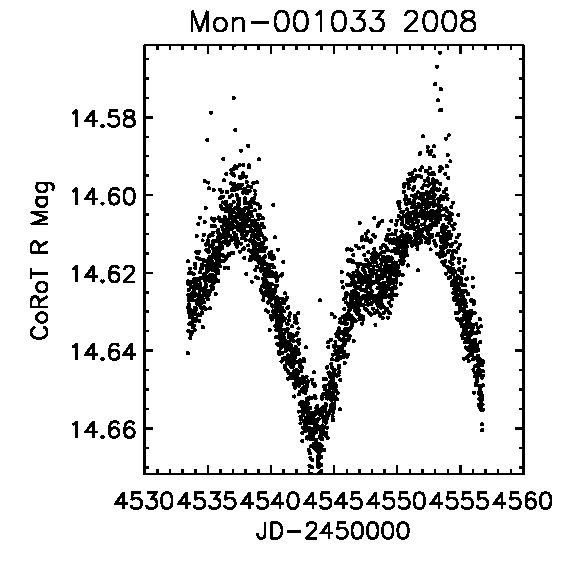}}\\
 \subfigure[]{\label{fig:NO2}\includegraphics[width=4.5cm]{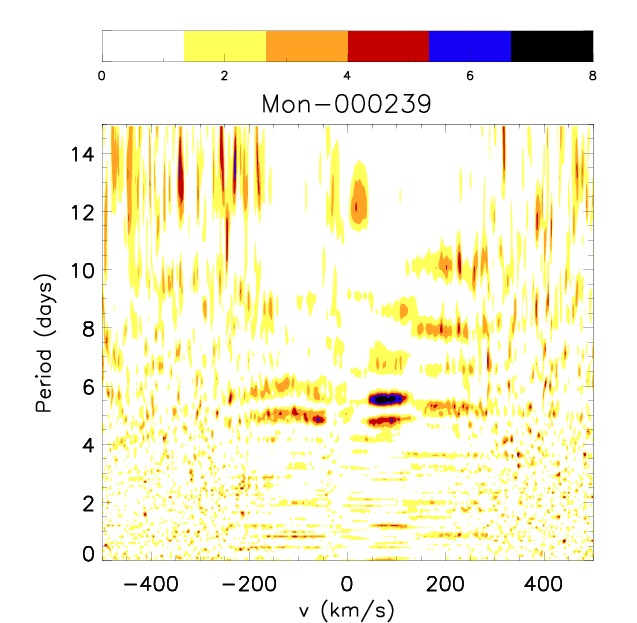}}
 \subfigure[]{\label{fig:S-NO2}\includegraphics[width=4.5cm]{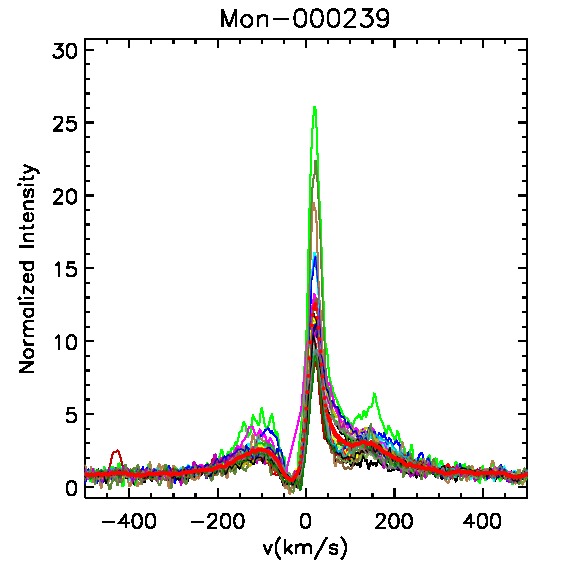}}
 \end{center}
 \caption{\label{fig:periodNPC}Periodic stars in the $\mathrm{H}\alpha$ line that were not observed by CoRoT in 2011. 
Left column: Bidimensional periodograms of the $\mathrm{H}\alpha$ line. The color code represents the power of the periodogram, varying from zero (white) to the maximum power intensity (black). Middle column: $\mathrm{H}\alpha$ line profiles. Different colors correspond to different nights of observation, and the thick red line is the average line profile. Right column: CoRoT light curve from 2008, when available.}
\end{figure*}

Two stars that show $\mathrm{H}\alpha$ line periodicity were not observed by CoRoT in 2011. 
The star Mon-001033, Fig. \ref{fig:NO}, although not observed by CoRoT in 2011, 
shows a possible period of $14$ days measured in 2011 with the \textit{Spitzer} (IRAC) light curve in the CSI2264 
campaign \citep{2014AJ....147...82C}. This star was observed by CoRoT in 2008, and its light curve was 
classified as spot-like, 
with a period of $14.15\,\mathrm{days}$, which agrees with the IRAC period, but
is more than twice that obtained with the $\mathrm{H}\alpha$ line ($\sim 6.30\,\mathrm{days}$). 
The periodic region in $\mathrm{H}\alpha$ corresponds to a narrow velocity range, where a 
redshifted absorption is present in the spectrum. 

The other star that presented a periodicity in $\mathrm{H}\alpha$, but
was not observed by CoRoT, is Mon-000239. This star does not have 
any period information in the literature and is not periodic with {\it Spitzer} 
photometry. The detected period in $\mathrm{H}\alpha$ occurs in a very limited profile region associated with a 
small redshifted absorption feature, like the one observed in Mon-001033. 

\subsection{$\mathrm{H}\alpha$ periodicity in spot-like stars} \label{sec:SpotHa}

\begin{figure*}
  \begin{center}
 \includegraphics[width=4.45cm]{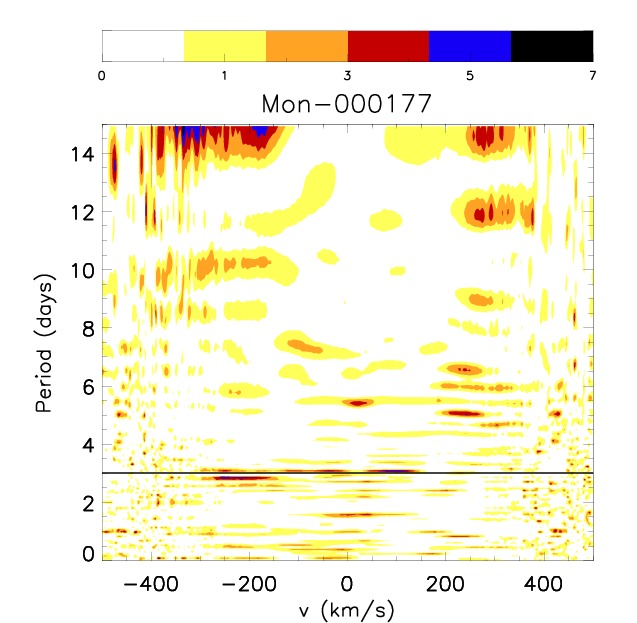}
 \includegraphics[width=4.45cm]{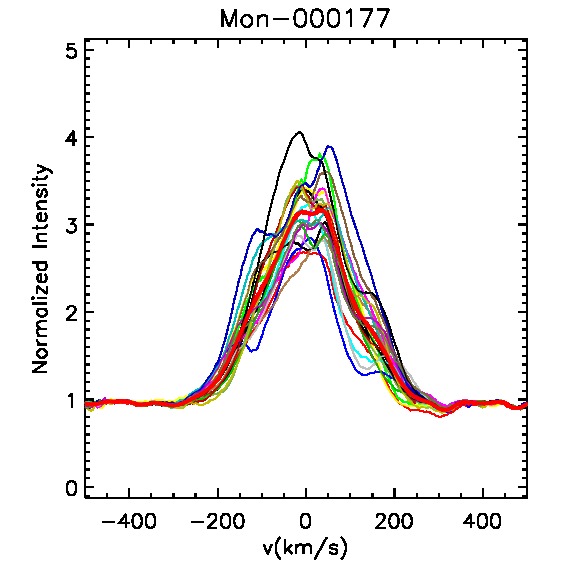}
 \includegraphics[width=4.45cm]{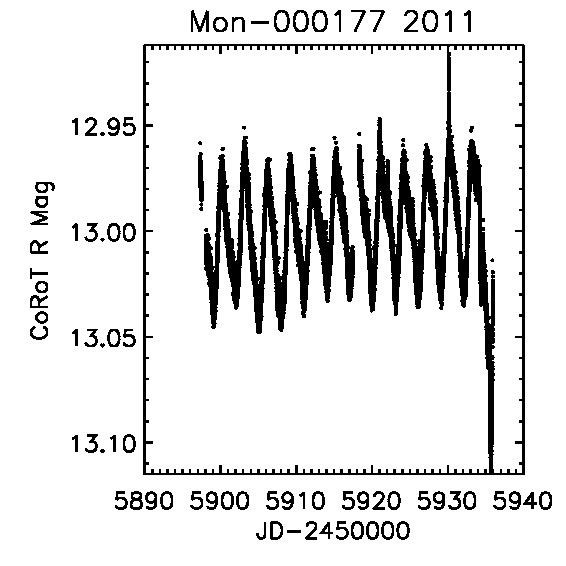}\\
 \includegraphics[width=4.45cm]{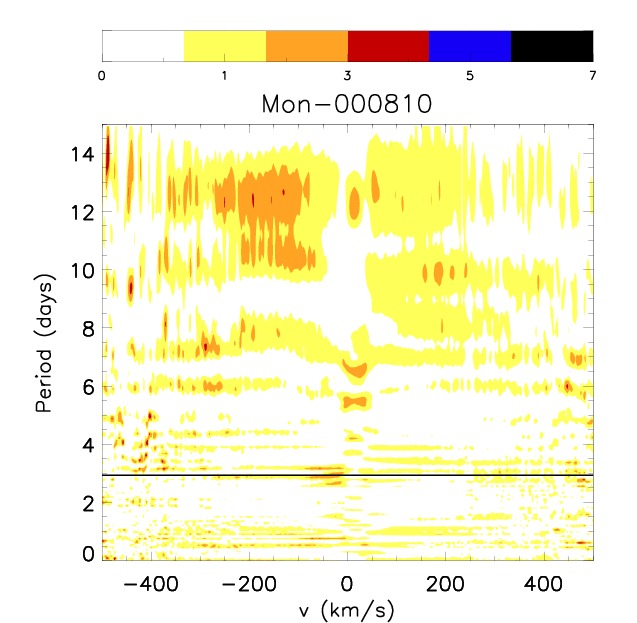}
 \includegraphics[width=4.45cm]{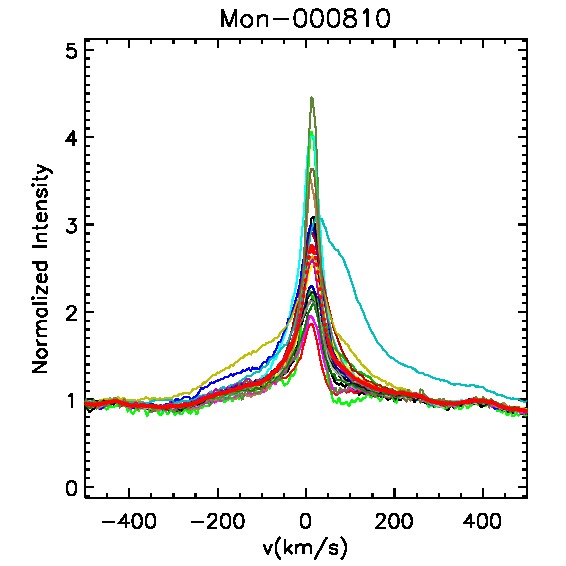}
 \includegraphics[width=4.45cm]{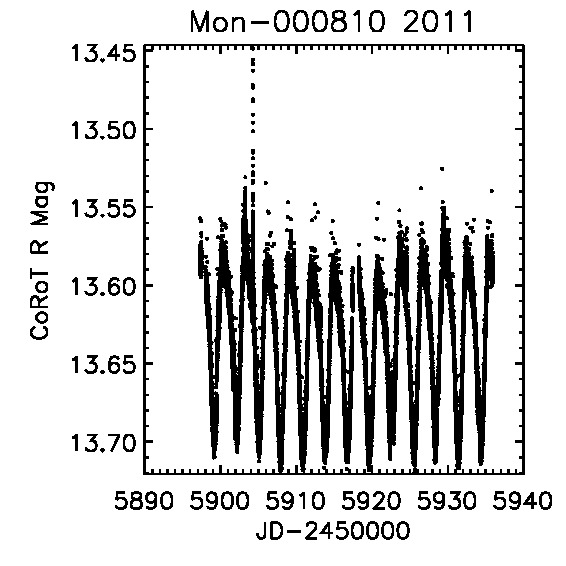}
  \end{center}
 \caption{\label{fig:NperHa} Spectroscopic and photometric variability of spot-like systems. Left column: Bidimensional periodogram of the $\mathrm{H}\alpha$ 
 line of two spot-like systems. The color code represents the power of periodogram, varying from zero (white) to the maximum power intensity (black). 
 Black lines correspond to the photometric period of 
 the stellar light curve observed with CoRoT in 2011. 
 Middle column: $\mathrm{H}\alpha$ line profiles. Different colors correspond to different nights 
 of observation, and the thick red line is the average line profile. Right column: CoRoT light 
 curve from 2011.
 None of the spot-like systems observed with FLAMES showed any periodicity
 in the $\mathrm{H}\alpha$ line.} 
 \end{figure*}

Eight CTTSs classified as spot-like with the 2011 CoRoT light curves were observed with
FLAMES. However, despite being periodic in the photometry, none of them showed any periodicity 
in the $\mathrm{H}\alpha$ line. Two examples are presented in Fig. \ref{fig:NperHa}. The photometric 
periodicity of spot-like stars is mostly due to cold spots on the photosphere, while the 
$\mathrm{H}\alpha$ line is mainly formed in the accretion 
funnel. Results from spectropolarimetry have shown that the main hot 
and cold spots tend to coincide in accreting stars \citep{2010MNRAS.409.1347D,2011MNRAS.412.2454D,2011MNRAS.417..472D}, and 
we would therefore expect the cold spot and the main accretion funnel, which ends in the hot spot, 
to present the same periodicity. At the same time, the $\mathrm{H}\alpha$ line profiles of spot-like systems
tend to be weak and may therefore have a significant chromospheric contribution. 
In some spot-like systems, the lack of $\mathrm{H}\alpha$ line periodicity could 
then be due to a rather uniform chromospheric emission contributing significantly to the profile, 
which would dilute the $\mathrm{H}\alpha$ modulation from the accretion funnel.

\section{Periodograms of the $\mathrm{HeI}\,6678\mathring{\mathrm{A}}$ emission line}\label{sec:heI}

The HeI $6678\,\mathring{\mathrm{A}}$ singlet line is often found in emission in stars with a 
moderate to high accretion rate \citep{1998AJ....116..455M,beristain01,2003ApJ...590..348L}.
Like the HeI $5876\,\mathring{\mathrm{A}}$ triplet line, it may present a narrow (NC) and a broad (BC)
component, with different kinematic characteristics. In the spectra of CTTSs, the NC is found to be slightly 
redshifted and circularly polarized and is interpreted to arise from gas in the post-shock region near
the stellar surface \citep{beristain01, edwards03}.
The BC is more complex. It can be found to be redshifted, but is most commonly observed blueshifted
in CTTSs.  The current interpretation is that it may come from the accretion funnel or the base of
a hot accretion-powered stellar wind \citep{beristain01,edwards06}.

Depending on the star-disk inclination with respect to our line of sight, the hot spot may be hidden by the inner disk
and remain invisible most of the time. In that case, we would not expect to see the
HeI line in emission, even at high accretion rates. This is probably the case of
the AA Tau-like system Mon-000660, which is seen at high inclination \citep[$79^\circ\,\pm\,11^\circ$,][]{Fonsecapreparation}
and does not present UV excess or $\mathrm{HeI}\, 6678\,\mathring{\mathrm{A}}$ in emission,
despite being actively accreting, as shown by its extended $\mathrm{H}\alpha$ emission profile.

The HeI NC is formed close to the stellar surface, therefore its 
periodicity, whenever present, should correspond closely to the stellar rotation period. 
We expect to detect periodic variability if accretion is in a stable regime, 
which creates a main accretion funnel and consequently a main 
hot spot in each hemisphere. 

We found that $24$ of the $58$ CTTSs observed with FLAMES 
clearly presented the $\mathrm{HeI}\, 6678\,\mathring{\mathrm{A}}$ line in emission. Another set of $11$ CTTSs
also show HeI in emission, but the S/N is too low in the line to allow a reliable analysis of line profile variability. 
According to \cite{beristain01}, the HeI NC is characterized by line widths smaller than 
$50\,\mathrm{km/s}$. In our sample, the width of the mean HeI $6678\,\mathring{\mathrm{A}}$ line profile
ranged from $29$ to $47\,\mathrm{km/s}$, depending on the target, with a mean value of 
$39.3\,\mathrm{km/s}$. Therefore, in our sample, all of the stars that showed HeI $6678\,\mathring{\mathrm{A}}$ 
in emission only presented an NC. Close to the HeI $6678\,\mathring{\mathrm{A}}$ line, there is a 
photospheric absorption FeI line (at $6677\,\mathring{\mathrm{A}}$). We used residual spectra to calculate the 
HeI $6678\,\mathring{\mathrm{A}}$ line equivalent widths because of this line. The 
FeI $6677\,\mathring{\mathrm{A}}$ line was subtracted from these
spectra, using as template the spectrum of a WTTS of the 
same spectral type of each star.

\begin{figure*}[ht]
 \begin{center}
\subfigure[]{\label{fig:AAHe}\includegraphics[width=4.4cm]{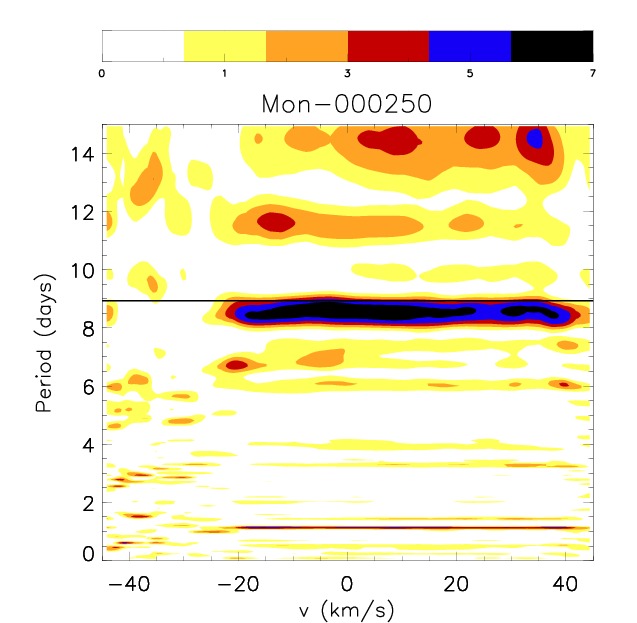}}
\subfigure[]{\label{fi:AAHeSp}\includegraphics[width=4.4cm]{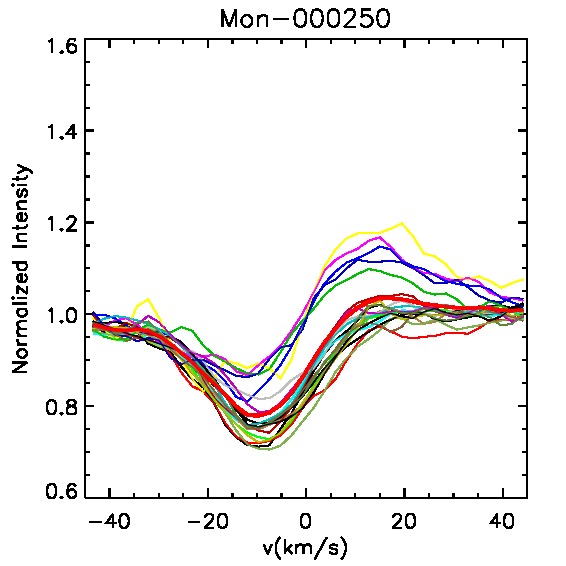}}
\subfigure[]{\label{fig:AA2He}\includegraphics[width=4.4cm]{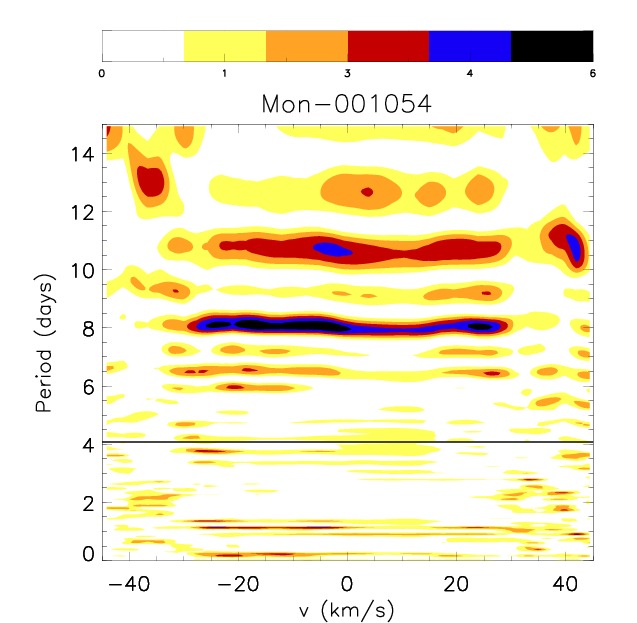}}
\subfigure[]{\label{fig:AA2He1Sp}\includegraphics[width=4.4cm]{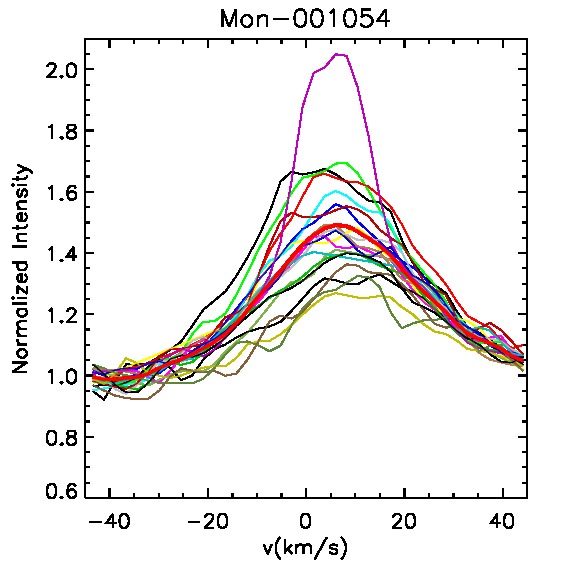}}\\
\subfigure[]{\label{fig:AA3He}\includegraphics[width=4.4cm]{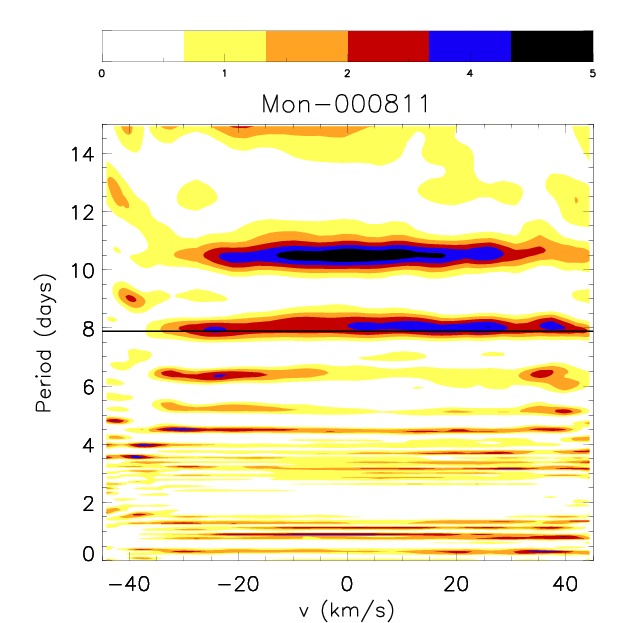}}
\subfigure[]{\label{fig:AA3HeSp}\includegraphics[width=4.4cm]{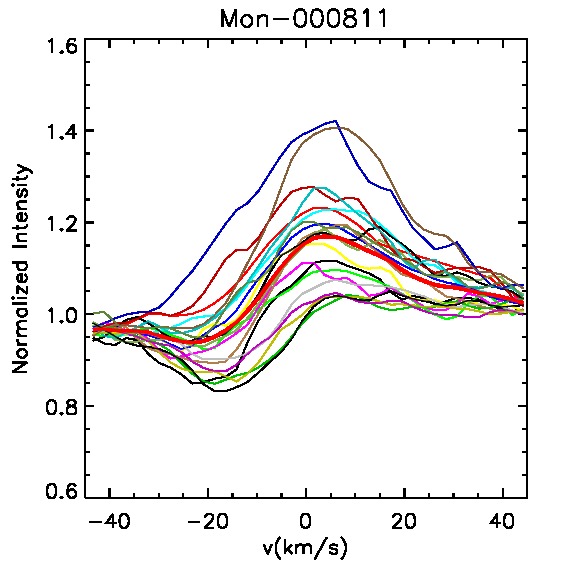}}
\subfigure[]{\label{fig:SpotHe}\includegraphics[width=4.4cm]{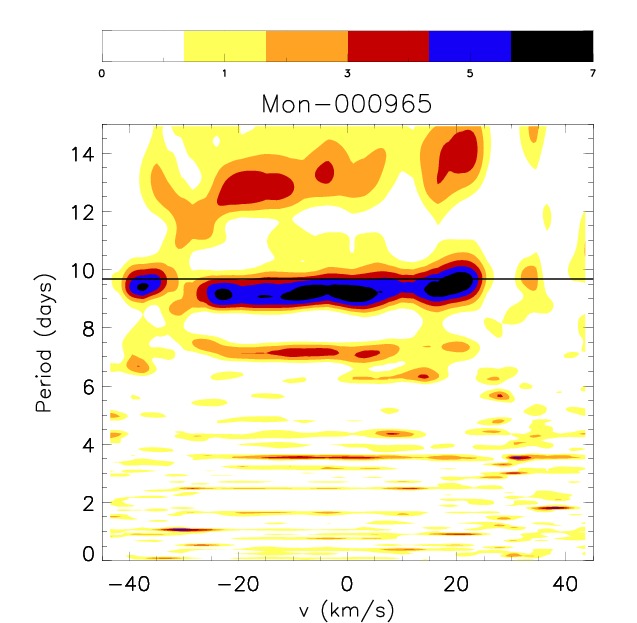}}
\subfigure[]{\label{fig:SpotHeSp}\includegraphics[width=4.4cm]{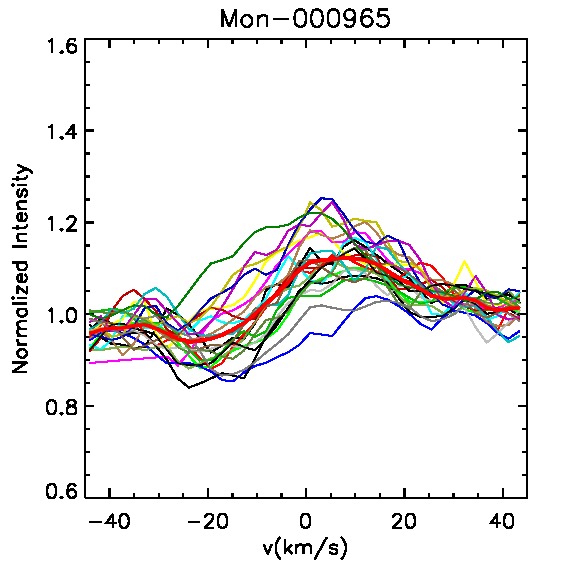}}
 \end{center}
\caption{\label{fig:PerHeI}Periodic stars in the HeI $6678\,\mathring{\mathrm{A}}$ that also showed
a periodicity in the CoRoT light curves. Bidimensional periodograms of the $\mathrm{HeI}$ line 
(first and third columns). The color code represents the power of the periodogram, varying from 
zero (white) to the maximum power intensity (black). The black 
horizontal lines correspond to the period of the 2011 CoRoT light curve. The second and fourth
columns present the HeI $6678\,\mathring{\mathrm{A}}$ lines used to calculate the periodograms. 
Different colors correspond to different nights of observation, and the thick red line is the average line profile.}
\end{figure*}

We calculated the HeI line periodograms as was done for the $\mathrm{H}\alpha$ line (see Sect. \ref{sec:per_halpha}). The periodograms of 
the HeI lines are shown in Figs. \ref{fig:PerHeI} and \ref{fig:PerHeI2}.
Of the $24$ CTTSs with clear HeI emission, $12$ stars showed no periodicity in this line, and
they were also non-periodic in the CoRoT photometry.
The non-periodicity of half of the HeI emission stars may indicate the intermittent formation of several
short-lived hot spots near the stellar surface, instead of one well-organized hot spot in each hemisphere.

Of the $12$ systems that showed a periodicity in the HeI line, $\text{two}$  presented more than one period in this line (as seen in Table \ref{tab:CttsFlames} and Figs. \ref{fig:He-3} and \ref{fig:He-5}), which may be because of the formation of several long-lived (compared to the time span of our observations) 
hot spots at different latitudes on the stellar surface, rather than a main hot spot on each hemisphere. 
In the next subsections, we present the analysis of the HeI line periodicity compared to the CoRoT light-curve morphology
discussed in Sect. \ref{sec:corot}.

\subsection{HeI line in AA Tau like stars}\label{sec:AAHe}

AA Tau-like stars are seen at high inclination with respect to our line of sight, and the 
inner disk is expected to hide part of the stellar photosphere as the system rotates. Depending on the
system geometry, the hot spot, and consequently the region where the HeI NC forms, may also be hidden 
in these systems. 
Only three of eight AA Tau-like stars observed with FLAMES showed the HeI line in 
emission, and these three systems presented a periodicity in this line. The HeI line
periodograms of Mon-000250, Mon-001054, and Mon-000811 are shown in Fig. \ref{fig:PerHeI}
and the measured periods are presented in Table \ref{tab:CttsFlames}.

Mon-000250 shows the same periodicity in HeI and $\mathrm{H}\alpha$ lines (Figs. 
\ref{fig:AAHe} and \ref{fig:AA1}), and also presents the same period in the CoRoT light curve, 
which confirms that the structure responsible for the stellar occultation is located close to the 
co-rotation radius of the star-disk system. 

Mon-001054 (Fig. \ref{fig:AA2He}) shows a HeI period of $\sim 8.1\,\mathrm{days}$, that is, 
twice the CoRoT photometric period from 2011. The photometric period, calculated with the 
modified Scargle periodogram, as explained in Sect. \ref{sec:corot}, presents a low detection power. 
Using a different period search method, such as the auto-correlation function \citep{2014AJ....147...82C,2013MNRAS.432.1203M}, we find 
a $\sim8.2\,\mathrm{day}$ period in the CoRoT light curve, but the detection significance
is again very low. From the literature, Mon-001054 has a photometric period of $\sim 7.8\,\mathrm{days}$ 
\citep{2004A&A...417..557L}, which agrees with our HeI period
and could correspond to the stellar rotation period.

Mon-000811 (Fig. \ref{fig:AA3He}) is periodic
in HeI ($\sim 10.5$, higher power and $\sim 8\,\mathrm{days}$ lower power) and $\mathrm{H}\alpha$ ($\sim 12.5\,\mathrm{days}$) 
(Figs. \ref{fig:AA3He}, \ref{fig:AA3}) and shows periodicity in the CoRoT light curve ($\sim 7.88\,\mathrm{days}$). 
However, these periods are different from each other. As discussed above, this star is classified as AA Tau-like, and
we would therefore expect accretion to occur in a stable regime, if the inner disk
warp is associated with the base of the accretion column, and forming
a major hot spot in each hemisphere at the top of the accretion column.
One possible scenario to explain this complex system is that the HeI line $\text{eight}$-day period
corresponds to the stellar rotation period, which is close to the photometric period,
and therefore the inner disk warp that causes the photometric variability is located
close to the co-rotation radius. The $\mathrm{H}\alpha$ $\sim 12.5$-day period is observed 
in the blue wing of the emission profiles and could then correspond
to emission or absorption from a wind located outside the corotation radius.  

\subsection{HeI line periodicity in spot-like stars} \label{sec:SpotHe}

Because of the high stability of the spot-like light curves on the timescale of our observations
and the low mass-accretion rates of these systems,
we initially assumed that large cold spots were the main cause of the observed photometric variability of spot-like systems.
If, instead, hot spots were causing the photometric variability, we might expect to observe the same type
of variability in the HeI line.
We observed $\text{eight}$ spot-like systems with FLAMES, and only one, Mon-000965, has the HeI line
in emission, as seen in Table \ref{tab:CttsFlames}.
The HeI periodogram of this star is shown in Fig. \ref{fig:SpotHe}, and it presents a clear periodicity
that coincides with the period obtained with the CoRoT photometry. Mon-000965 is not clearly periodic
in $\mathrm{H}\alpha$ (Sect. \ref{sec:SpotHa}), but a faint periodic signal (low power in the 
periodogram) is seen at the photometric period.
The fact that this star has the HeI NC line in emission indicates that the 
hot spot is prominent, which suggests that its light-curve variability might be dominated by a stable 
hot spot instead of a cold spot, as initially assumed during our 2011 campaign. 

\subsection{Periodic stars in the HeI line that are not periodic in the CoRoT photometry or were not observed by CoRoT in 2011}

\begin{figure*}[ht]
 \begin{center}
\subfigure[]{\label{fig:He-1}\includegraphics[width=4.4cm]{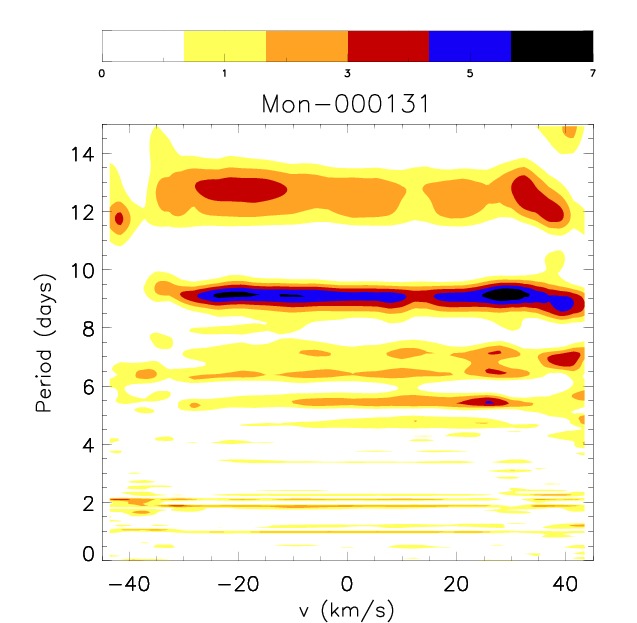}}
\subfigure[]{\label{fi:He-1sp}\includegraphics[width=4.4cm]{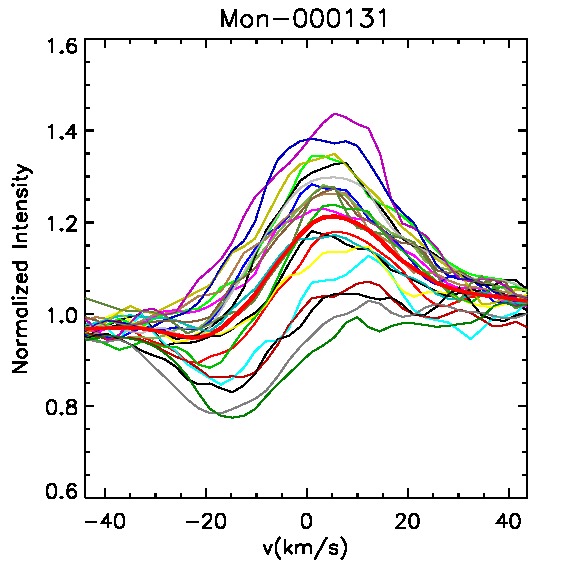}}
\subfigure[]{\label{fig:He-2}\includegraphics[width=4.4cm]{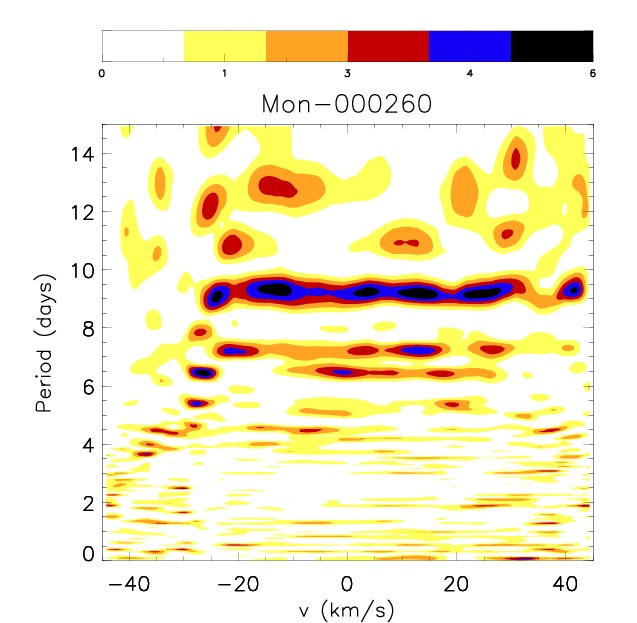}}
\subfigure[]{\label{fig:He_2sp}\includegraphics[width=4.4cm]{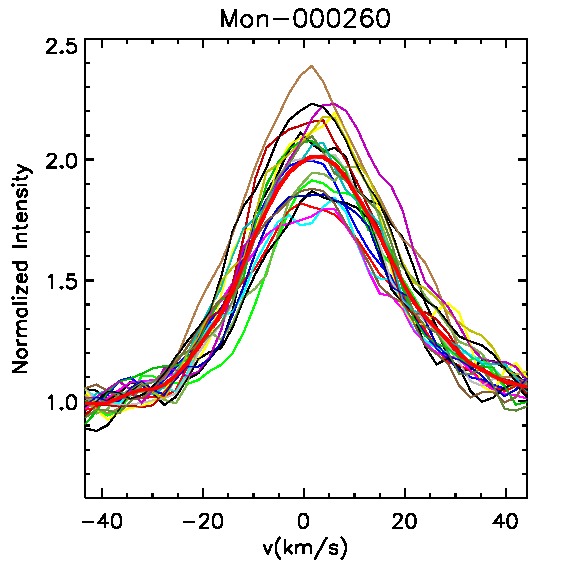}}
\subfigure[]{\label{fig:He-3}\includegraphics[width=4.4cm]{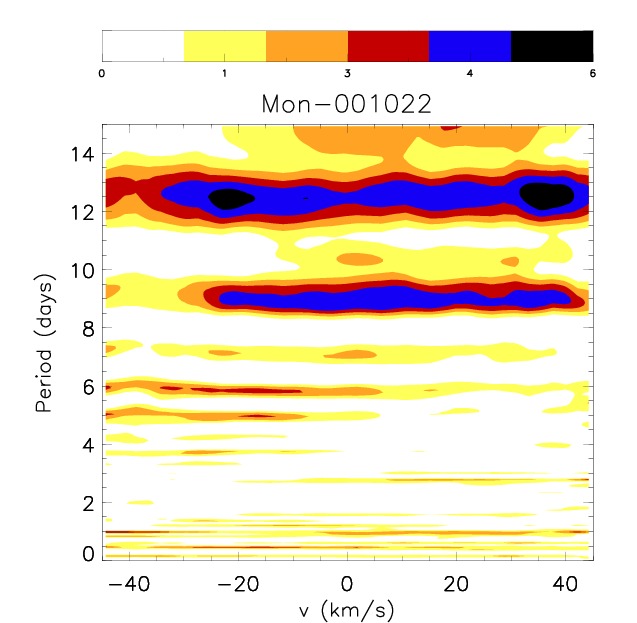}}
\subfigure[]{\label{fig:He-3sp}\includegraphics[width=4.4cm]{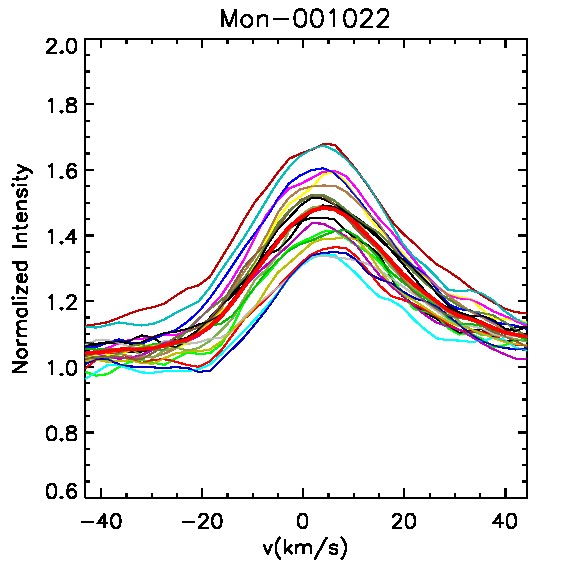}}
\subfigure[]{\label{fig:He-4}\includegraphics[width=4.4cm]{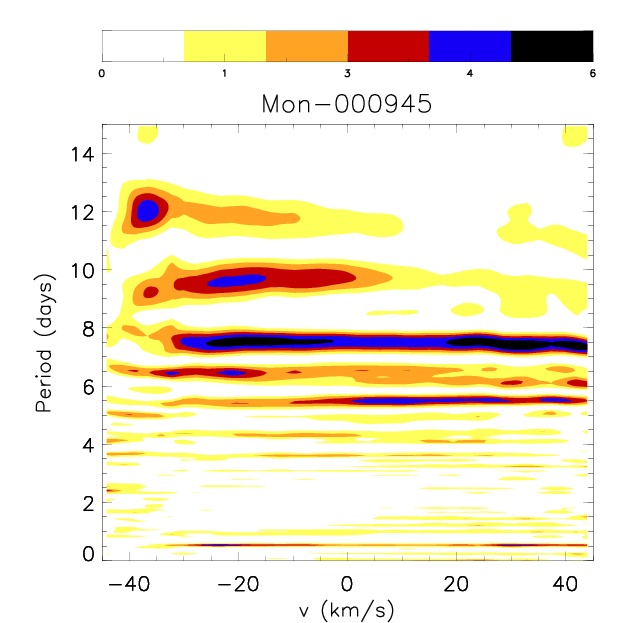}}
\subfigure[]{\label{fig:He-4sp}\includegraphics[width=4.4cm]{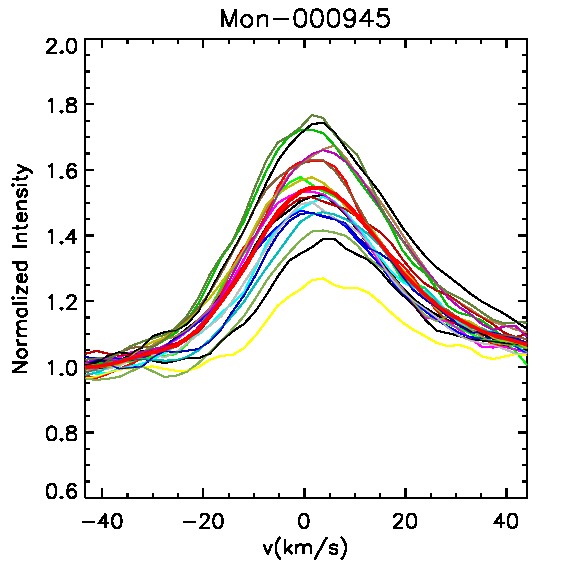}}
\subfigure[]{\label{fig:He-5}\includegraphics[width=4.4cm]{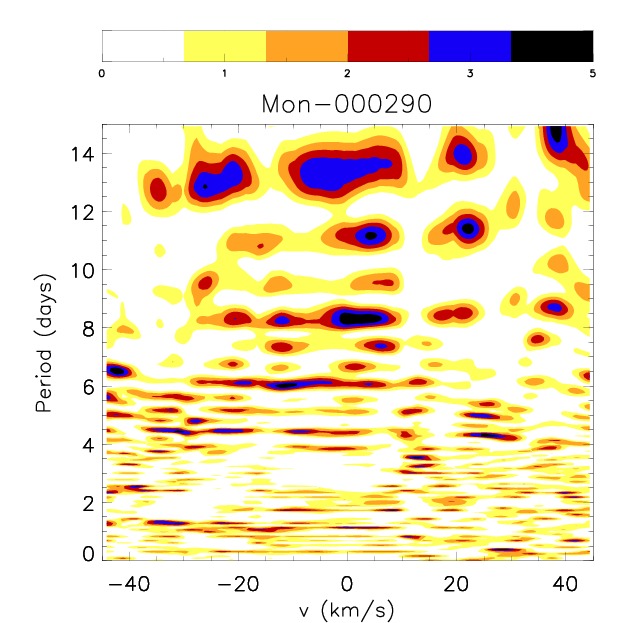}}
\subfigure[]{\label{fig:He-5sp}\includegraphics[width=4.4cm]{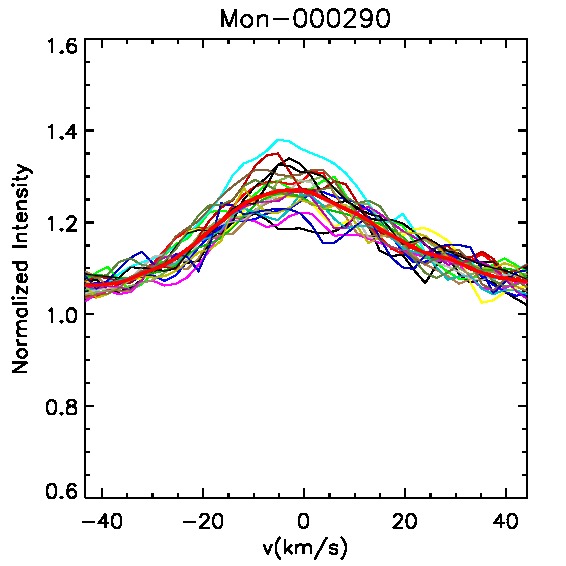}}
\subfigure[]{\label{fig:He-6}\includegraphics[width=4.4cm]{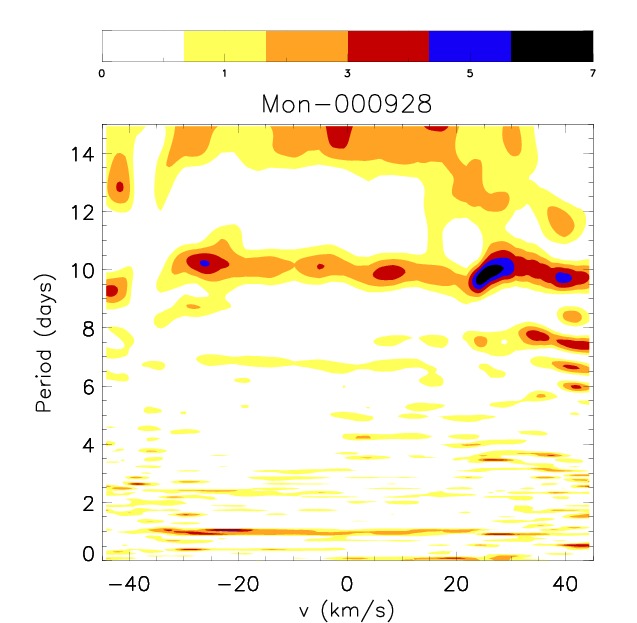}}
\subfigure[]{\label{fig:He-6sp}\includegraphics[width=4.4cm]{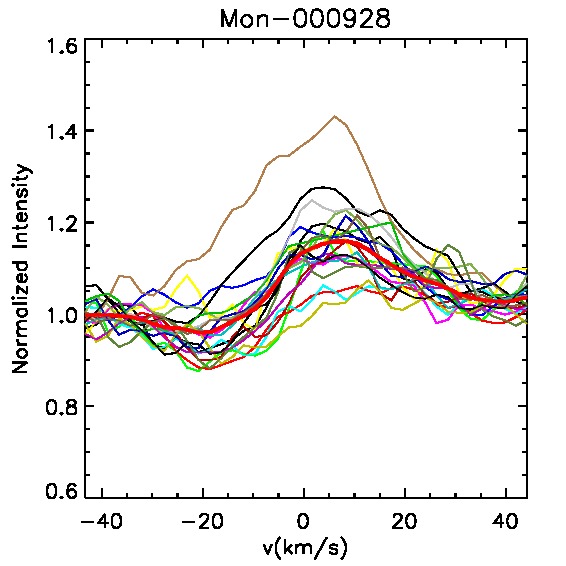}}
\subfigure[]{\label{fig:NPHe}\includegraphics[width=4.4cm]{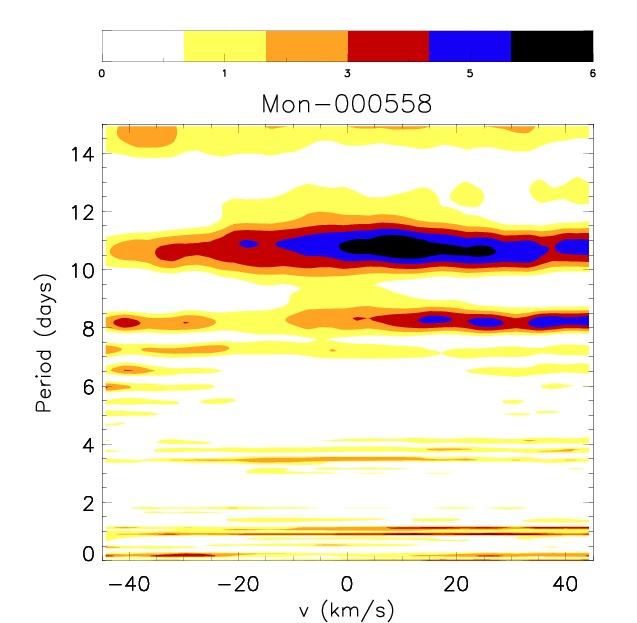}}
\subfigure[]{\label{fig:NPHeSp}\includegraphics[width=4.4cm]{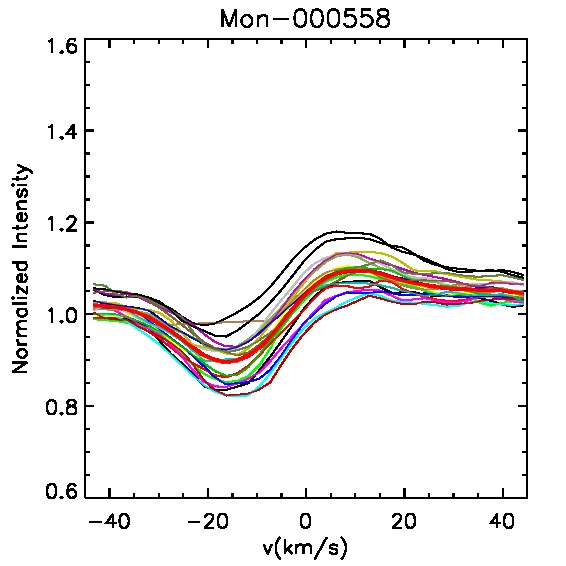}}
\subfigure[]{\label{fig:Acc2He}\includegraphics[width=4.4cm]{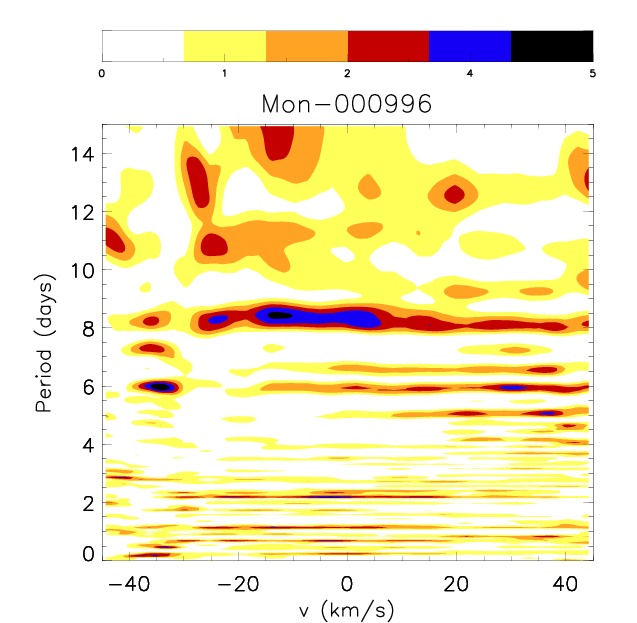}}
\subfigure[]{\label{fig:Acc2HeSp}\includegraphics[width=4.4cm]{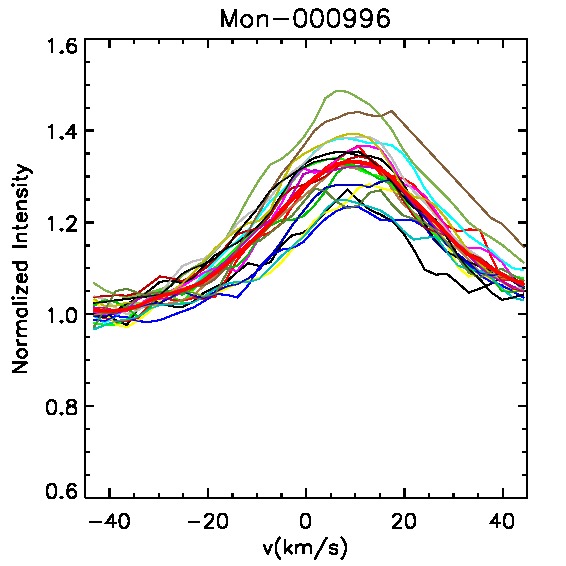}}
 \end{center}
\caption{\label{fig:PerHeI2}
Periodic stars in the HeI $6678\,\mathring{\mathrm{A}}$ line that did not show a periodicity in the CoRoT light curves.
The first and third columns show bidimensional periodograms of the HeI line. The color code represents the power of the periodogram, varying from zero (white) to the maximum power intensity (black).
The second and fourth columns show the HeI $6678\,\mathring{\mathrm{A}}$ lines used to calculate the periodograms.
Different colors correspond to different nights of observation, and the thick red line is the average line profile.}
\end{figure*}

Half of the $24$ stars that showed HeI in emission presented periodicity in this line, but 
only $4$ (Sect. \ref{sec:AAHe} and \ref{sec:SpotHe}) were also periodic 
in the CoRoT photometry, as seen in Fig. \ref{fig:PerHeI}. 
The other $8$ stars that showed a periodicity in the HeI line were not found to be
periodic in the CoRoT light curves of 2011 ($6$ stars) or were not observed with CoRoT in 2011 ($2$ stars), as seen in Fig. \ref{fig:PerHeI2}.

All of the  accretion burst systems observed by FLAMES ($\text{five}$ in total)
are included in the stars that showed HeI in emission, which is reasonable, 
since they all presented high mass-accretion rates and had light curves
dominated by hot spot variability \citep{2014AJ....147...83S}. Three of these five stars (Mon-000996, Mon-001022 and Mon-000945)
presented a periodicity in the HeI line, despite showing no periodic signal in the CoRoT
photometry. Mon-000996 (\ref{fig:Acc2He}) also presented similar periodicities
in both the HeI and $\mathrm{H}\alpha$ lines, as shown in \ref{fig:Acc2}. 

\section{Stable and unstable accretion} \label{sec:Inst}

CTTSs are photometrically variable on different timescales, from seconds to even 
decades \citep{2008MNRAS.391.1913R}, depending on the various physical phenomena 
that affect the stellar and circumstellar environment. Variable accretion is one of the
main sources of variability in this evolutionary phase. As discussed by 
\citet{2013MNRAS.431.2673K}, accretion can occur in stable and unstable regimes. 
The unstable regime may be triggered by the Rayleigh-Taylor instability, which 
takes place in the inner disk region. This type 
of variability produces non-periodic light curves that are dominated by 
accretion bursts, as observed in some stars in NGC 2264 \citep[see][]{2014AJ....147...83S}.  
This variability also produces detectable spectral features in emission lines, such as 
$\mathrm{H}\beta$, $\mathrm{H}\gamma$, and $\mathrm{Pa}\beta$ \citep{2013MNRAS.431.2673K}. 

According to theoretical predictions, stars in a stable accretion regime will present 
redshifted absorption in just a few rotational phases, when the main hot spot is seen 
projected through the accretion funnel in our line of sight. In this situation, 
photons emitted by the hot spot will be absorbed by the high-velocity gas 
in the accretion funnel, causing the observed redshifted absorption components. 
In the unstable accretion regime, redshifted absorption may be seen at any rotational phase because of 
the large number of accretion funnels, and consequently, hot spots on the stellar
surface.
The accretion burst stars are all expected to be in the unstable accretion regime.
Unfortunately, only $\text{five}$ accretion burst 
systems were included in our sample of stars observed with FLAMES, and none of them showed redshifted absorption in $\mathrm{H}\alpha$. Their
$\mathrm{H}\alpha$ line profiles are dominated by emission due to accretion (Reipurth type I 
profile) that typically does not present redshifted absorption (see Fig. \ref{fig:Hatype}).
To investigate the occurrence of redshifted absorption in the emission lines of accretion 
burst systems, we need to analyze higher order Balmer lines, as suggested by \cite{2013MNRAS.431.2673K},
or the HeI ($10830\,\mathring{\mathrm{A}}$) line, which often shows both
redshifted and blueshifted absorption in CTTS spectra, as discussed by \cite{edwards06} and \cite{2014ApJ...797..112C}. 

In the simulations by \cite{2013MNRAS.431.2673K}, the redshifted absorption components 
are not visible in the $\mathrm{H}\alpha$ line either because its emission is very intense
and hides the shallow absorption.
Redshifted absorption is most commonly seen observationally in weaker Balmer emission 
lines such as $\mathrm {H}\beta$ and $\mathrm{H}\gamma$. We do not have the higher Balmer lines in our spectra,
but $\text{ten}$ stars show a clear redshifted absorption component in the $\mathrm{H}\alpha$ line 
(marked in Col. 8 of Table \ref{tab:CttsFlames}).
Despite the small number of targets, we tried to verify how the redshifted 
absorption component varies with stellar rotational phase and compare our results 
with the prediction of stable and unstable accretion simulations.

\begin{figure}
 \begin{center}
\includegraphics[width=9.3cm]{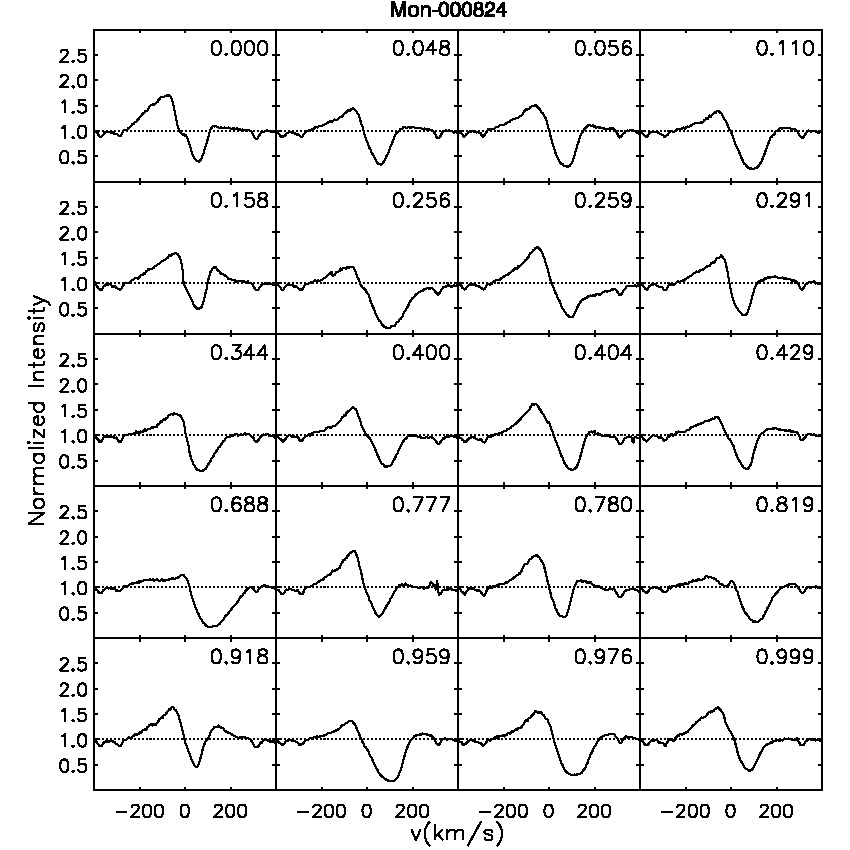}
 \end{center}
\caption{\label{fig:specfaseInst}$\mathrm{H}\alpha$ line of Mon-000824 as a function of rotational 
phase, indicated in each plot. This star was not observed by CoRoT in 2011, and we used 
the photometric period ($7.05\,\mathrm{days}$) obtained in 2008, when the star
presented an AA Tau-like light curve, to calculate the rotational phases. This star
is a candidate to be in the unstable accretion regime in 2011.} 
\end{figure}

Of our $\text{ten}$ selected targets, only Mon-000824 is possibly in the unstable accretion regime, where the redshifted 
absorption is present in the line profile at all rotational phases. This star was 
classified as AA Tau-like in 2008, but was not observed by CoRoT in 2011 (see 
Fig. \ref{fig:specfaseInst}), which means that in principle we
would not know its photometric behavior when
the spectroscopic data were obtained. As seen in Sect. \ref{sec:corot}, CTTSs 
are highly variable and in a few years may change the morphology of the their light 
curve from AA Tau to non-periodic and vice versa. Although Mon-000824 was not observed with CoRoT in 2011,
during the CSI campaign about 900 epochs of I-band photometry were obtained from November 2011 to March 2012 of 
the central region of NGC 2264, including Mon-000824, with the US Naval 40'' telescope. As discussed by
\cite{2015A&A...577A..11M}, the simultaneous I-band and CoRoT photometry typically
match very well. We therefore inspected the I-band light curve of Mon-000824 
in 2011 and found that it was indeed not periodic, implying that this star probably changed from a stable to an unstable accretion 
regime from 2008 to 2011.

\begin{figure}
 \begin{center}
\includegraphics[width=9.3cm]{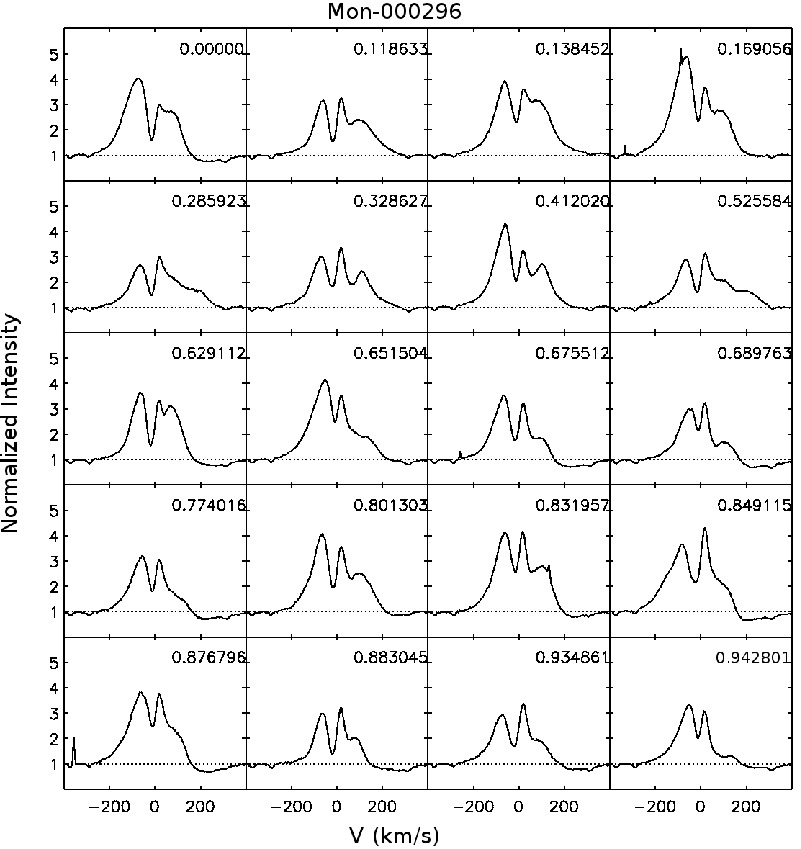}
 \end{center}
\caption{\label{fig:specfaseSt} $\mathrm{H}\alpha$ line of Mon-000296 as a function of rotational 
phase, indicated in each plot. The central peak is the nebular line that was not entirely removed. 
This star presents an AA Tau-like light curve with a
$3.91\,\mathrm{day}$ period and is a candidate to be in the stable accretion regime in 2011. } 
\end{figure}

The other $\text{nine}$ selected stars ($\text{two}$ spot-like, $\text{five}$ AA Tau-like, $\text{one}$ aperiodic extinction variables 
and $\text{one}$ not observed by CoRoT in 2011) are apparently in a stable accretion 
regime, showing redshifted absorptions in only a few rotational phases. In 
Fig. \ref{fig:specfaseSt}, we show the spectra of Mon-000296 as
an example. 
 
The predictions that stable and unstable accretion regimes present a different rotational
phase distribution of redshifted absorption components are based on
the theoretical line profiles computed by \cite{2013MNRAS.431.2673K}
for a system with an inclination of $60^\circ$ with respect to the line of sight. 
However, in systems with lower inclinations, hot spots will always be in view 
\citep{2008MNRAS.385.1931K}, hence, even in a stable accretion regime
the redshifted absorption may be present in the line profile at all rotational phases. 
This is, however, not the case of Mon-000824 (Fig. \ref{fig:specfaseInst}), since its inclination 
with respect to our line of sight is $72^\circ \pm5^\circ$  \citep{2015A&A...577A..11M}.

We would expect stars in a stable accretion regime to show periodic light curves
and stars in an unstable accretion regime to show non-periodic light curves 
\citep{2013MNRAS.431.2673K,2008MNRAS.386..673K,2009MNRAS.398..701K}. 
This is because the stable accretion regime generates an organized circumstellar
environment that shows periodicity as the star-disk system rotates. 
Of the $\text{nine}$ stars we classified as being in a stable regime, 
using this criterion of redshifted absorption, only $\text{two}$ did not present periodic light 
curves, as seen in Table \ref{tab:CttsFlames}. Instead their light curves were classified as aperiodic extinction variables. 
The stable accretion regime classification therefore agrees with the observation of periodicity 
in the CoRoT light curves for most stars with redshifted absorption in just a few
rotational phases. 

In the stable accretion regime, the variability of the redshifted absorption is 
expected to be periodic. Of the $\text{nine}$ stars we classified as being in a stable regime, $\text{five}$ 
are periodic in the $\mathrm{H}\alpha$ line, agreeing with the theoretical predictions 
of \cite{2013MNRAS.431.2673K}.
This means that only about half of the stars we classified as being in a stable 
accretion regime are periodic in the $\mathrm{H}\alpha$ line.
The only star in the unstable accretion regime (Fig. \ref{fig:specfaseInst}) 
is also not periodic in the $\mathrm{H}\alpha$ line, as predicted by \cite{2013MNRAS.431.2673K}.

The classification of a stable accretion regime based on                                                
the presence of redshifted absorption at well-defined rotational phases agrees 
only sometimes with the type of observed variability in the CoRoT light curves. Neither is a star that shows redshifted absorption at all rotational phases  
necessarily accreting in the unstable regime. Other factors, such as the inclination 
of the system, may influence the presence of the redshifted absorption in the Balmer lines. 
Moreover, we must keep in mind that the $\mathrm{H}\alpha$ line is not the best choice 
for this type of analysis, since other lines, such as $\mathrm{H}\beta$, 
$\mathrm{H}\gamma$, and $\mathrm{Pa}\beta$, are weaker, and the redshifted 
absorption is more commonly visible in the line profiles.

\section{Correlation matrices} \label{sec:Matcorr}

The $\mathrm{H}\alpha$ emission line can be formed in several different regions, 
in the accretion funnel, the chromosphere, and/or the wind. The analysis of 
the correlation of different parts of the $\mathrm{H}\alpha$ line profile 
can help us investigate if the line comes from
different regions and if these regions are correlated. 

One method of checking the correlation across a line profile is to calculate 
correlation matrices. We divided each $\mathrm{H}\alpha$ profile into small 
velocity intervals of $1.5 \, \mathrm{km/s}$ and calculated the linear correlation 
coefficient, $r(i,j)$, of the line intensity at each velocity interval $i$ with 
each velocity interval $j$,
as described by \cite{1995AJ....109.2800J}. 
The correlation coefficient ranges from $-1$ to $1$, $1$ corresponding to a perfect 
correlation, $0$ to no correlation, and $-1$ to a perfect anticorrelation. 
When $i=j$, $r(i,j)=1$, and for all values of $i$ and $j$, $r(i,j)=r(j,i)$, implying that 
the matrix is symmetrical relative to the diagonal. We present the result
as a color-map plot of the correlation coefficients, which is useful to 
visually identify regions of the profile that are correlated or anticorrelated. 
This has been used in the literature to analyze correlations across the 
$\mathrm{H}\alpha$ line and other lines such as $\mathrm{H}\beta$ 
\citep{1995ApJ...449..341J,2000A&A...362..615O,2002ApJ...571..378A,2005MNRAS.358..671K}.

All of the $58$ CTTSs that have FLAMES spectra present some correlation signal of 
the red wing with itself. Some stars show this correlation just in the redshifted 
absorption position and others just in the redshifted emission. We find that $\sim74\%$ of the CTTSs
show a very good correlation across the entire $\mathrm{H}\alpha$ red wing, which
then varies coherently. Almost all the CTTSs observed by FLAMES 
also exhibited some correlation of the blue 
wing with itself, $\sim74\%$ presenting a substantial correlation. The blue wing 
of $\mathrm{H}\alpha$ also varies coherently. This shows 
that the variability in each wing of the $\mathrm{H}\alpha$ profile is coherent, which 
indicates that each line wing (red or blue) is formed in a region dominated by a given 
physical process.
 
We find that $\sim34\%$ of the CTTSs observed with FLAMES show no sign of correlation between
the red and the blue wings, implying that they vary independently in those systems.
At the same time, $\sim66\%$ CTTSs show some correlation of the red wing with the blue wing and
among these, $\sim55\%$ show substantial correlation across the entire wings, while $\sim45\%$ show
correlations that occur only in some specific velocity range. 

Only $7$ of the $58$ CTTSs observed with FLAMES present 
anticorrelated variations in the $\mathrm{H}\alpha$ line profile. 
The anticorrelation seen in $4$ of these $6$ stars is associated with a blue-
or redshifted absorption that varies in antiphase with the rest of the profile. 
The other three stars present an anticorrelation associated with a blueshifted 
emission and the other profile parts. 

Overall, the $\mathrm{H}\alpha$ line profiles present variations that cannot be explained by continuum variability alone that
is due, for example, to hot spots. If excess continuum emission from hot 
spots were the main cause of emission line variability, the entire profile would be correlated, 
and the $\mathrm{H}\alpha$ line would not vary in shape, but only in intensity. Other factors
such as wind emission or absorption, accretion variability, and non-symmetrically distributed circumstellar material
must influence the emission line variability, often simultaneously, since many stars do not present 
a very good correlation across the entire $\mathrm{H}\alpha$ line profile. 
It is possible, however, that in some cases the $\mathrm{H}\alpha$ line may be formed mainly in one specific 
region, like the wind or the accretion column. We would then expect the line profile to be dominated 
by the same physical process and the variability of the entire line profile to be correlated. 

\begin{figure}[!ht]
 \begin{center}
\subfigure[]{\label{fig:a}\includegraphics[width=4.4cm]{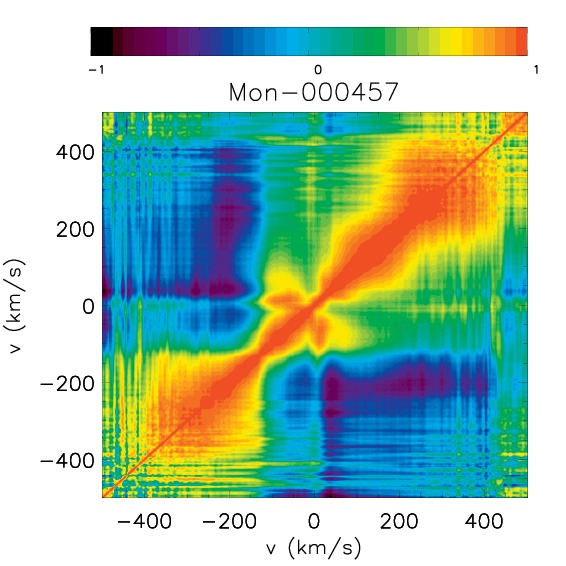}} 
\subfigure[]{\label{fig:a1}\includegraphics[width=4.4cm]{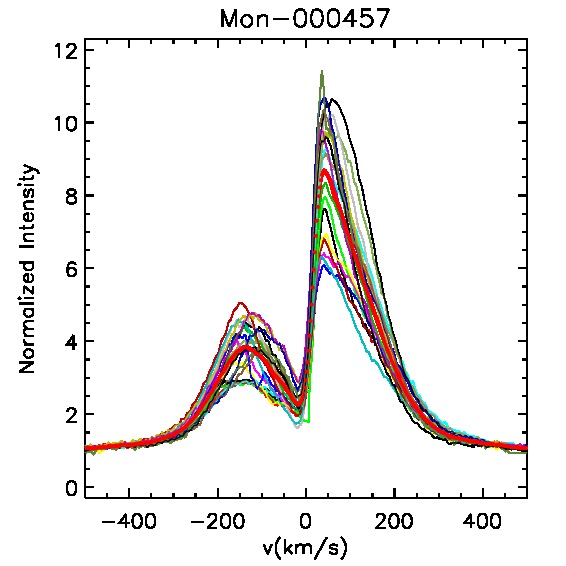}}\\
\subfigure[]{\label{fig:b}\includegraphics[width=4.4cm]{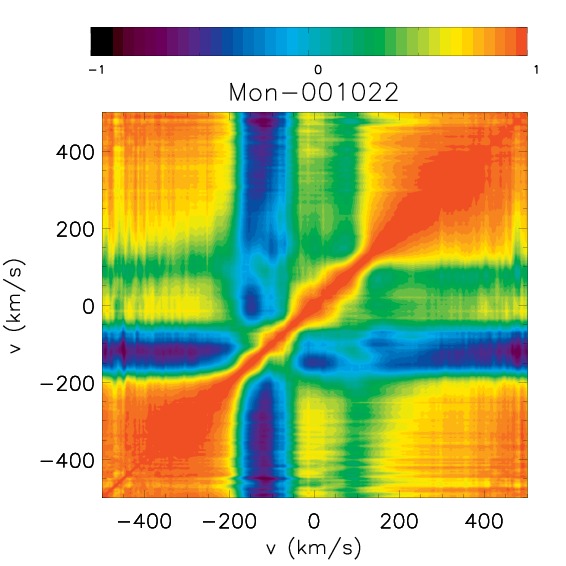}}
\subfigure[]{\label{fig:b1}\includegraphics[width=4.4cm]{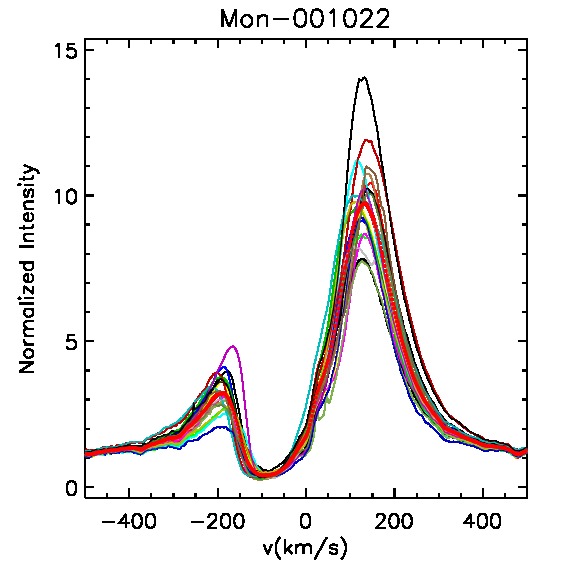}}
\end{center}
\caption{\label{fig:MatcHa}Examples of correlation matrices (left) of the 
$\mathrm{H}\alpha$ line of two CTTSs and the corresponding line profiles (right).
In panels (a) and (c), the color range corresponds to the value of the linear 
correlation coefficient between the different velocity bins of the $\mathrm{H}\alpha$ 
line profiles. Perfect anticorrelation corresponds to -1 (black), no correlation to 0 
(light blue), and a perfect correlation to 1 (orange).
In panels (b) and (d), different colors correspond to spectra observed in different nights, and 
the thick red line is the average line profile. The correlation matrices and $\mathrm{H}\alpha$ line profiles of all the CTTSs in our 
sample observed with FLAMES are presented in the Appendix.}
\end{figure}

\begin{figure}[!ht]
 \begin{center}
\subfigure[]{\label{fig:d}\includegraphics[width=4.4cm]{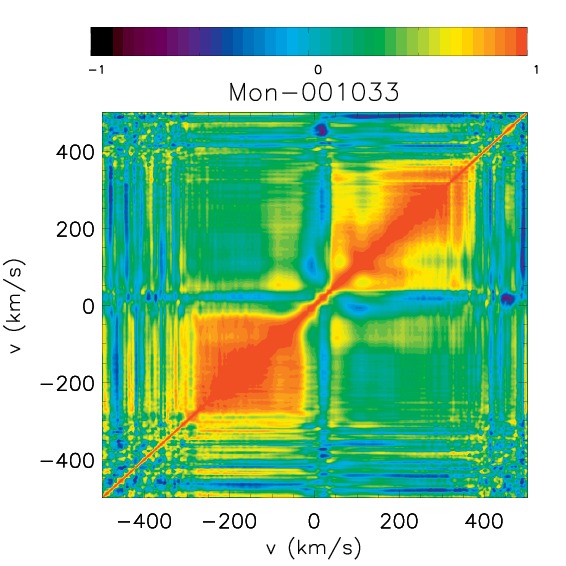}}
\subfigure[]{\label{fig:d1}\includegraphics[width=4.4cm]{fig/Spec_500007379}}\\
\subfigure[]{\label{fig:c}\includegraphics[width=4.4cm]{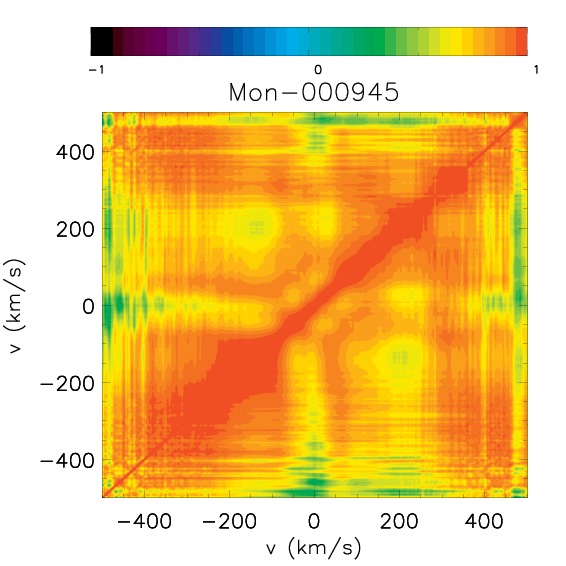}} 
\subfigure[]{\label{fig:c1}\includegraphics[width=4.4cm]{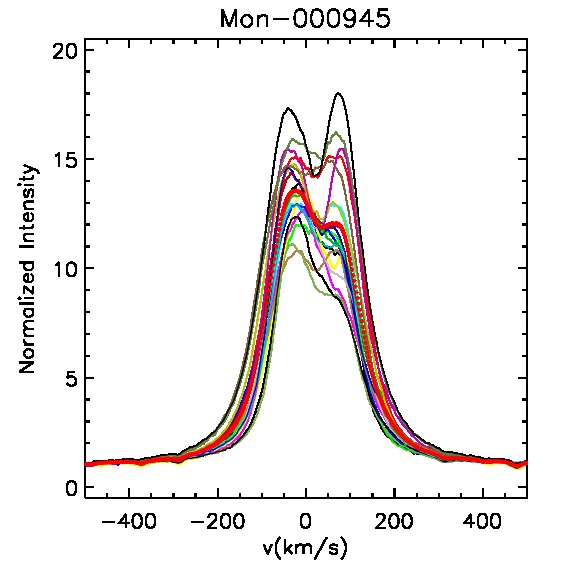}} 
 \end{center}
\caption{\label{fig:MatcHa2}Examples of correlation matrices (left) of the 
$\mathrm{H}\alpha$ line of two CTTSs and the corresponding line profiles (right). 
The color code is the same as in Fig. \ref{fig:MatcHa}.}
\end{figure}

Figures \ref{fig:MatcHa} and \ref{fig:MatcHa2} show examples of correlation matrices 
for some CTTSs, together with their observed $\mathrm{H}\alpha$ line profiles. These figures 
are representative of the different types of correlation matrices we obtained. In these plots, 
the correlation coefficients are represented by colors ranging from black ($r=-1$, maximum 
anticorrelation) to light blue ($r=0$, no correlation) and to orange ($r=1$, maximum correlation). 
 The correlation matrices and $\mathrm{H}\alpha$ line profiles of all the CTTSs in our 
sample observed with FLAMES are presented in the Appendix.

Figure \ref{fig:a},b shows that for Mon-000457 the red wing of $\mathrm{H}\alpha$ correlates well 
with itself in the velocity range of $50 \,\mathrm{km/s} < \mathrm{v} < 350\,\mathrm{km/s}$, and
the blue wing shows a good correlation with itself at  $-350\,\mathrm{km/s} < \mathrm{v} < -75\, \mathrm{km/s}$. 
We can also see a region of correlated variability in the blue wing at $-100 \,\mathrm{km/s} < \mathrm{v} < 0\,\mathrm{km/s}$,
which corresponds to the blueshifted absorption from a wind that varies quite independently
of the rest of the profile.
The blue wing is anticorrelated
with the red wing for velocities in 
the range of $\sim-100$ to $\sim-350\,\mathrm{km/s}$ and $\sim0$ to $\sim350\,\mathrm{km/s}$. 
The line profiles show that these regions correspond to the blue and red emissions
that vary in antiphase with each other. This indicates a strong influence of different
physical processes in each $\mathrm{H}\alpha$ profile wing, the red wing is probably dominated by accretion
and the blue wing by the wind. The accretion process and the wind may contribute to the entire
profile, but can have a stronger contribution in one particular velocity range,
as seen for example in theoretical line profiles calculated by 
\citet{2006MNRAS.370..580K} and \citet{2010A&A...522A.104L},
which include both the wind and accretion contributions.
In the case of Mon-000457, if the blue wing is mostly influenced by the wind and
accretion is probably the main process acting on the red wing, the anticorrelation
between the two wings could be explained as being due to rotational modulation
of a CTTS with an inclined magnetosphere, as also observed in SU Aur
\citep{1994ASPC...64..190J}. When the main accretion funnel faces the observer, 
accretion is strong and ejection is low, and half a rotation cycle later, the opposite occurs.
In an ideal case, these variations should be periodic, but the $\mathrm{H}\alpha$ 
line variability of this star does not show very clear signs of periodicity.

Figure \ref{fig:b} shows a strong correlation of the red wing with 
itself and with the blue wing of the $\mathrm{H}\alpha$ profile at high velocities 
($|v| > 200 \,\mathrm{km/s}$) for Mon-001022 . At the same time, the strong blueshifted absorption 
($-200 \,\mathrm{km/s} <\mathrm{v}< 0\,\mathrm{km/s}$) is anticorrelated with the 
blue and red emission regions of $\mathrm{H}\alpha$. 
This feature is due to photons from the accretion columns and a hot spot absorbed 
by the outflowing disk wind material in our line of sight.  Stars that present very strong 
blueshifted absorption in $\mathrm{H}\alpha$ normally present emission profiles dominated
by the wind, as can be seen in the theoretical profiles calculated by 
\citet{2006MNRAS.370..580K} and \citet{2010A&A...522A.104L}. When the emission due to the wind 
increases, photons are even more strongly absorbed by the wind, increasing the depth of 
the blueshifted absorption. This causes the blueshifted absorption to become 
anticorrelated with the rest of the emission line. 

Figure \ref{fig:d} shows the correlation matrix and the $\mathrm{H}\alpha$ 
line profiles of Mon-001033. The blue and red wings of the $\mathrm{H}\alpha$ 
profile correlate well with themselves, but show no correlation or anticorrelation
with each other. The emission component
is quite variable, not very intense, and suffers strong influence from absorption
in the blue and red wings. Blueshifted absorption is associated with winds and redshifted
absorption with the accretion funnel. The combination of faint emission and strong absorption
that probably comes from different physical processes created a
variability pattern where the variations of one wing are not correlated with the other wing.

Theoretical correlation matrices are rare in the literature. In \citet{2012A&A...541A.116A},
$\mathrm{H}\alpha$ correlation matrices were calculated from theoretical line profiles that
included only the magnetospheric contribution. Their theoretical correlation matrices 
showed a good correlation of the entire $\mathrm{H}\alpha$ line profile, but outflows were not
taken into account. Our observed
CTTSs include stars such as  Mon-000945, whose $\mathrm{H}\alpha$ emission line profiles show 
no hint of blueshifted absorption and should be dominated by accretion. Figure \ref{fig:c},d 
shows the $\mathrm{H}\alpha$ profiles of Mon-000945 and its corresponding 
correlation matrix, which indeed is well correlated across the entire profile, as predicted
theoretically by magnetospheric accretion models. However, the $\mathrm{H}\alpha$ profile of most CTTSs cannot be explained by magnetospheric accretion alone, and it would be very interesting to be able to compare
the observed correlation matrices with theoretical matrices calculated from profiles
that include both accretion and outflows.

Veiling variations would also produce a correlation across the entire line profile, as 
observed in the $\mathrm{H}\alpha$ line of Mon-000945.
However, the $\mathrm{H}\alpha$ line profile presents changes in its morphology, 
not only in its intensity (see Fig. \ref{fig:c1}), 
as would be expected for variations that are only due to the extra veiling continuum. 
The type of correlated variability and $\mathrm{H}\alpha$ profile morphology 
observed in Mon-000945 is also seen in the CTTSs T Tau 
\citep{1995ApJ...449..341J} and TW Hya  \citep{2002ApJ...571..378A}, for which it has also been suggested that the 
$\mathrm{H}\alpha$ line is mostly formed in the accretion columns. 

The analysis of the correlation matrices shows that the $\mathrm{H}\alpha$ line
variability can be associated with different physical processes. 
The variability of the red wing often does not affect the 
variability of the blue wing. This is confirmed by the lack of correlation of the blue wing with the 
red wing, seen in $\sim34\%$ of the $\mathrm{H}\alpha$ correlation matrices. 
This lack of correlation disagrees with theoretical models that only include 
the magnetospheric accretion contribution to the emission line profile. It would be very interesting 
to compare observational correlation matrices to matrices calculated from hybrid magnetosphere and wind 
models, such as those of \citet{2006MNRAS.370..580K,2011MNRAS.416.2623K}, and
determine which circumstellar environment would be able to reproduce the observational matrices.  
At the same time, the red wing is almost always well correlated with itself, as is the blue wing with itself, 
indicating that the variability of each wing is dominated by the same physical phenomenon. 

\section{Interesting objects}\label{sec:IntObj}

Some objects are interesting enough to be analyzed individually in future work. We found three spectroscopic
binaries with the FLAMES data. The CTTS Mon-000448 is a double-line spectroscopic binary,
while the WTTS Mon-000497 and the CTTS Mon-000804 are single-line spectroscopic binaries. These stars 
do not present eclipses in the CoRoT light curves, however.

The AA Tau-like star Mon-00811 is a CTTS that has different periods in different accretion and circumstellar
features. One possible interpretation is that the different periods come from different regions of the 
star-disk system. It has a photometric period ($\sim 7.9\,\mathrm{days}$) due to occultation of the star by 
the inner disk, a stellar rotation period ($\sim 8$ days) obtained from the HeI line periodogram, 
a $10.5$-day period from the HeI blueshifted emission, and a period of $\sim 13\,\mathrm{days}$ 
in the $\mathrm{H}\alpha$ line blue wing that we associate with a disk wind. 

Some WTTSs have strong IR excesses, but do not show a clear sign of accretion 
(Mon-001157, Mon-000434, Mon-000279, Mon-000753, and Mon-000271). For some unknown reason, accretion has been shut off or is at a very
low level, despite the available material in their inner disks. It would be interesting to reobserve these systems 
to verify whether accretion features reappear in the future, of if it has indeed come to an end.

\section{Conclusions}\label{sec:conclusion}

We investigated accretion in a group of classical T Tauri stars 
that belong to the young cluster NGC 2264, using observational data obtained 
in the \textit{Coordinated Synoptic Investigation of NGC 2264} campaign. 
The main results of this work are listed below.

\begin{enumerate}

 \item The light curves observed by CoRoT were morphologically classified as spot-like, 
AA-Tau-like, or non-periodic. Of the non-periodic light curves, some are due to  
accretion bursts, others to circumstellar disk obscuration, but most non-periodic light curves 
are very complex and could not be associated with a major physical phenomenon. 
In the 2011 campaign, we found $\sim 13\,\%$ spot-like, $\sim12\,\%$ AA Tau-like, and 
$\sim70\,\%$ non-periodic systems.
Of the non-periodic ones, $\sim19\,\%$ were associated with accretion bursts, $\sim13\,\%$ 
with occultation by circumstellar dust, and $\sim68\,\%$ presented variability of unknown origin.
 
 \item Of the CTTSs observed by CoRoT, $\sim43\%$ and $\sim30\%$ showed periodic photometric variability in 2008 and 
2011, respectively. The number of periodic stars did not change significantly between the two runs, 
but a larger number of non-periodic stars were included in the 2011 campaign.

 \item Of the $84$ stars that were observed with the CoRoT in the two different epochs (2008 and 2011), 
$\sim30\%$ showed substantial variability in their light-curve morphology, changing to a 
different category in our light-curve classification scheme. These changes show how dynamic the CTTS circumstellar environment is in only a few years.
 
 \item When UV excesses are not available, mass-accretion rates can be obtained with good results from the $\mathrm{H}\alpha$ flux using the currently 
available calibrations from the literature. 
 
 \item Most of the CTTSs we observed showed no periodicity in the $\mathrm{H}\alpha$ line. Only $\text{eight}$ stars
in our sample of $58$ CTTSs observed with FLAMES were periodic in $\mathrm{H}\alpha$. 

\item The HeI $6678\,\mathring{\mathrm{A}}$ narrow component line variability can provide the rotation period of a CTTS
when the star is in a stable accretion regime, since in that case a major hot spot is expected to form in 
each stellar hemisphere. We found that $24$ of the $58$ stars observed by FLAMES had this line in emission. 
Of these, $12$ were periodic in the HeI line, and $4$ were also periodic photometrically.
 
\item None of the photometrically periodic spot-like systems that were observed spectroscopically presented a 
periodicity in $\mathrm{H}\alpha$, while spectropolarimetry results show that hot and cold spots 
tend to be coincident in CTTSs, and we might expect to observe the same periodicity in the cold spot 
and accretion funnel diagnostics. Spot-like systems tend to present low mass-accretion rates, and the lack of 
periodicity in the $\mathrm{H}\alpha$ line variability
of spot-like systems could be due to chromospheric emission contributing significantly to the emission profile
in these systems, which would dilute the $\mathrm{H}\alpha$ modulation from the spot.
 
 \item We analyzed which CTTSs were in stable or unstable accretion regimes. \cite{2013MNRAS.431.2673K}
suggested that CTTSs in a stable accretion regime should present redshifted absorption components 
in emission line profiles only in a few rotational phases, when photons from the hot spot are 
seen in our line of sight projected against the main accretion column. Systems in unstable accretion will
produce random accretion tongues and should present redshifted accretion components at any rotational phase.
We analyzed the $\text{ten}$ CTTSs that showed redshifted absorption components in $\mathrm{H}\alpha$ and for
which we had some information about the rotational period. Of these, $\text{nine}$ CTTSs fulfilled the stable
accretion criteria ($\text{five}$ stars were AA Tau-like, $\text{two}$ stars were spot-like, $\text{one}$ was non-periodic, and $\text{one}$ was 
not observed in 2011 by CoRoT), and in almost all the systems accretion in a stable regime agrees with the 
type of variability observed in the CoRoT light curves. 
Only one star of the $\text{ten}$ selected CTTSs was apparently in the 
unstable accretion regime, presenting redshifted absorption at all rotational
phases and showing no periodicity in $\mathrm{H}\alpha$, as expected.
 
 \item The analysis of correlation matrices showed that the variability in the $\mathrm{H}\alpha$ 
line can be associated with different physical processes, and the red and blue wings are sometimes 
formed in different regions, since the variability of one wing does not always correlate with the variability 
of the other. This is confirmed by the lack of correlation of the blue with the red wings, seen in 
$\sim34\%$ of the CTTSs. Each wing tends to be well correlated with itself, indicating that each wing
is dominated by a single physical phenomenon.

\end{enumerate}

\begin{acknowledgements}
APS and SHPA acknowledge support from CNPq, CAPES and Fapemig. JFG aknowledges support from FCT ref project UID/FIS/04434/2013.
\end{acknowledgements}


\bibliographystyle{aa}   
\bibliography{ref}

\begin{appendix}
\section{Correlation matrices and $\mathrm{H}\alpha$ line profiles} 

\begin{figure*}[!H]
 \begin{center}
\subfigure[]{\includegraphics[width=4.4cm]{fig/mcHaskc_500007369}}
\subfigure[]{\includegraphics[width=4.4cm]{fig/Spec_500007369}}
\subfigure[]{\includegraphics[width=4.4cm]{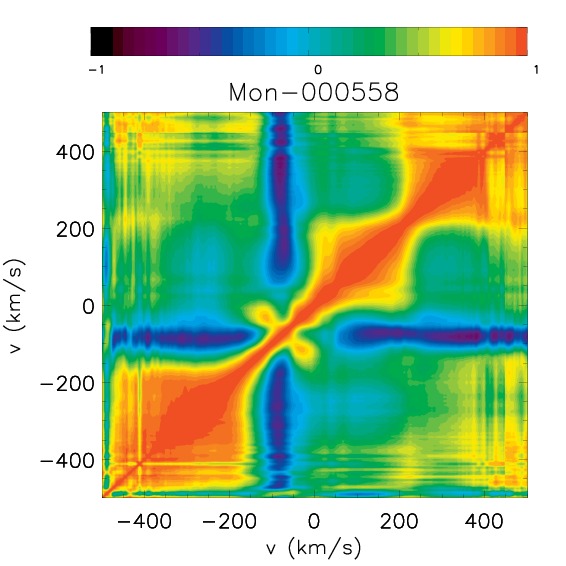}} 
\subfigure[]{\includegraphics[width=4.4cm]{fig/Spec_223990964}}
\subfigure[]{\includegraphics[width=4.4cm]{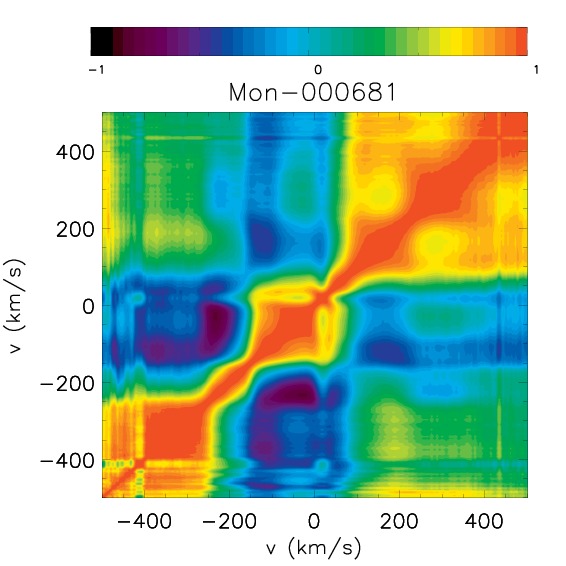}}
\subfigure[]{\includegraphics[width=4.4cm]{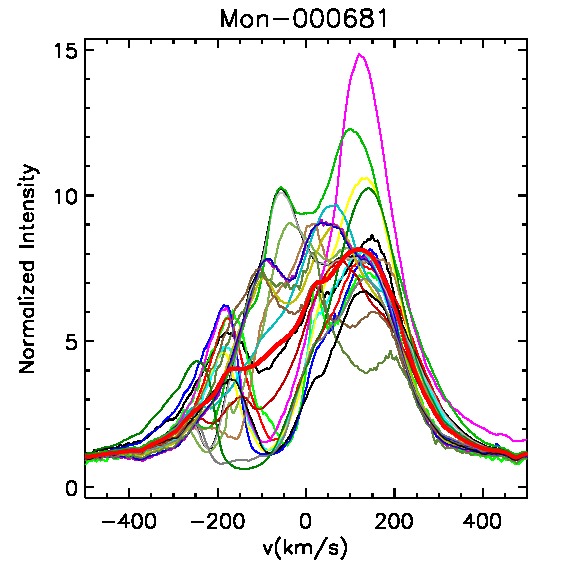}}
\subfigure[]{\includegraphics[width=4.4cm]{fig/mcHaskc_500007252}}
\subfigure[]{\includegraphics[width=4.4cm]{fig/Spec_500007252}}
\subfigure[]{\includegraphics[width=4.4cm]{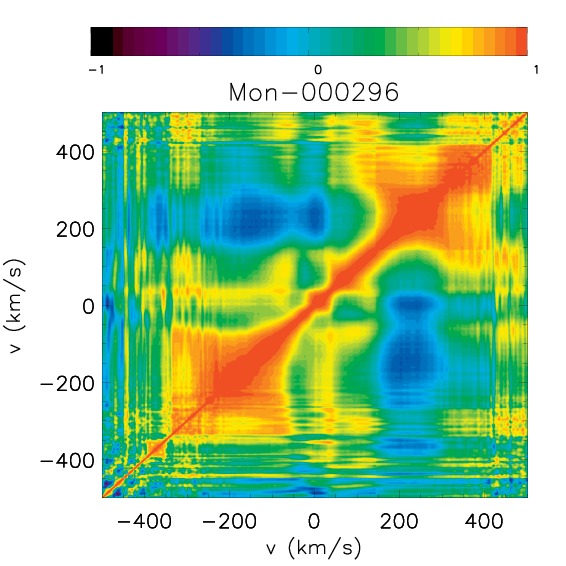}}
\subfigure[]{\includegraphics[width=4.4cm]{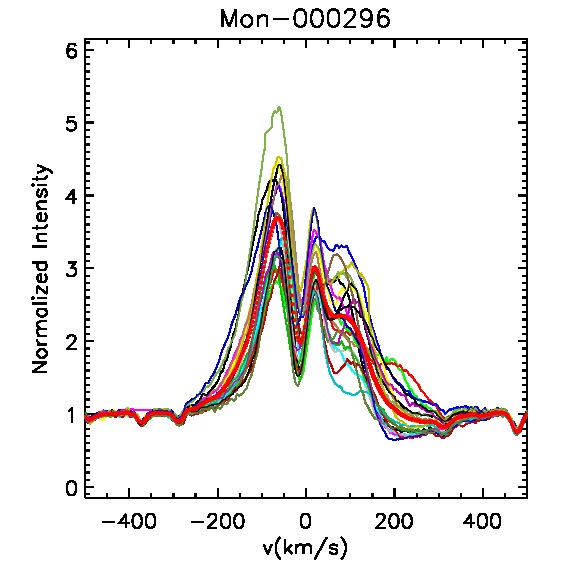}}
\subfigure[]{\includegraphics[width=4.4cm]{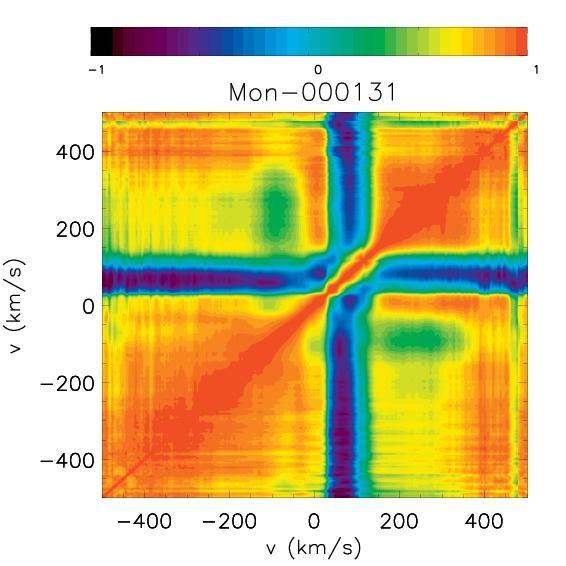}}
\subfigure[]{\includegraphics[width=4.4cm]{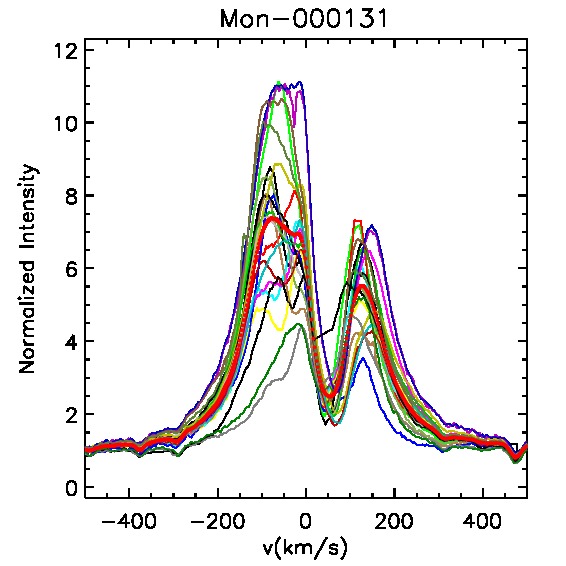}}
\subfigure[]{\includegraphics[width=4.4cm]{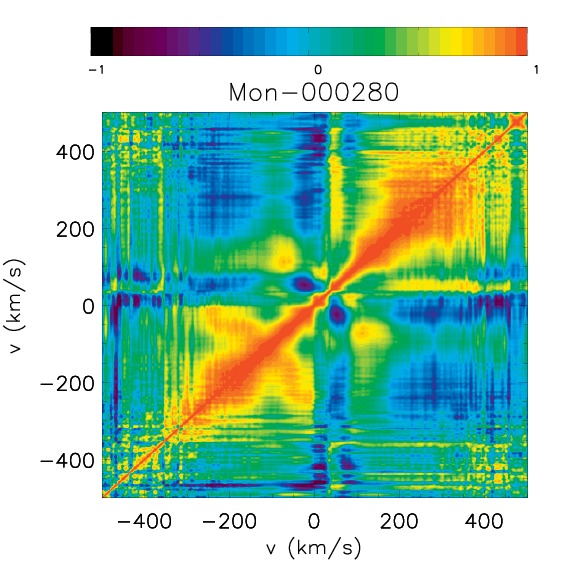}} 
\subfigure[]{\includegraphics[width=4.4cm]{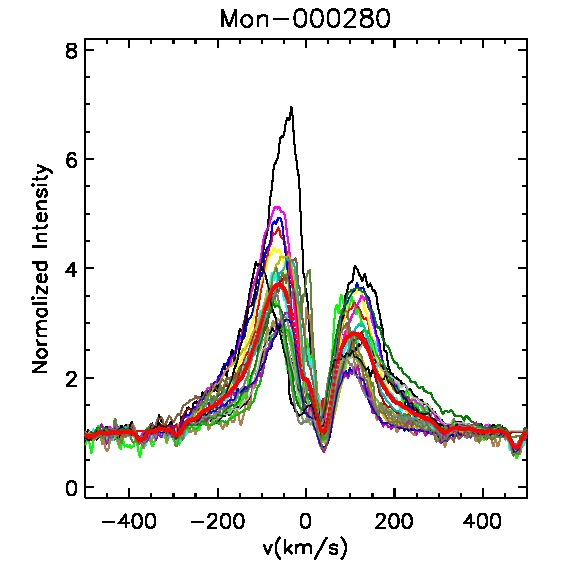}}
\end{center}
\caption{\label{fig:MatcHa1}Correlation matrices (left) of the 
$\mathrm{H}\alpha$ line of selected CTTSs and the corresponding line profiles (right).
In the left panels, the color range corresponds to the value of the linear 
correlation coefficient, $r(i,j)$, between different velocity bins ($i$ and $j$), of the $\mathrm{H}\alpha$ 
line profiles, as described in Sect. \ref{sec:Matcorr}. Perfect anticorrelation corresponds 
to -1 (black), no correlation to 0 (light blue), and a perfect correlation to 1 (orange).  
When $i=j$, $r(i,j)=1$, and for all values of $i$ and $j$, $r(i,j)=r(j,i)$, implying that 
the matrix is symmetrical relative to the diagonal.
In the right panels, different colors correspond to spectra observed in different nights, and 
the thick red line is the average line profile. The stars in this group present 
anticorrelation of some part of the line profile. The anticorrelation seen in these stars is 
associated with blue- or redshifted absorption or emission that varies in antiphase with 
the rest of the profile. 
For example, when in panel a (Mon-000457) the red wing emission increases in intensity, 
the blue wing emission decreases.
}
\end{figure*}

\begin{figure*}[!ht]
 \begin{center}
\subfigure[]{\includegraphics[width=4.4cm]{fig/mcHaskc_223977953}}
\subfigure[]{\includegraphics[width=4.4cm]{fig/Spec_223977953}}
\subfigure[]{\includegraphics[width=4.4cm]{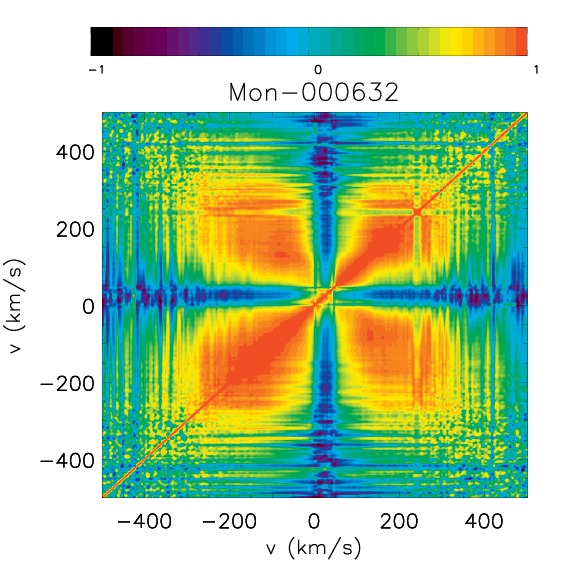}}
\subfigure[]{\includegraphics[width=4.4cm]{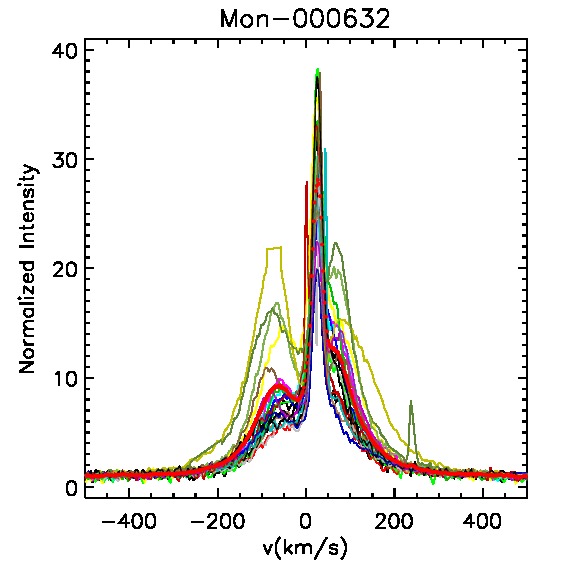}}
\subfigure[]{\includegraphics[width=4.4cm]{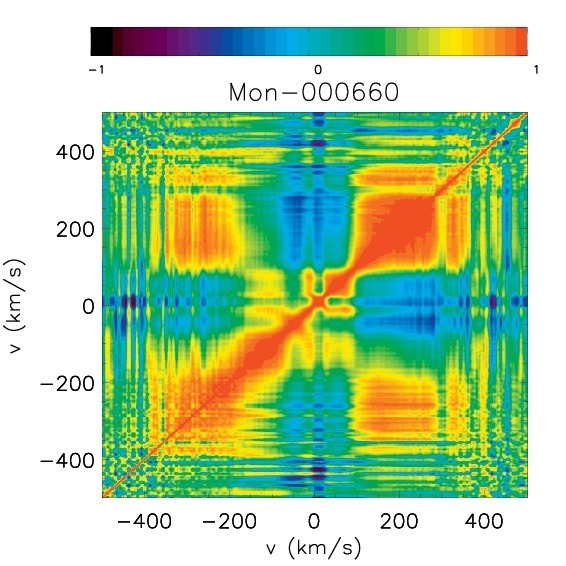}}
\subfigure[]{\includegraphics[width=4.4cm]{fig/Spec_223980693}}
\subfigure[]{\includegraphics[width=4.4cm]{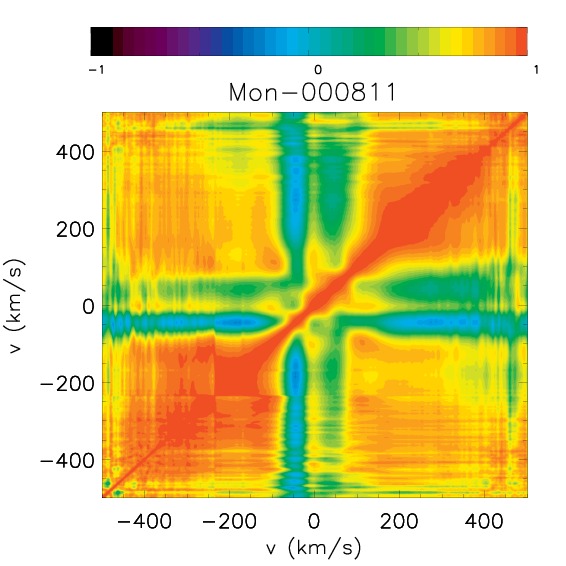}} 
\subfigure[]{\includegraphics[width=4.4cm]{fig/Spec_500007460}}
\subfigure[]{\includegraphics[width=4.4cm]{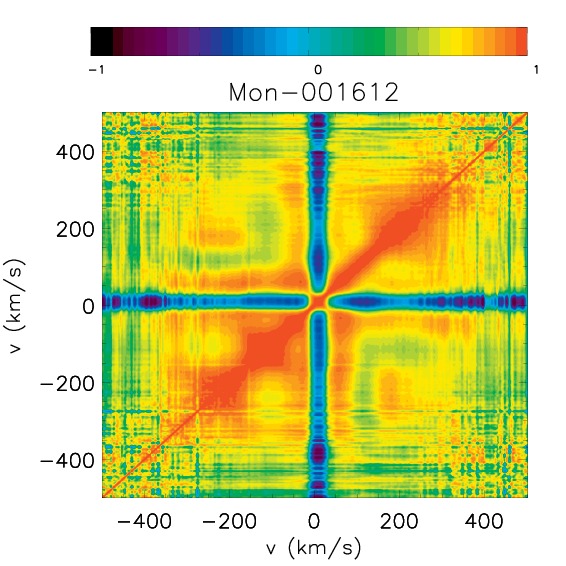}}
\subfigure[]{\includegraphics[width=4.4cm]{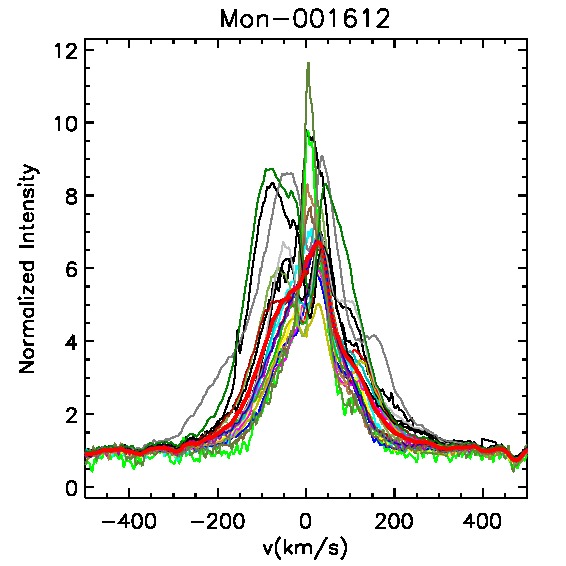}}
\subfigure[]{\includegraphics[width=4.4cm]{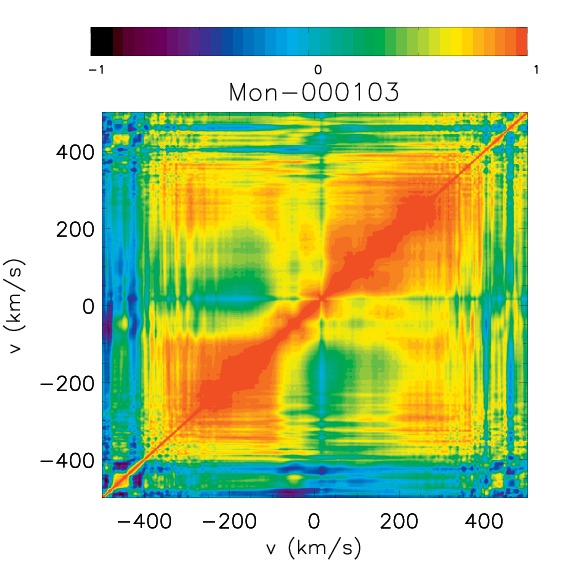}} 
\subfigure[]{\includegraphics[width=4.4cm]{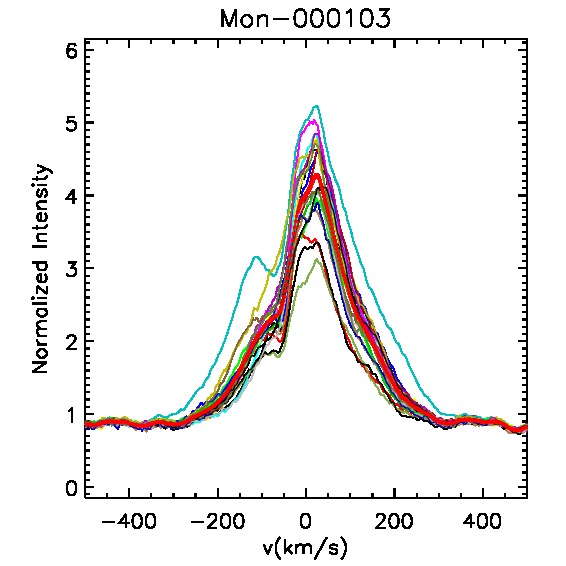}}
\subfigure[]{\includegraphics[width=4.4cm]{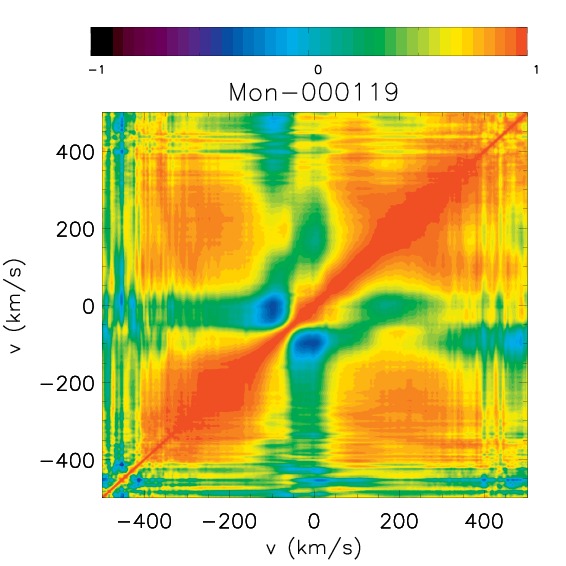}}
\subfigure[]{\includegraphics[width=4.4cm]{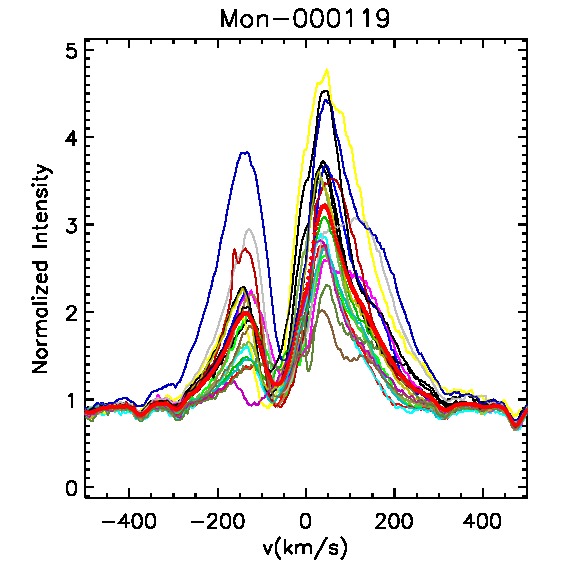}}
\subfigure[]{\includegraphics[width=4.4cm]{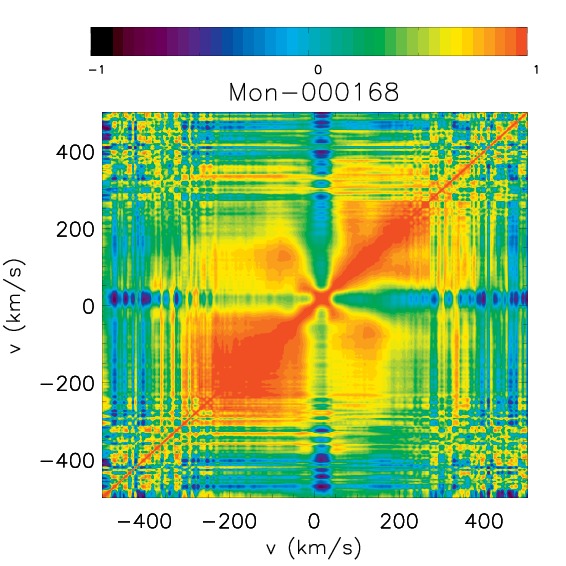}}
\subfigure[]{\includegraphics[width=4.4cm]{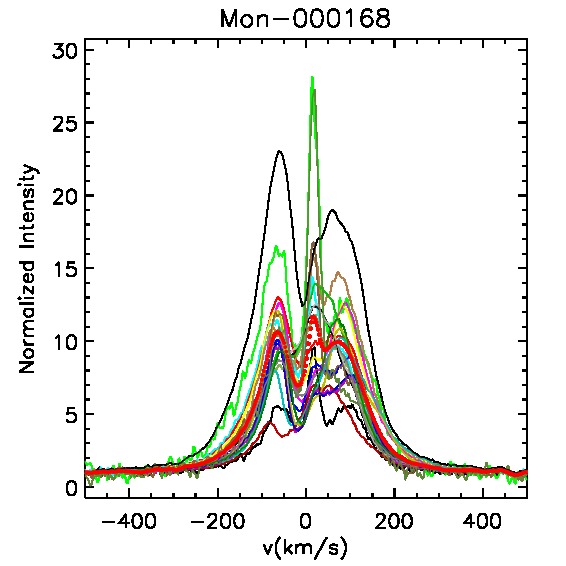}}
\subfigure[]{\includegraphics[width=4.4cm]{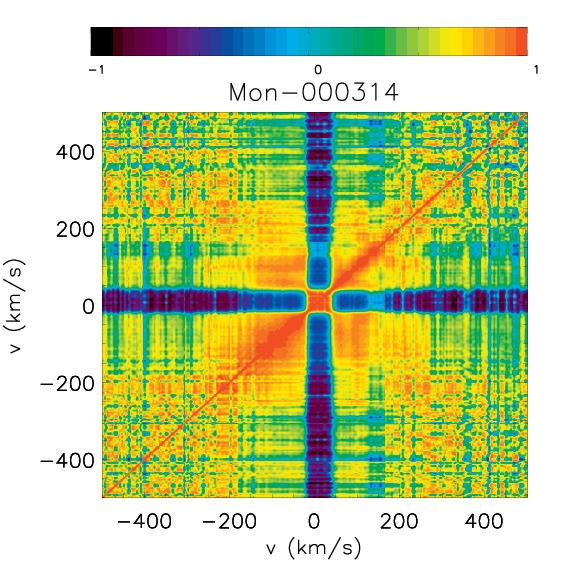}}
\subfigure[]{\includegraphics[width=4.4cm]{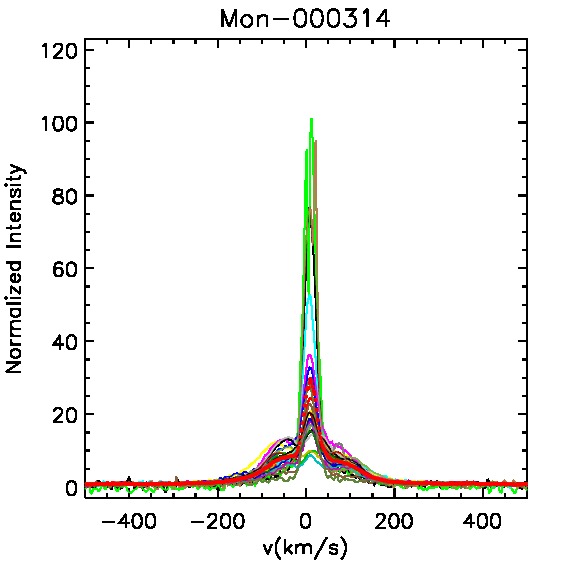}}
\subfigure[]{\includegraphics[width=4.4cm]{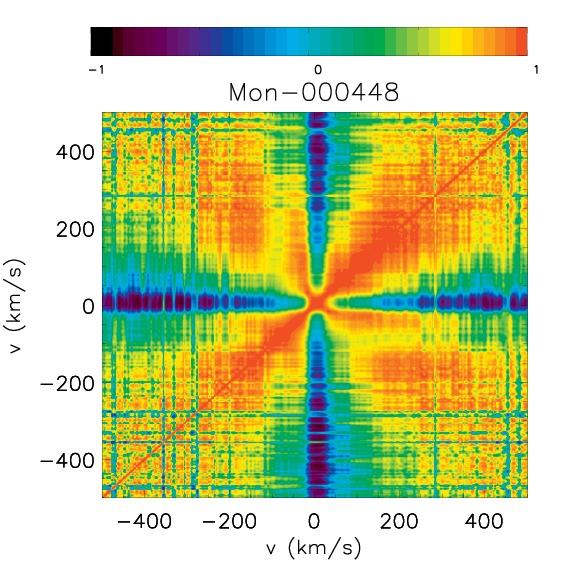}} 
\subfigure[]{\includegraphics[width=4.4cm]{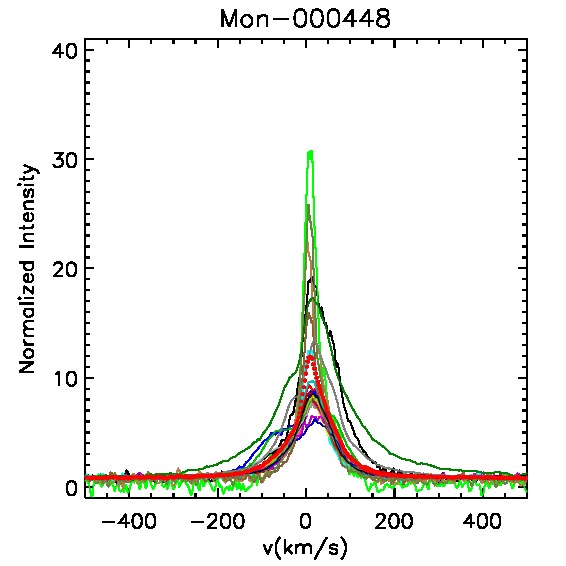}}
\end{center}
\caption{\label{fig:MatcHa22} 
Same as Fig. \ref{fig:MatcHa1}, but for stars that present strongly correlated
line profiles, as indicated by the positive correlation in almost the entire line profile. 
The narrow anticorrelation region  in the matrices of Mon-000632, Mon-001612, Mon-000119, 
Mon-000314, and Mon-000448 is due to the nebular contribution that was not entirely removed.}
\end{figure*}

\begin{figure*}[!ht]
 \begin{center}
\subfigure[]{\includegraphics[width=4.4cm]{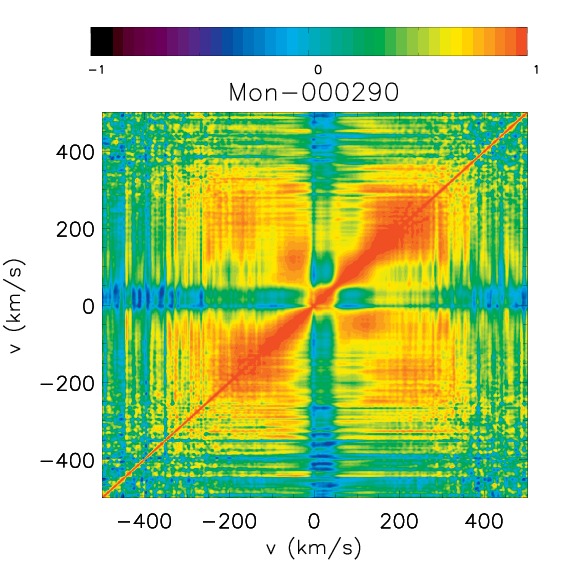}} 
\subfigure[]{\includegraphics[width=4.4cm]{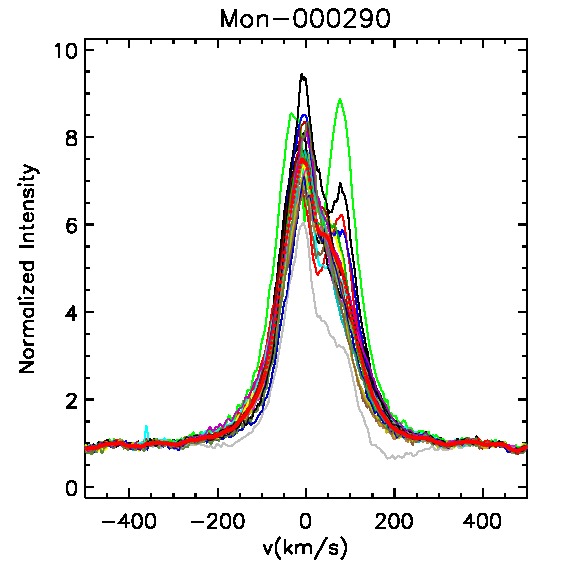}}
\subfigure[]{\includegraphics[width=4.4cm]{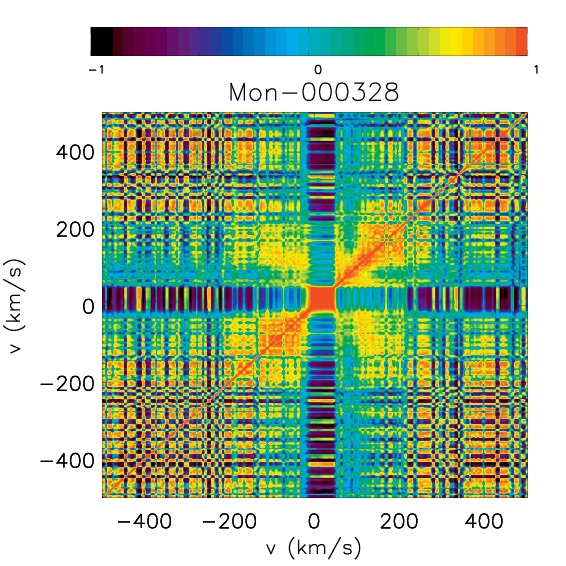}}
\subfigure[]{\includegraphics[width=4.4cm]{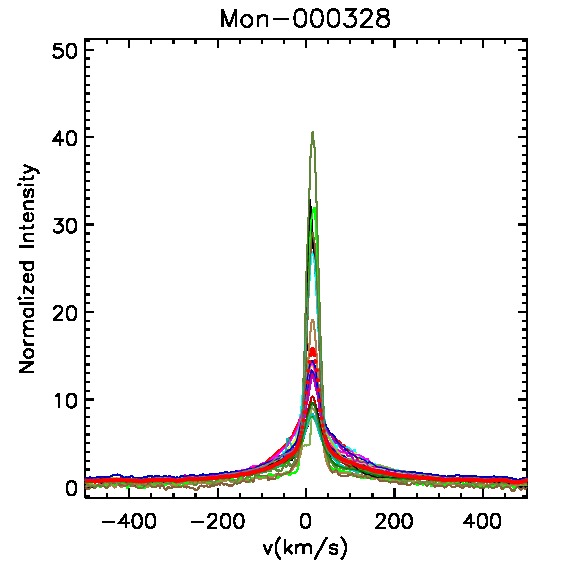}}
\subfigure[]{\includegraphics[width=4.4cm]{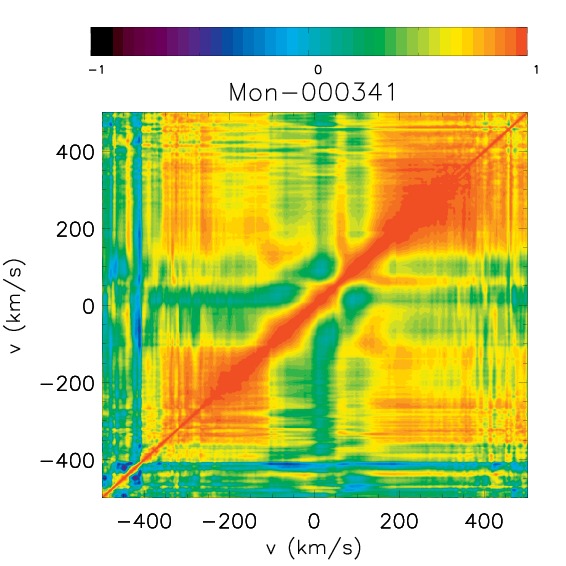}}
\subfigure[]{\includegraphics[width=4.4cm]{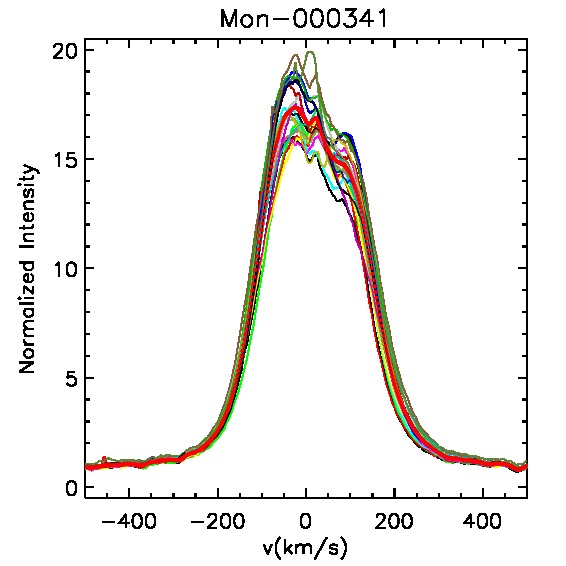}}
\subfigure[]{\includegraphics[width=4.4cm]{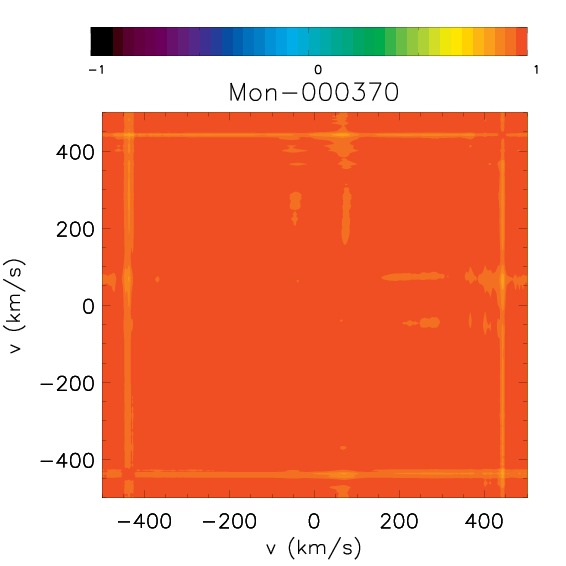}}
\subfigure[]{\includegraphics[width=4.4cm]{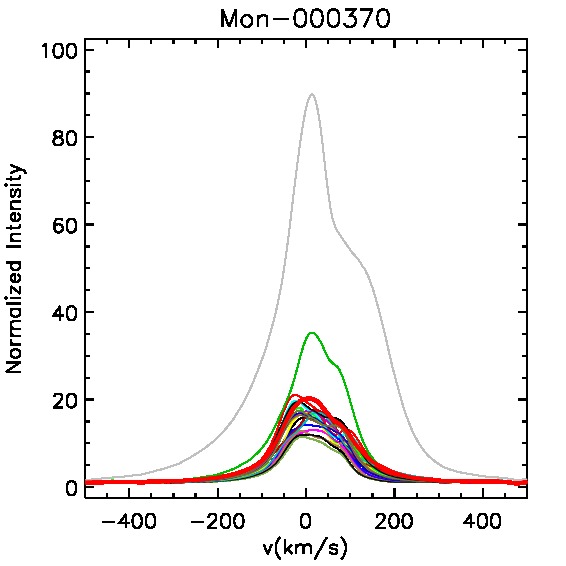}}
\subfigure[]{\includegraphics[width=4.4cm]{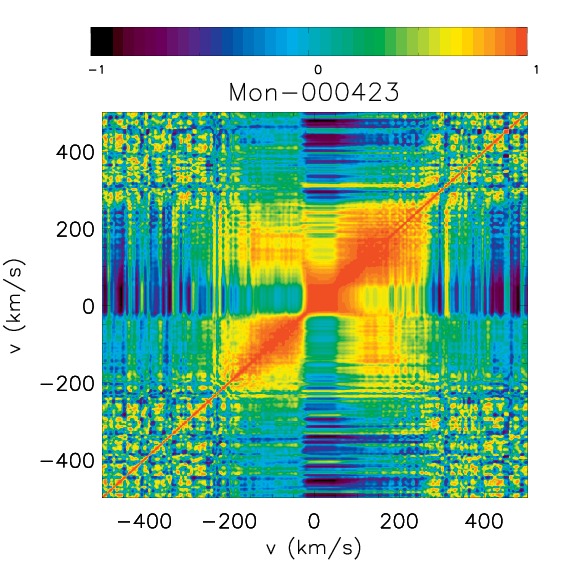}}
\subfigure[]{\includegraphics[width=4.4cm]{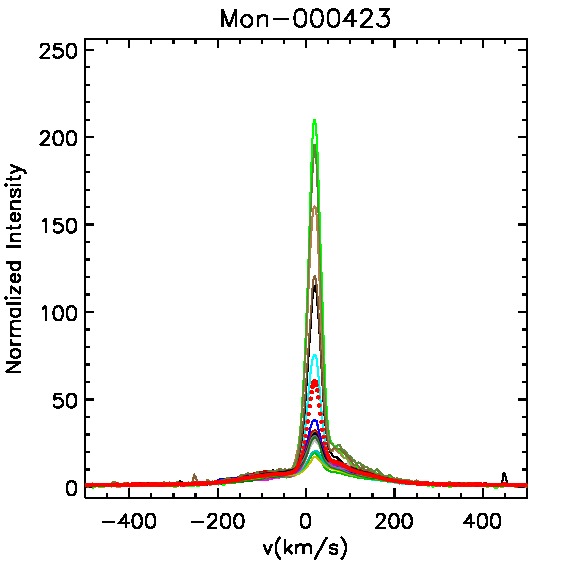}}
\subfigure[]{\includegraphics[width=4.4cm]{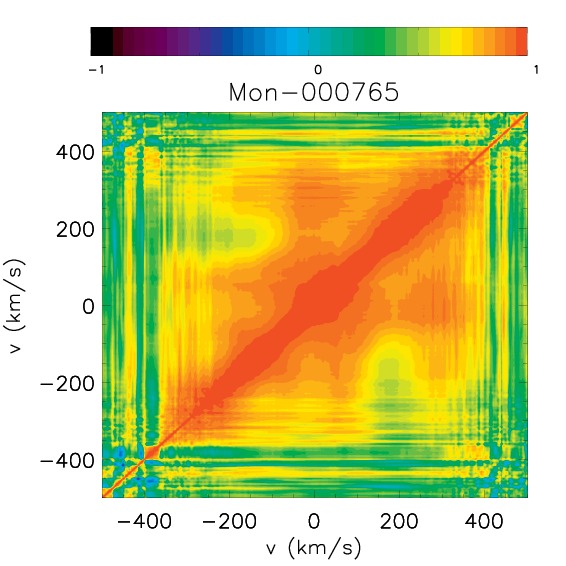}} 
\subfigure[]{\includegraphics[width=4.4cm]{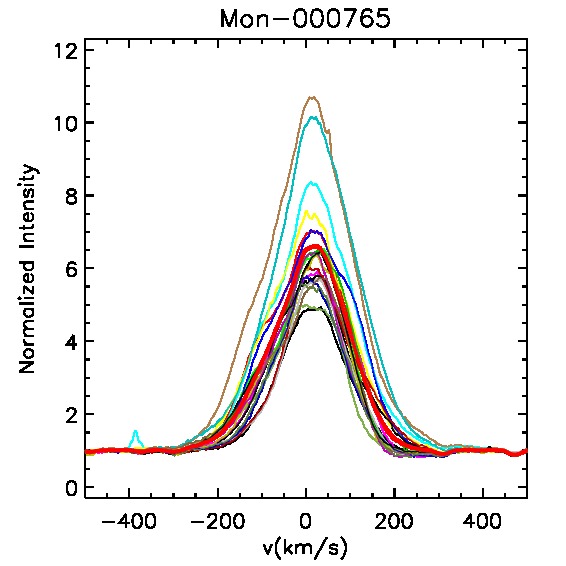}}
\subfigure[]{\includegraphics[width=4.4cm]{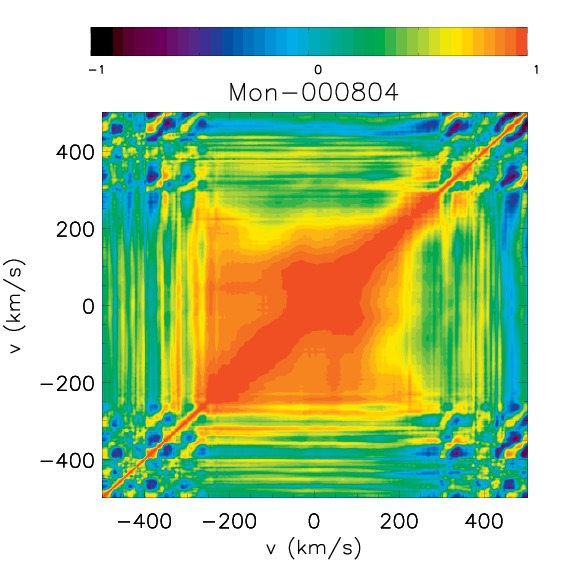}}
\subfigure[]{\includegraphics[width=4.4cm]{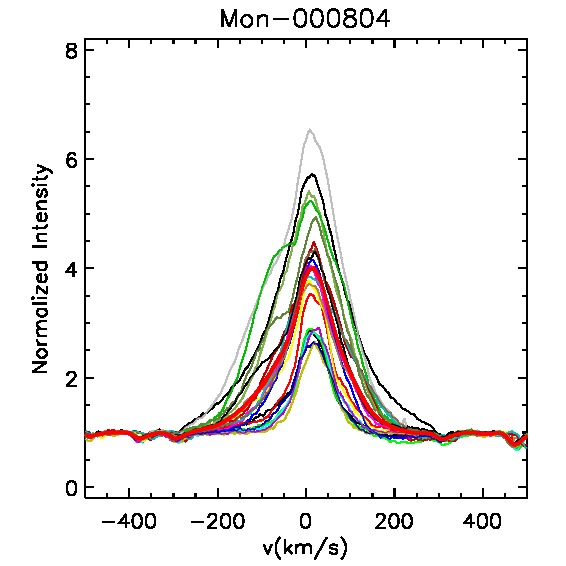}}
\subfigure[]{\includegraphics[width=4.4cm]{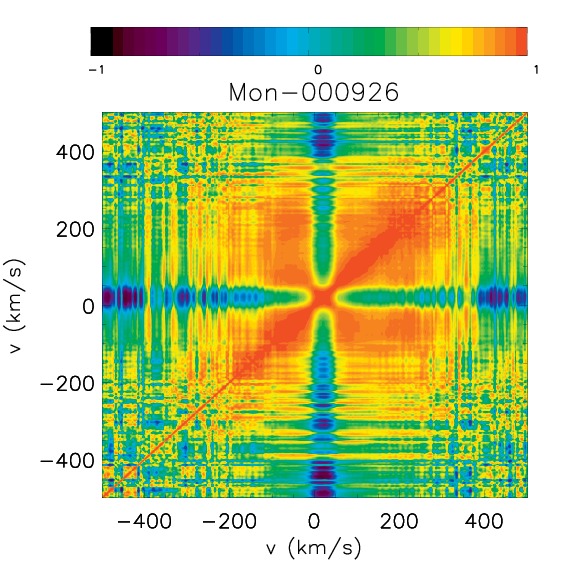}}
\subfigure[]{\includegraphics[width=4.4cm]{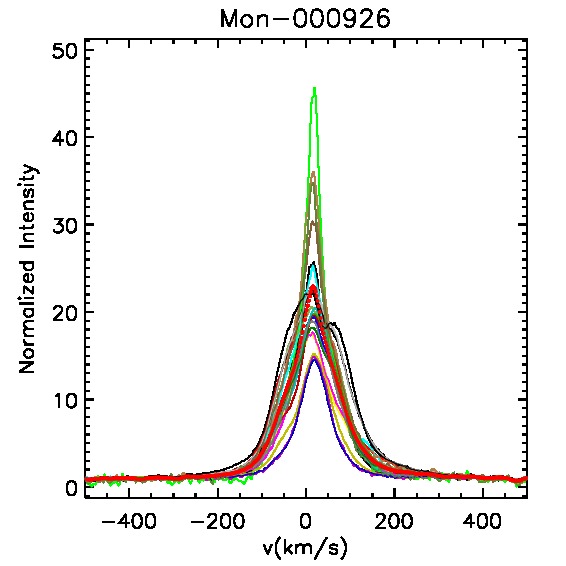}}
\subfigure[]{\includegraphics[width=4.4cm]{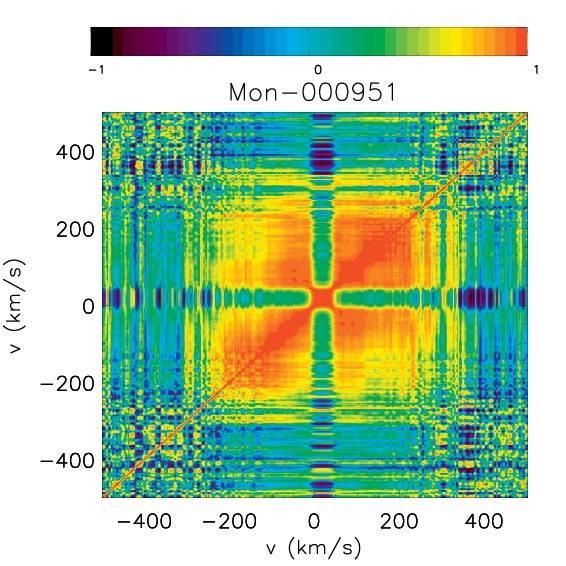}} 
\subfigure[]{\includegraphics[width=4.4cm]{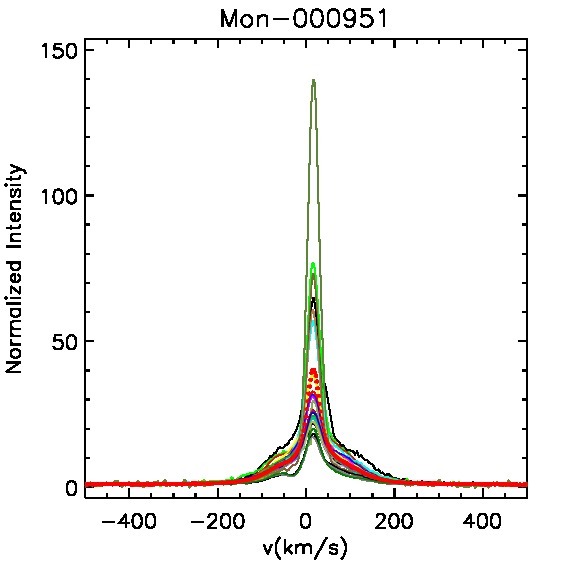}}
\subfigure[]{\includegraphics[width=4.4cm]{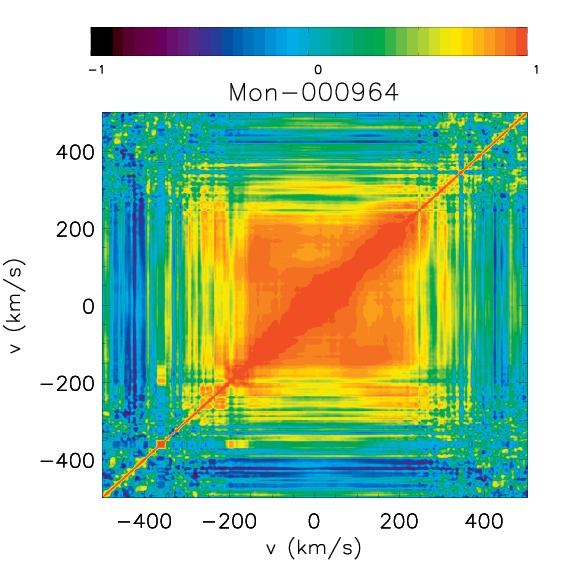}}
\subfigure[]{\includegraphics[width=4.4cm]{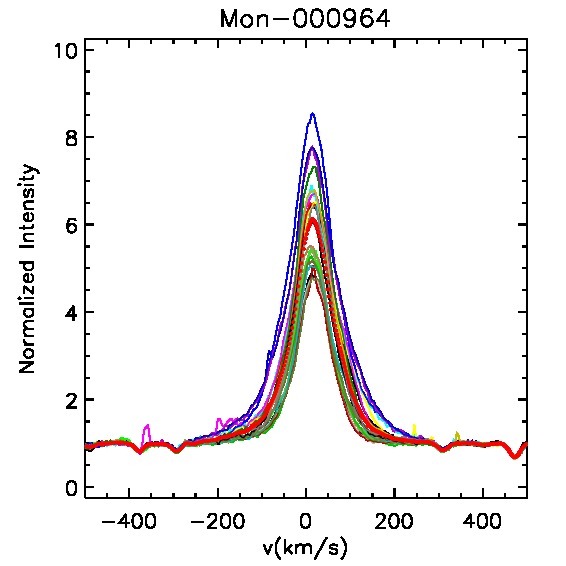}}
\end{center}
\caption{\label{fig:MatcHa3}
Same as Fig. \ref{fig:MatcHa22}}
\end{figure*}

\begin{figure*}[!ht]
 \begin{center}
\subfigure[]{\includegraphics[width=4.4cm]{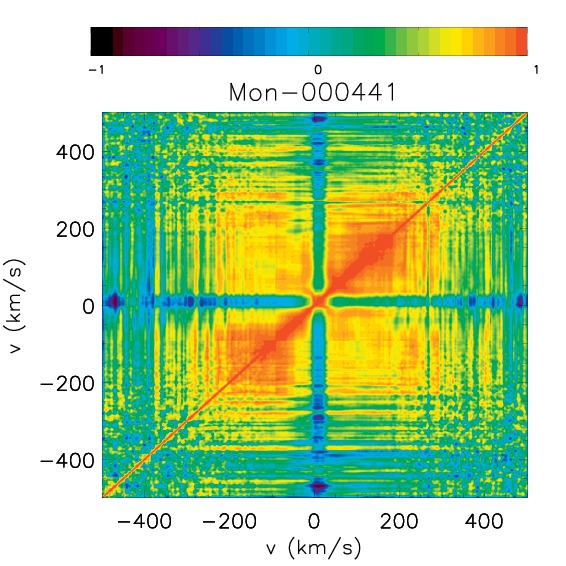}}
\subfigure[]{\includegraphics[width=4.4cm]{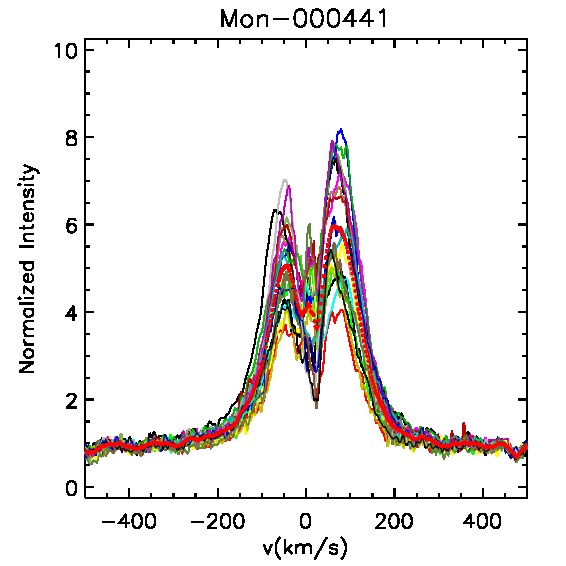}}
\subfigure[]{\includegraphics[width=4.4cm]{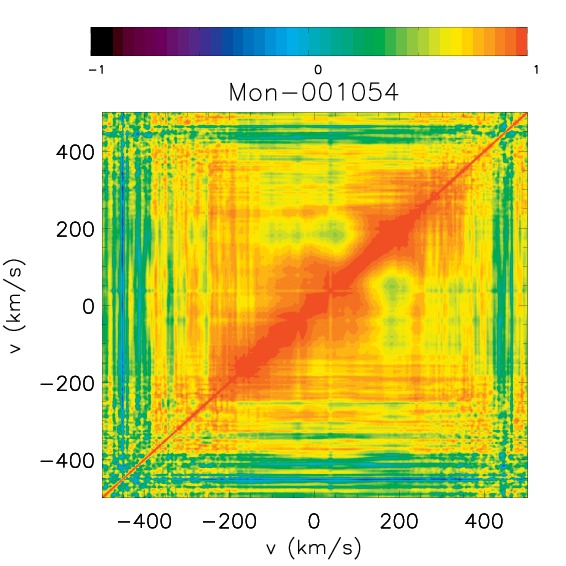}}
\subfigure[]{\includegraphics[width=4.4cm]{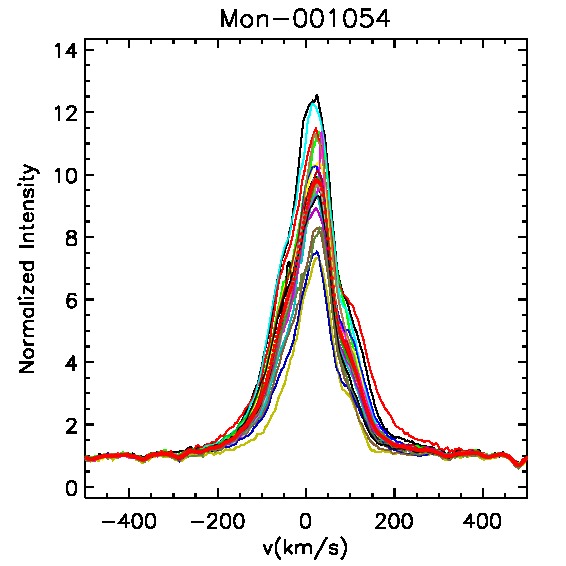}}
\subfigure[]{\includegraphics[width=4.4cm]{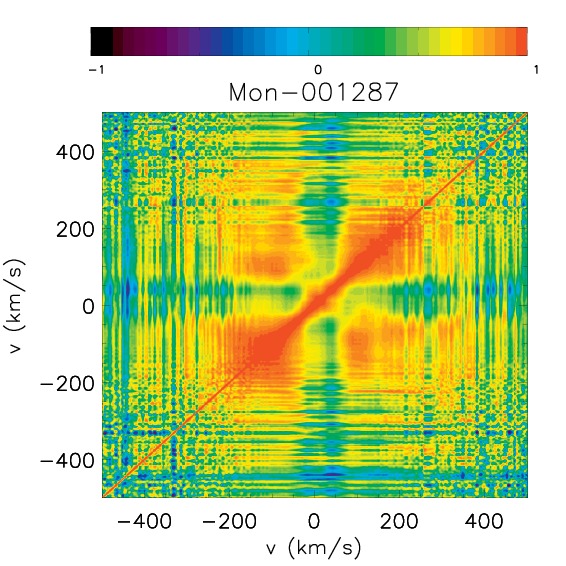}} 
\subfigure[]{\includegraphics[width=4.4cm]{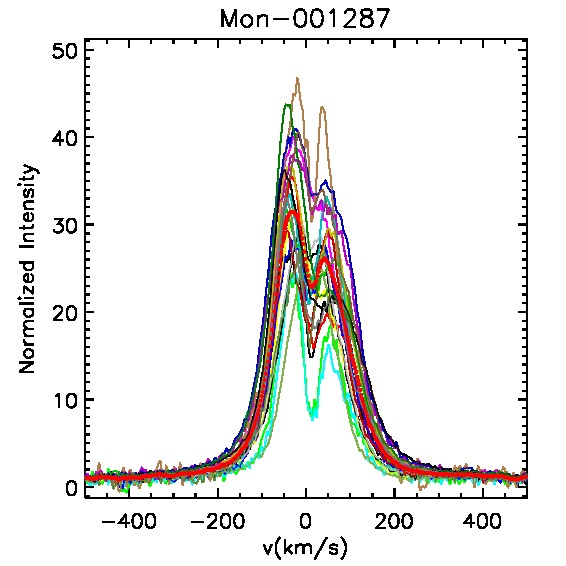}}
\end{center}
\caption{\label{fig:MatcHa4}
Same as Fig. \ref{fig:MatcHa22}}
\end{figure*}

\begin{figure*}[!ht]
 \begin{center}

\subfigure[]{\includegraphics[width=4.4cm]{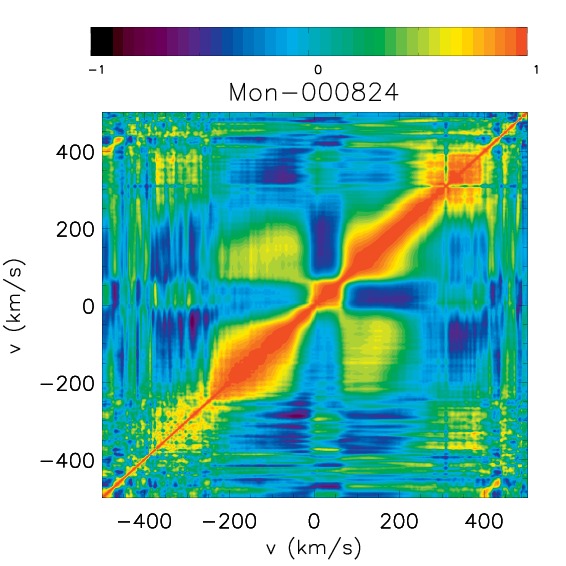}}
\subfigure[]{\includegraphics[width=4.4cm]{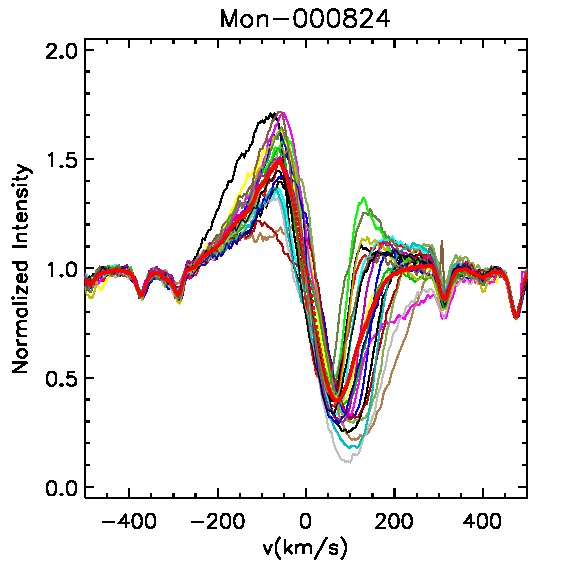}}
\subfigure[]{\includegraphics[width=4.4cm]{fig/mcHaskc_500007379}}
\subfigure[]{\includegraphics[width=4.4cm]{fig/Spec_500007379}}
\subfigure[]{\includegraphics[width=4.4cm]{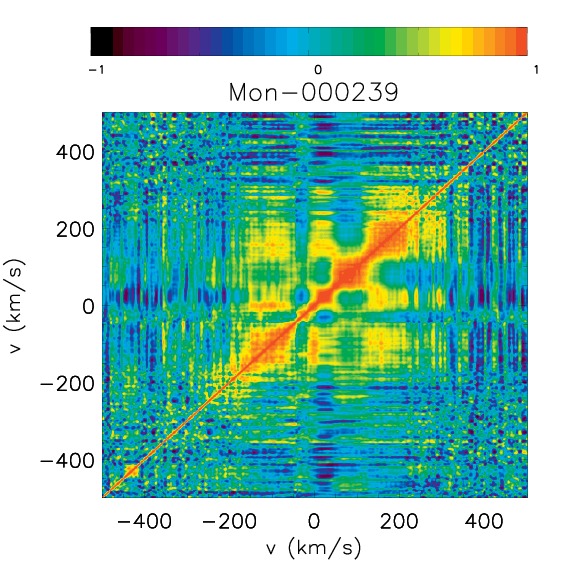}} 
\subfigure[]{\includegraphics[width=4.4cm]{fig/Spec_ngc2264067}}
\subfigure[]{\includegraphics[width=4.4cm]{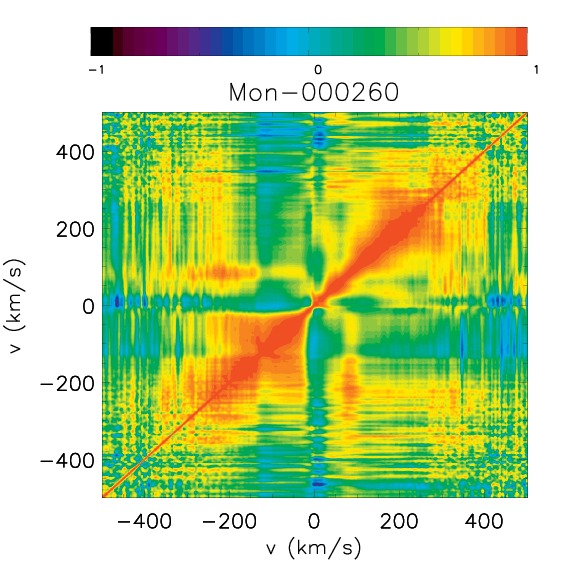}}
\subfigure[]{\includegraphics[width=4.4cm]{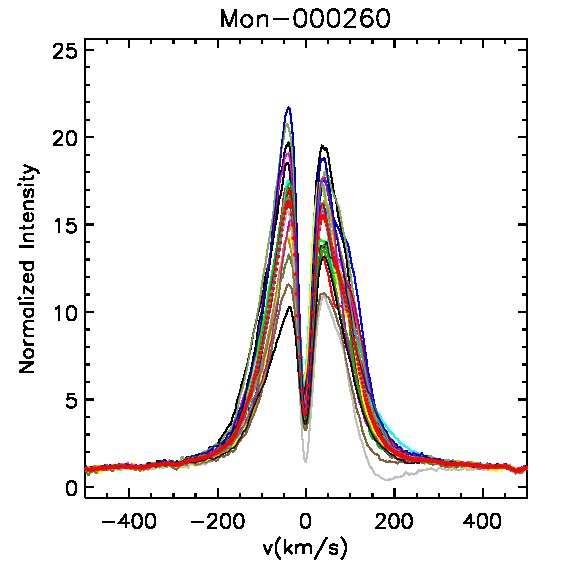}}
\subfigure[]{\includegraphics[width=4.4cm]{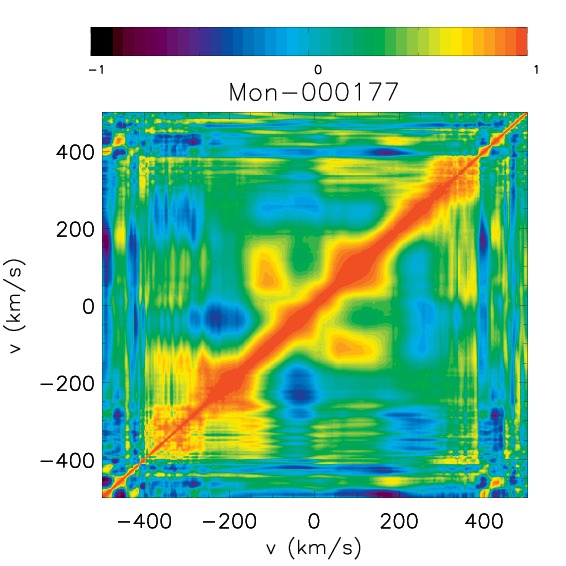}}
\subfigure[]{\includegraphics[width=4.4cm]{fig/Spec_223982136}}
\subfigure[]{\includegraphics[width=4.4cm]{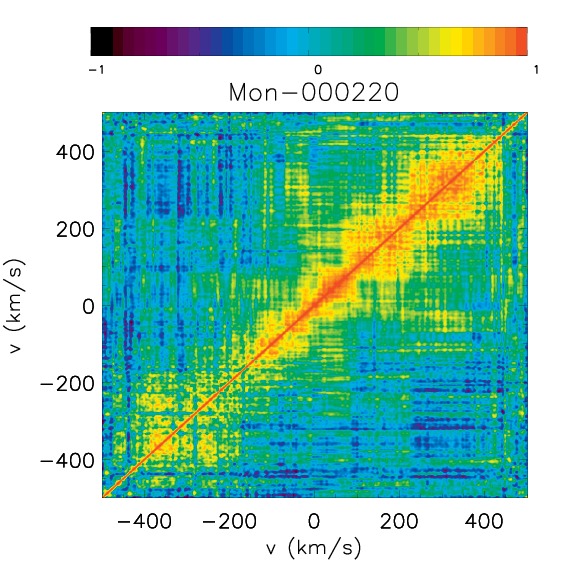}}
\subfigure[]{\includegraphics[width=4.4cm]{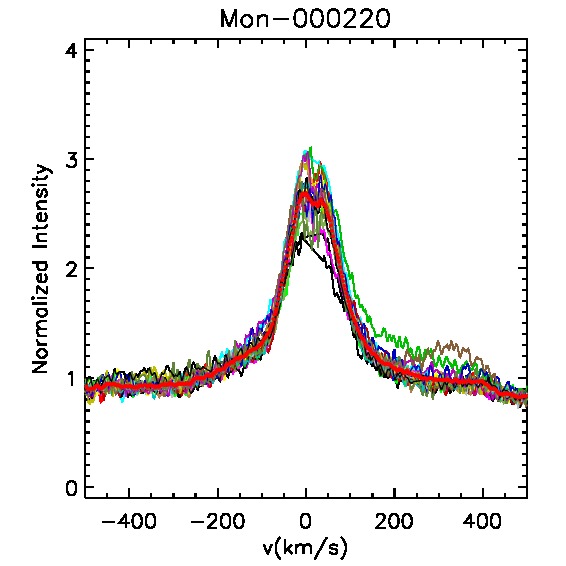}}
\subfigure[]{\includegraphics[width=4.4cm]{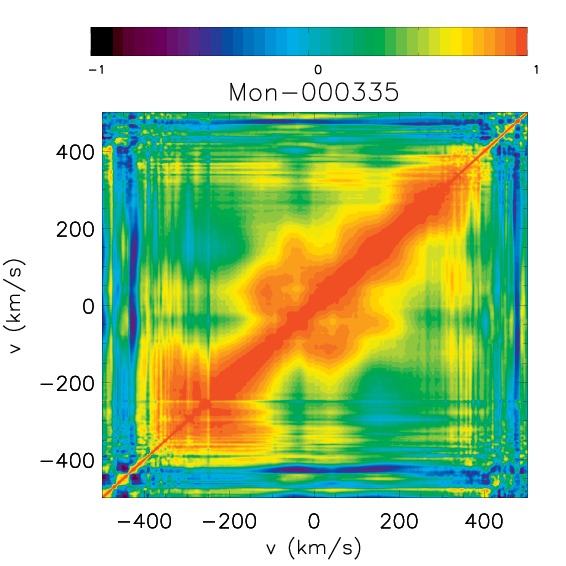}}
\subfigure[]{\includegraphics[width=4.4cm]{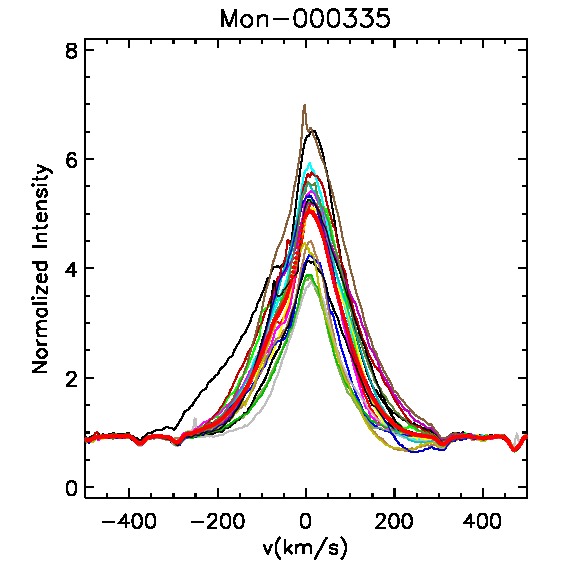}}
\subfigure[]{\includegraphics[width=4.4cm]{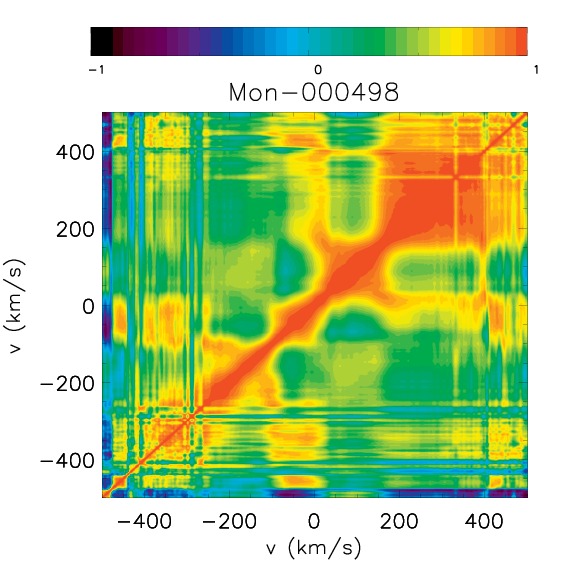}} 
\subfigure[]{\includegraphics[width=4.4cm]{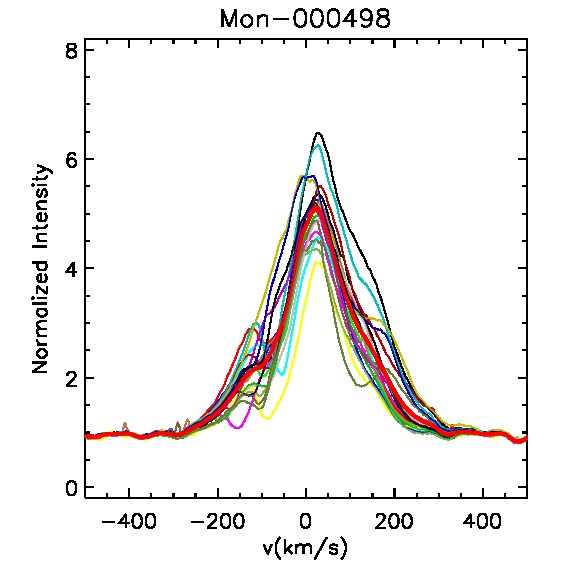}}
\subfigure[]{\includegraphics[width=4.4cm]{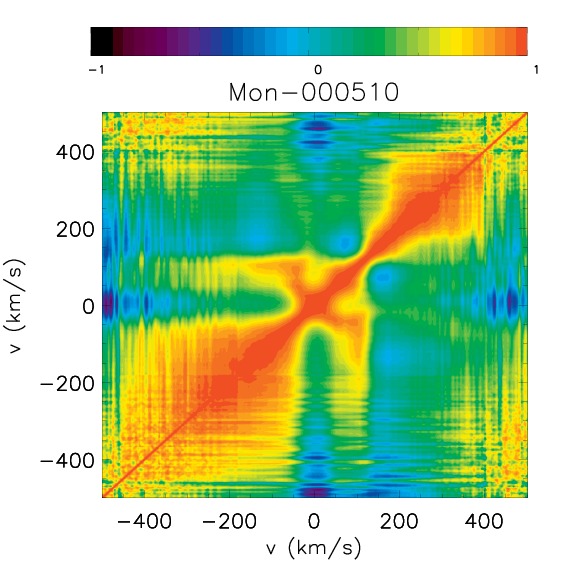}} 
\subfigure[]{\includegraphics[width=4.4cm]{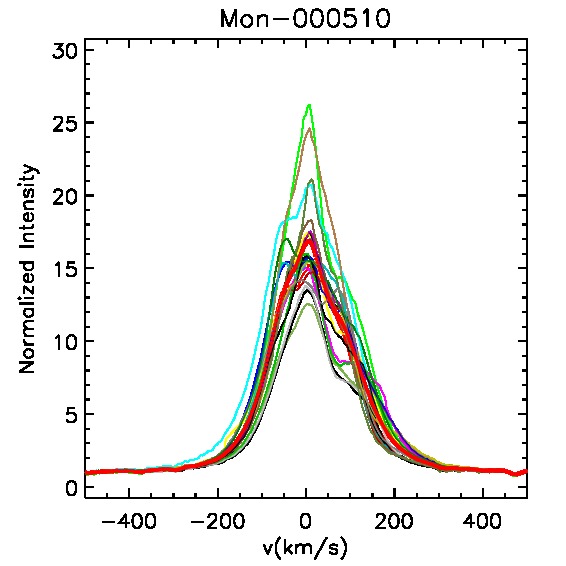}}
\subfigure[]{\includegraphics[width=4.4cm]{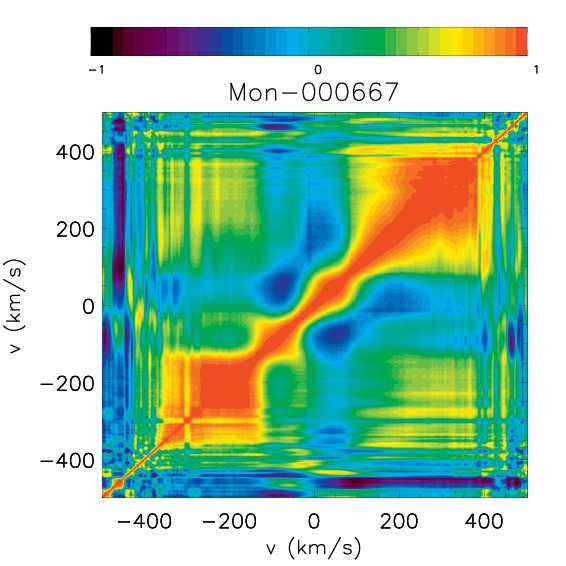}}
\subfigure[]{\includegraphics[width=4.4cm]{fig/Spec_223987997}}
\end{center}
\caption{\label{fig:MatcHa5}
Same as Fig. \ref{fig:MatcHa1}, but for stars that show no sign of correlation between
the red and blue wings of the $\mathrm{H}\alpha$ line profile. The narrow anticorrelation 
 in the matrices of Mon-000824, Mon-000177, and Mon-000667 is due to the nebular contribution 
that was not entirely removed.}
\end{figure*}

\begin{figure*}[!ht]
 \begin{center}
\subfigure[]{\includegraphics[width=4.4cm]{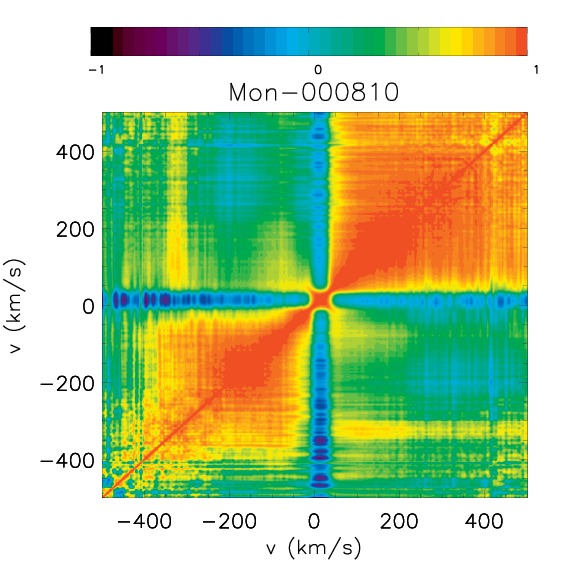}} 
\subfigure[]{\includegraphics[width=4.4cm]{fig/Spec_500007137}}
\subfigure[]{\includegraphics[width=4.4cm]{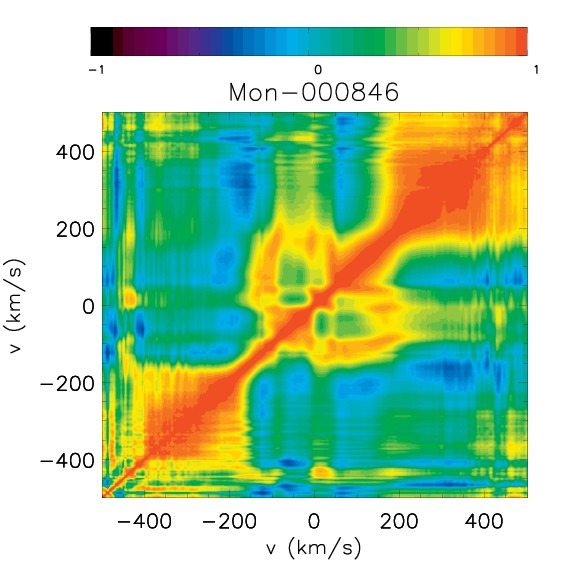}} 
\subfigure[]{\includegraphics[width=4.4cm]{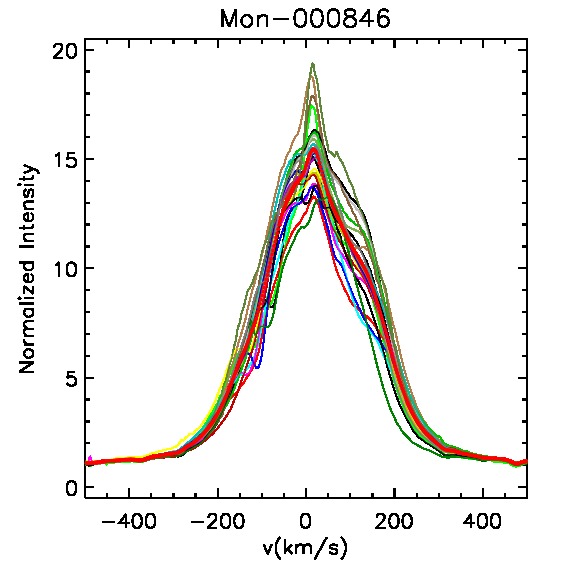}}
\subfigure[]{\includegraphics[width=4.4cm]{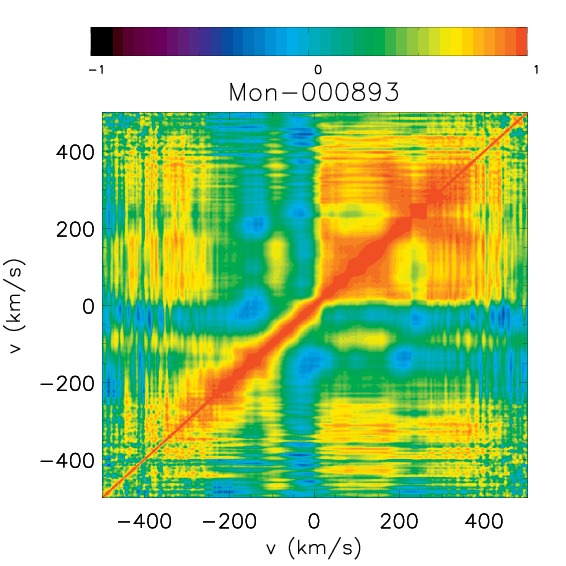}}
\subfigure[]{\includegraphics[width=4.4cm]{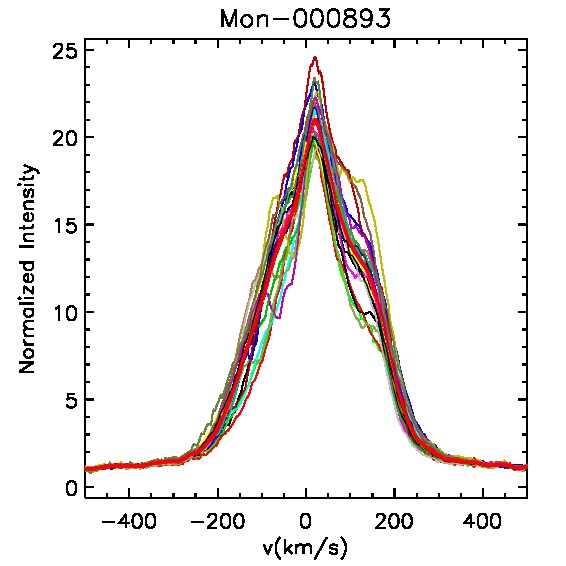}}
\subfigure[]{\includegraphics[width=4.4cm]{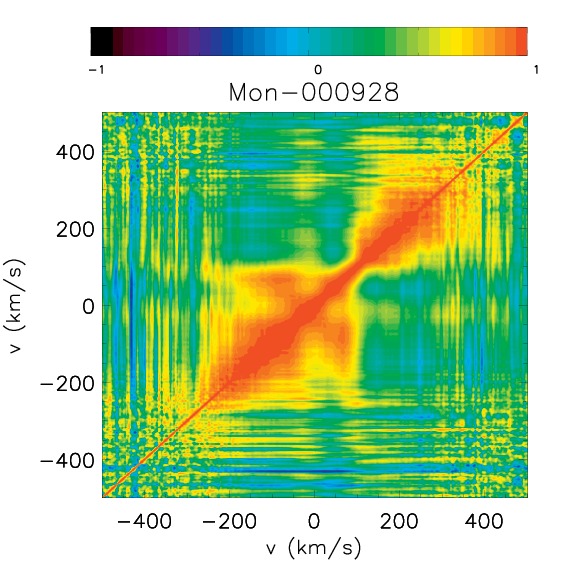}}
\subfigure[]{\includegraphics[width=4.4cm]{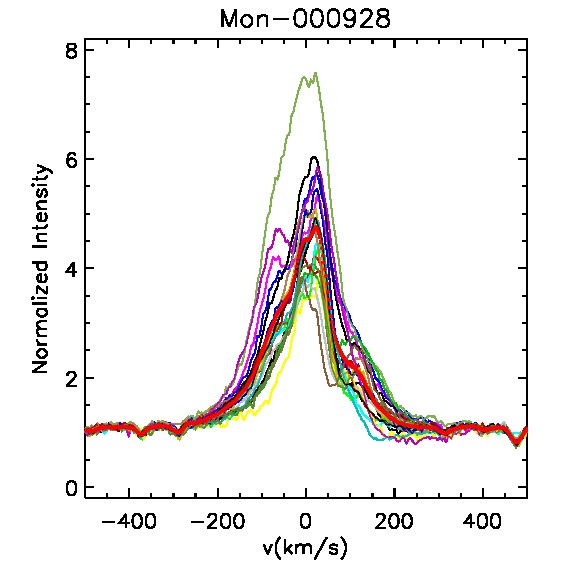}}
\subfigure[]{\includegraphics[width=4.4cm]{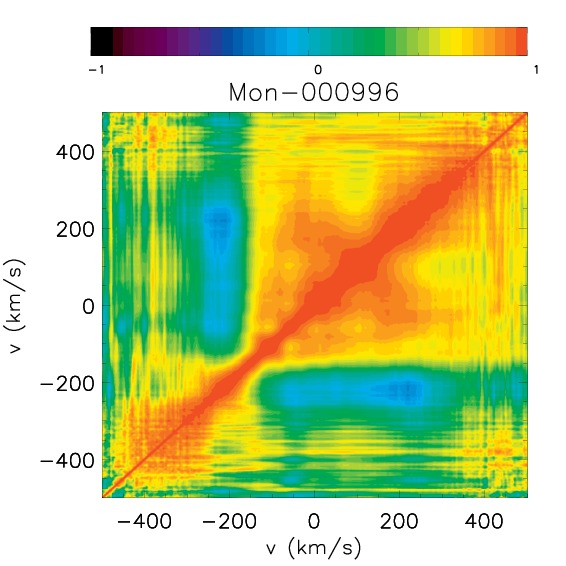}} 
\subfigure[]{\includegraphics[width=4.4cm]{fig/Spec_500007315}}
\subfigure[]{\includegraphics[width=4.4cm]{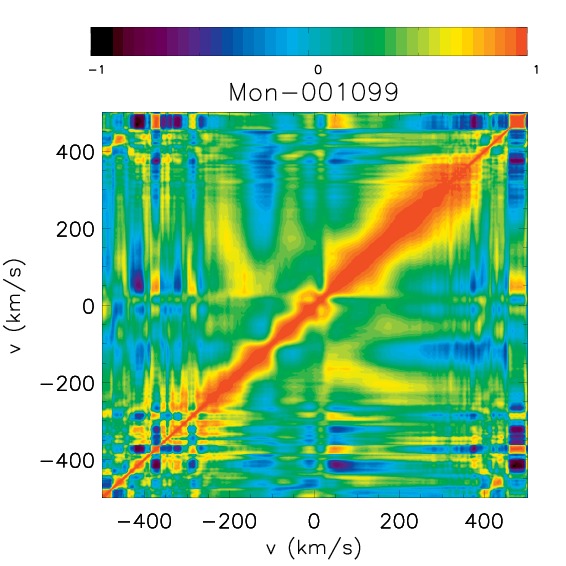}}
\subfigure[]{\includegraphics[width=4.4cm]{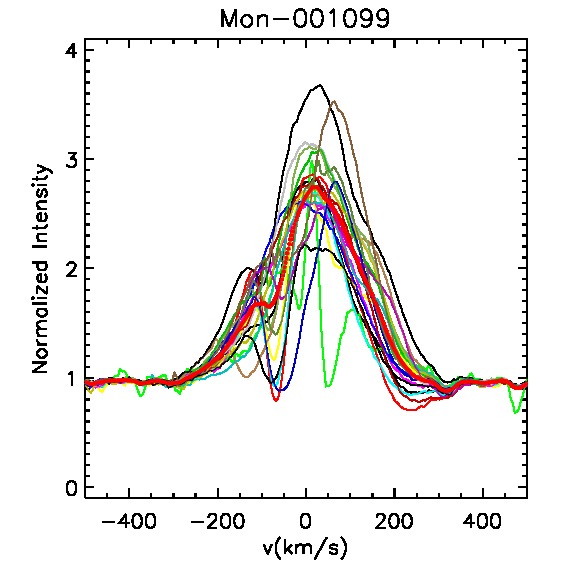}}
\subfigure[]{\includegraphics[width=4.4cm]{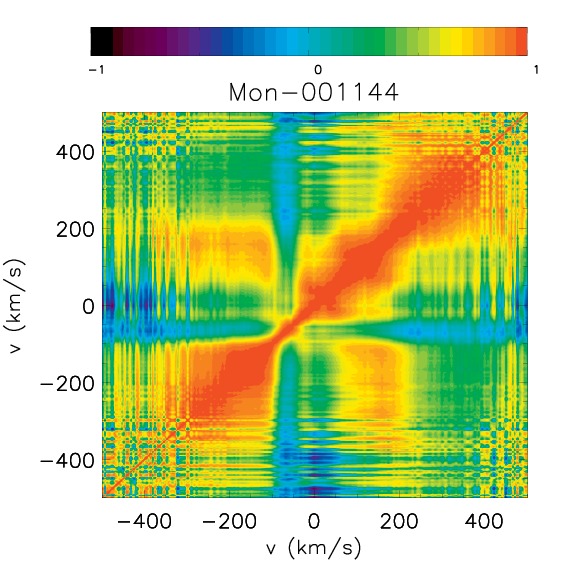}}
\subfigure[]{\includegraphics[width=4.4cm]{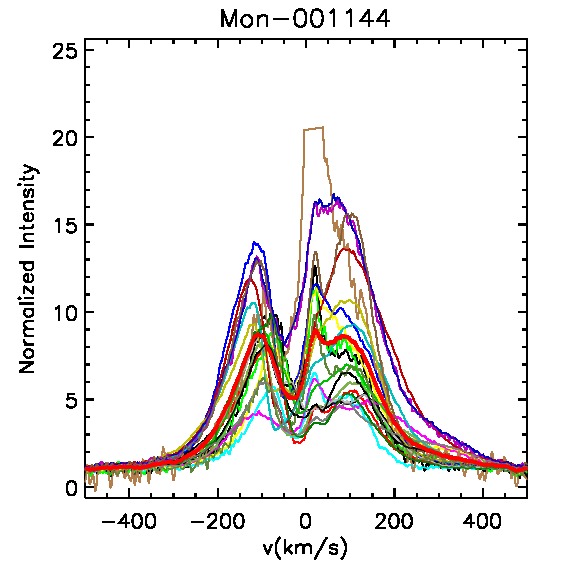}}
\subfigure[]{\includegraphics[width=4.4cm]{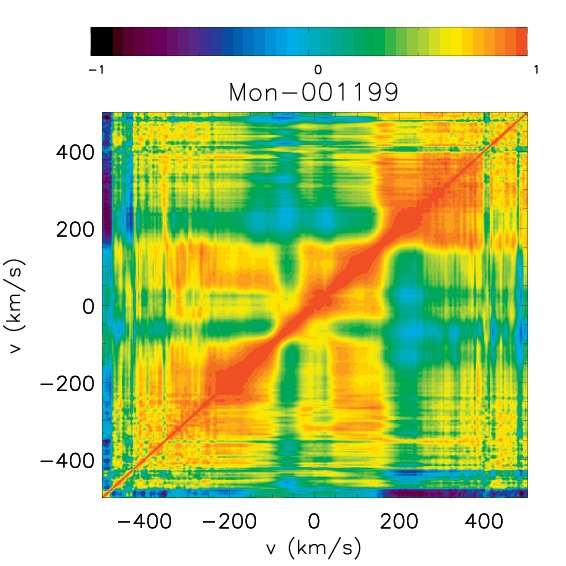}} 
\subfigure[]{\includegraphics[width=4.4cm]{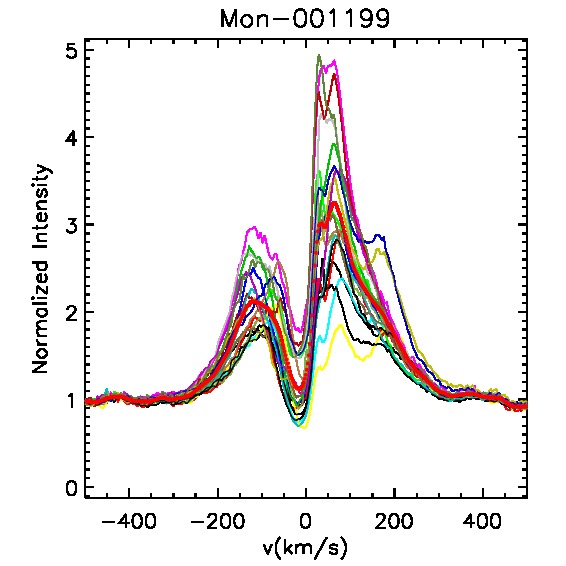}}
\subfigure[]{\includegraphics[width=4.4cm]{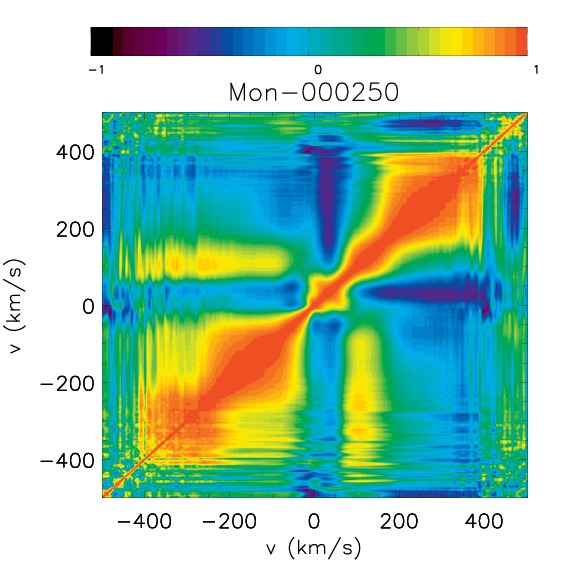}} 
\subfigure[]{\includegraphics[width=4.4cm]{fig/Spec_223980688}}
\subfigure[]{\includegraphics[width=4.4cm]{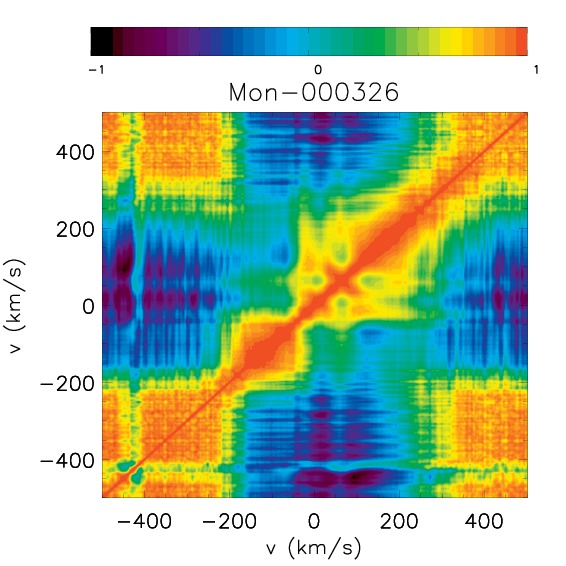}}
\subfigure[]{\includegraphics[width=4.4cm]{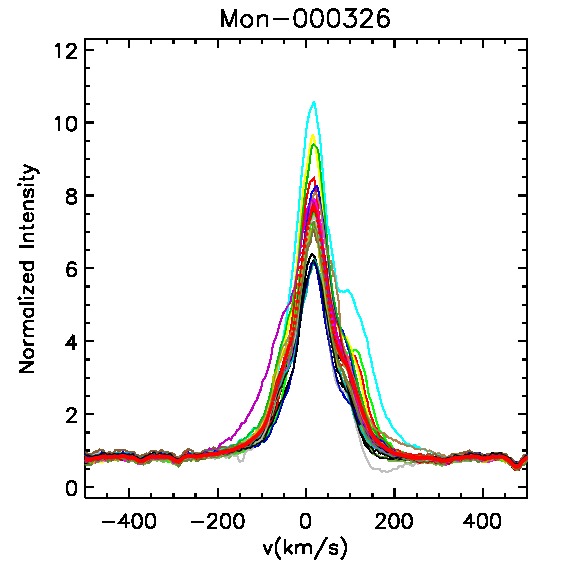}}
\end{center}
\caption{\label{fig:MatcHa6}
Same as Fig. \ref{fig:MatcHa5}. The anticorrelation in the matrice of Mon-000250 is due to the nebular contribution that was not entirely removed.}
\end{figure*}

 We present in this appendix the correlation matrices and $\mathrm{H}\alpha$ line profiles
of all the CTTSs we observed with the FLAMES spectrograph during the CSI 2264 campaign. 
We organized the correlation matrices according to the three main morphologies discussed 
in Sect \ref{sec:Matcorr}. Some correlations matrices are difficult
to fit in a single morphological class, and in these cases, we classified the star according
to the most representative feature present in the correlation matrix.
In Fig. \ref{fig:MatcHa1}, we grouped all the CTTSs that show anticorrelation 
in some part of the $\mathrm{H}\alpha$ line profile, like the stars in Fig. \ref{fig:MatcHa}. 
A few stars that were not included in this first group present correlation matrices 
with anticorrelation only in the narrow nebular emission region, near zero velocity. 
This is due to difficulties in removing the nebular contribution, and the
anticorrelation is not related to an emission or absorption region of the star-disk system.  
In Figs. \ref{fig:MatcHa22}, \ref{fig:MatcHa3} and \ref{fig:MatcHa4}, we organized the
stars that show correlation in almost all of the profile, like the matrix in Fig. \ref{fig:c}. 
Finally, in Figs. \ref{fig:MatcHa5} and \ref{fig:MatcHa6}, we present the stars that show no 
sign of correlation between the red and the blue wings in most of the profile, 
like the matrix in Fig. \ref{fig:d}.

\begin{figure*}[!ht]
 \begin{center}
\subfigure[]{\includegraphics[width=4.4cm]{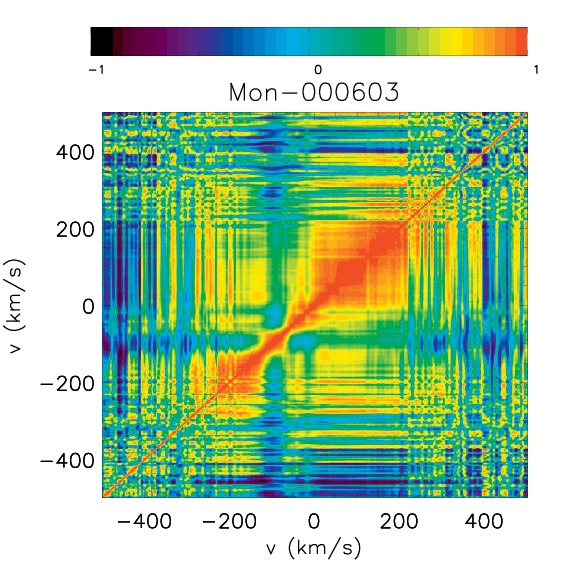}}
\subfigure[]{\includegraphics[width=4.4cm]{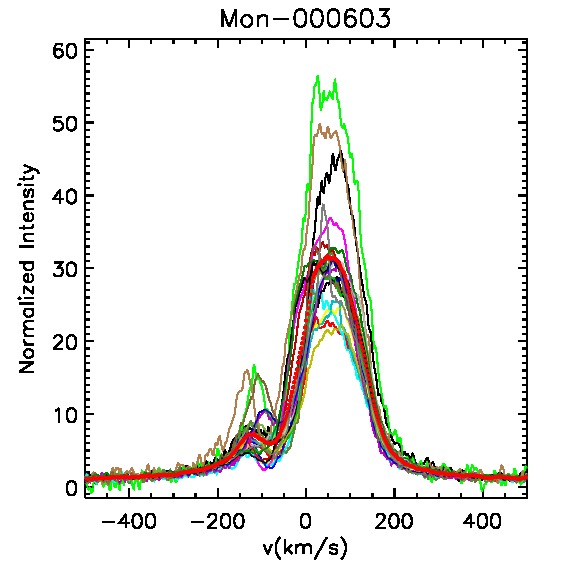}}
\subfigure[]{\includegraphics[width=4.4cm]{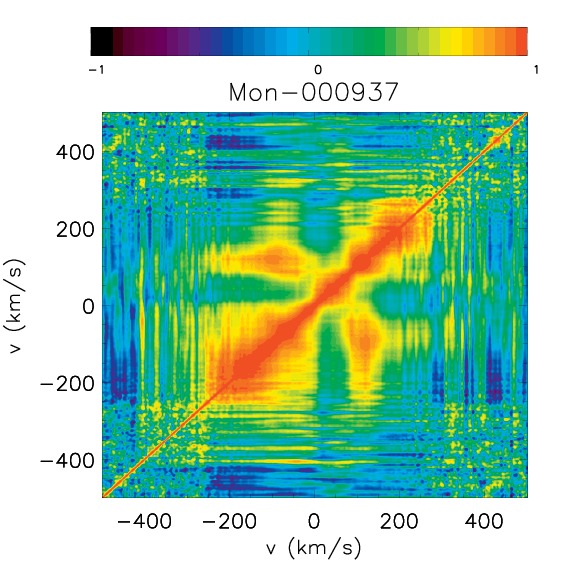}}
\subfigure[]{\includegraphics[width=4.4cm]{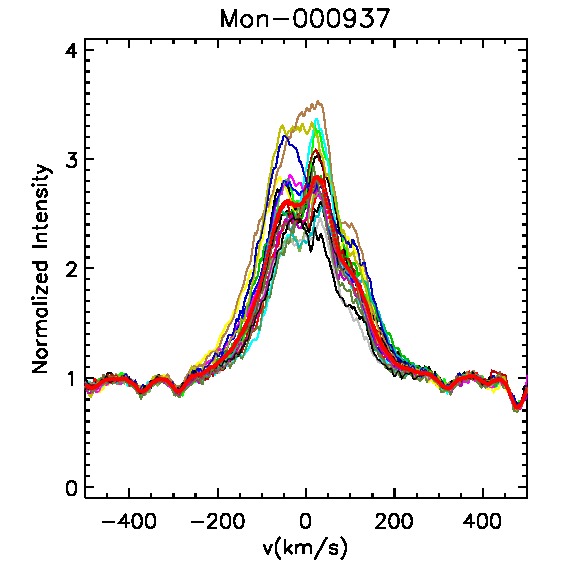}}
\subfigure[]{\includegraphics[width=4.4cm]{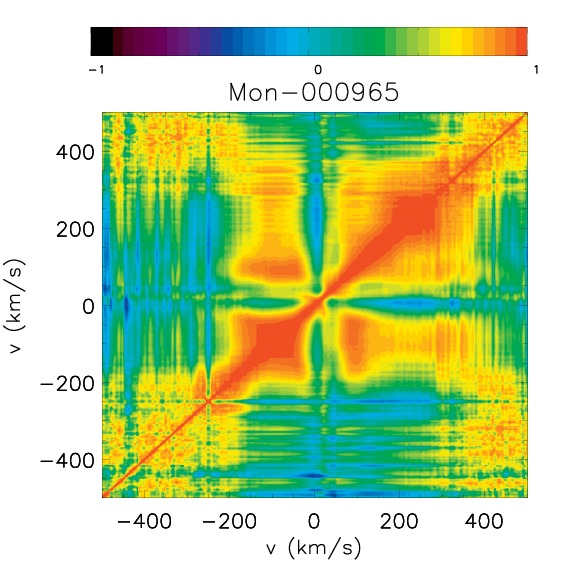}}
\subfigure[]{\includegraphics[width=4.4cm]{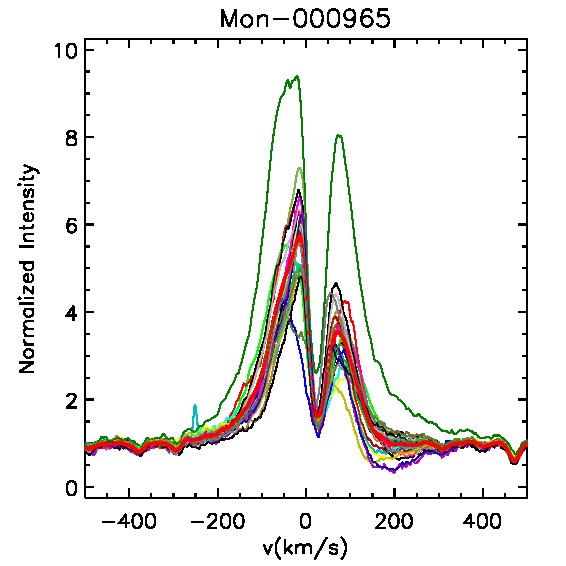}}
\subfigure[]{\includegraphics[width=4.4cm]{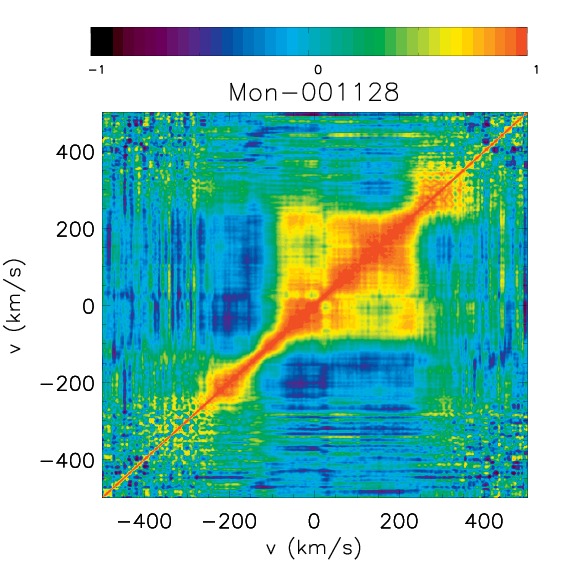}} 
\subfigure[]{\includegraphics[width=4.4cm]{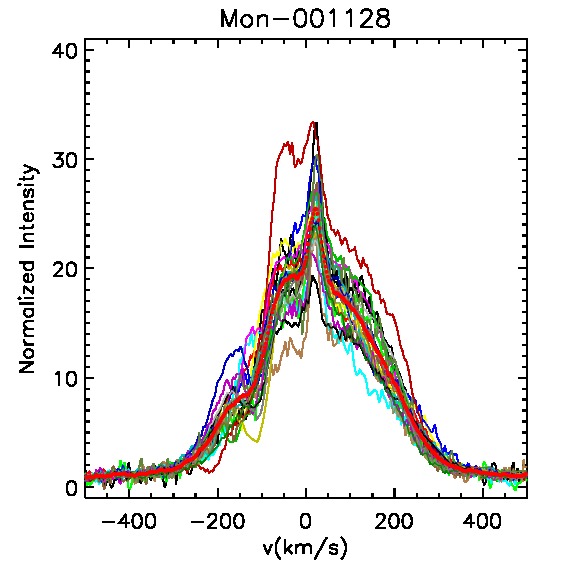}}
\subfigure[]{\includegraphics[width=4.4cm]{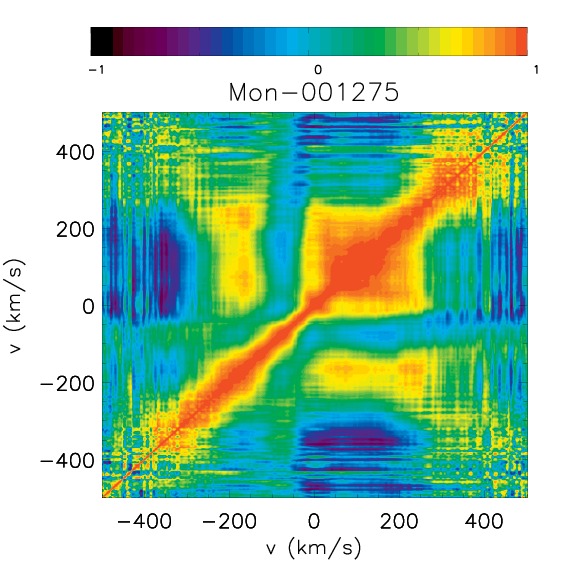}}
\subfigure[]{\includegraphics[width=4.4cm]{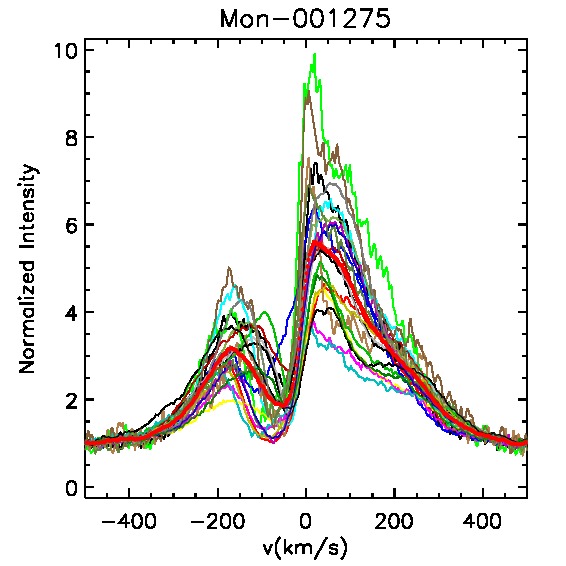}}
\subfigure[]{\includegraphics[width=4.4cm]{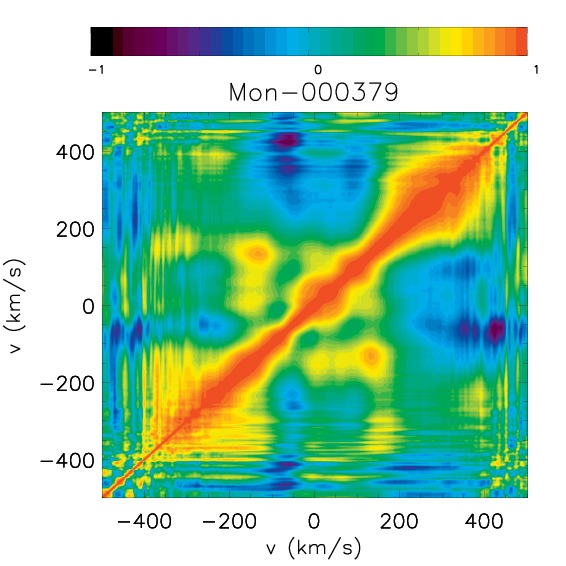}} 
\subfigure[]{\includegraphics[width=4.4cm]{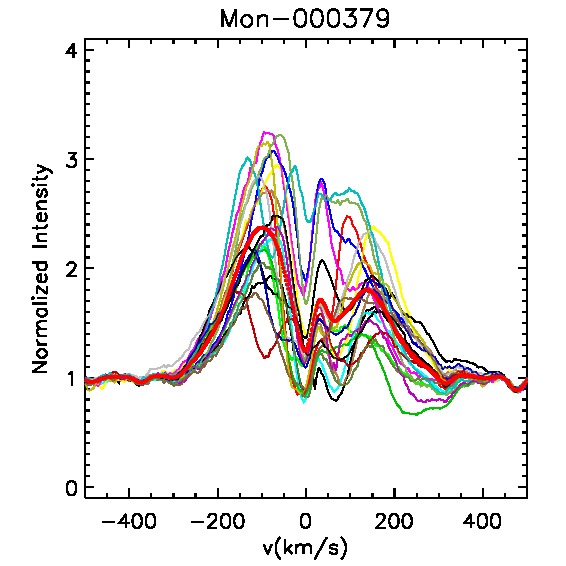}}
\subfigure[]{\includegraphics[width=4.4cm]{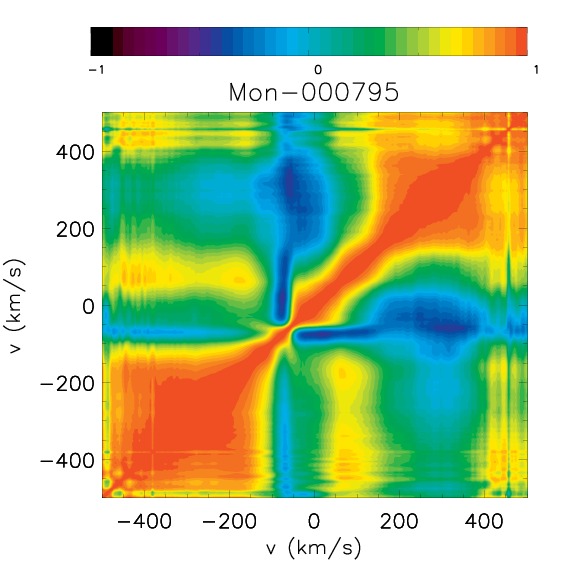}}
\subfigure[]{\includegraphics[width=4.4cm]{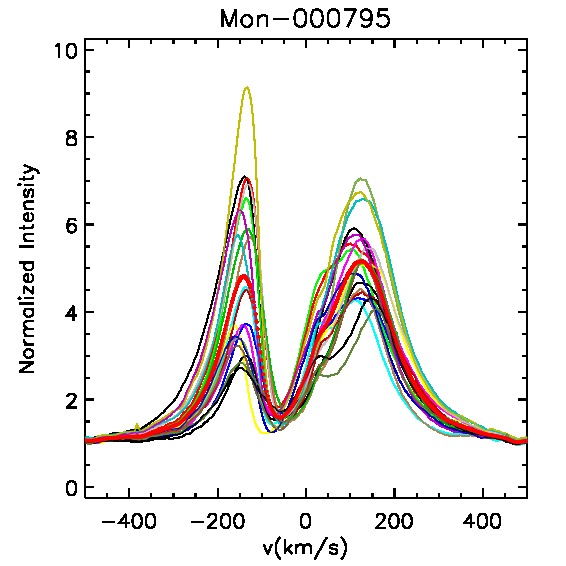}}
\subfigure[]{\includegraphics[width=4.4cm]{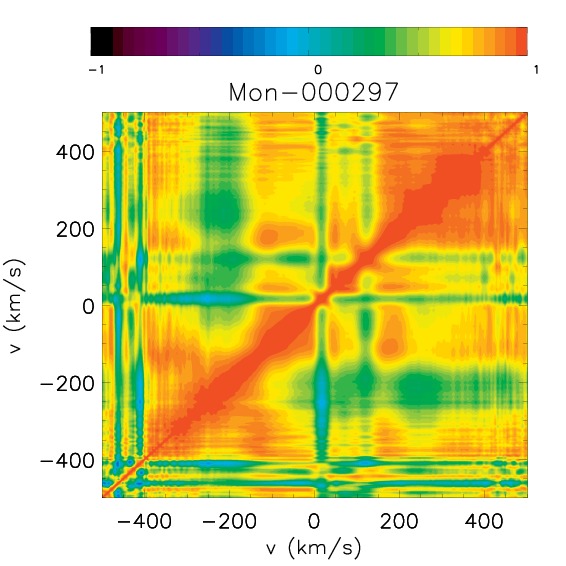}}
\subfigure[]{\includegraphics[width=4.4cm]{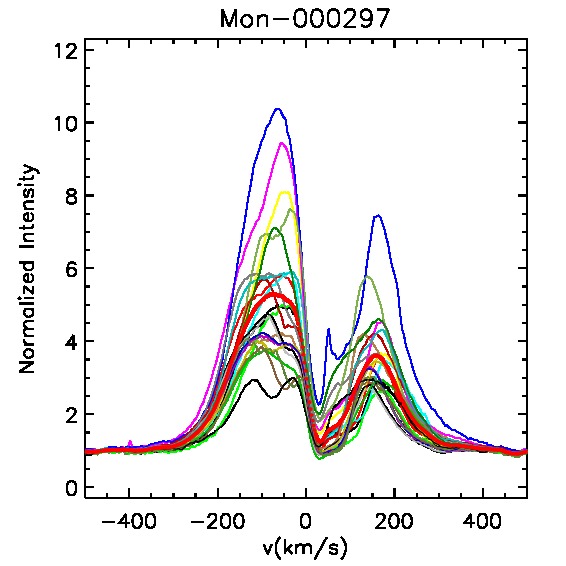}}
\end{center}
\caption{\label{fig:MatcHa7}
Same as Fig. \ref{fig:MatcHa5}. The anticorrelation in the matrice of Mon-000795 is due to the nebular contribution that was not entirely removed.}
\end{figure*}

\end{appendix}

\end{document}